\documentclass[aps,physrev,preprint,superscriptaddress]{revtex4-2}

\usepackage[utf8]{inputenc}
\usepackage[T1]{fontenc}
\usepackage{amsmath,amssymb}
\usepackage{bm}
\usepackage{graphicx}
\usepackage{booktabs}
\usepackage{enumitem}
\usepackage{subcaption}
\usepackage[dvipsnames]{xcolor}
\usepackage[normalem]{ulem}
\usepackage[colorlinks=true,linkcolor=blue,citecolor=teal,urlcolor=magenta]{hyperref}
\usepackage{braket}
\renewcommand{\arraystretch}{0.9}
\newcommand{\C}{\mathbb{C}}

\newcommand{\rr}{\mathbf{r}}
\newcommand{\one}{\mathbf{1}}
\newcommand{\zerov}{\mathbf{0}}
\newcommand{\dg}{\dagger}
\newcommand{\eff}{\mathrm{eff}}
\DeclareMathOperator{\diag}{diag}
\DeclareMathOperator{\Imag}{Im}
\DeclareMathOperator{\Realp}{Re}
\newcommand{\Bop}{\mathrm{Gram}}
\newcommand{\SY}{\mathrm{Spec}}
\newcommand{\SBf}{\mathrm{Herm}}
\newcommand{\BBop}{\mathrm{Gram}^{\mathrm{iso}}}
\newcommand{\SBBf}{\mathrm{Herm}^{\mathrm{iso}}}
\newcommand{\Ydiag}{Y_{\mathrm{diag}}}

\newlength{\panelW}
\newcommand{\pnl}[1]{\includegraphics[width=\panelW]{#1}}
\newcommand{\rlbl}[1]{\rotatebox{90}{\small\textbf{#1}}}
\newcommand{\lbl}[2]{\begin{minipage}{\panelW}\centering\pnl{#1}\\[-1pt]{\scriptsize #2}\end{minipage}}

\begin{document}

\title{Localization Landscapes based on the Green's Matrix of Non-Hermitian Optical Scattering Systems}

\author{Luca Dal Negro}
\email{dalnegro@bu.edu}
\affiliation{Department of Electrical \& Computer Engineering, Boston University}
\affiliation{Department of Physics, Boston University}
\affiliation{Division of Materials Science, Boston University}

\author{Hari Sundar}
\email{hari@cs.tufts.edu}
\affiliation{Department of Computer Science, Tufts University}

\date{\today}

\begin{abstract}
The Filoche--Mayboroda (FM) localization landscape predicts
where the spatially localized modes of a self-adjoint elliptic operator reside
without diagonalizing it, but the construction fails for the complex,
non-normal Green's matrix $G$ of an open scattering array of electric dipoles. We propose
three real-valued, positive landscapes built from the Feinberg--Zee
hermitization of $G$: a regularized Gramian inverse ($\Bop$), the
diagonal Schur block that carries the Metz--Neri--Rogers spectral
density ($\SY$), and the full hermitized inverse contracted against a
uniform source (TM) or an isotropic polarization pair (TE)
($\SBf$). The three form a
progression from spectrum-blind to fully hermitization-faithful, agree
in the limit $z\to 0$, and diverge where it matters: at nonzero
spectral shift $\SY$ resolves modes that $\Bop$ collapses, and $\SBf$
resolves modes that $\SY$ collapses. We demonstrate the progression on
random and deterministic aperiodic scattering arrays with a wide range of structural correlations, extend to the vectorial case of transverse electric polarization (TE), and provide rigorous pointwise envelope bounds for the Spec and Herm landscapes, a physical argument for the spatial compactness of long-lived modes, and a complete numerical recipe in the appendices.
\end{abstract}

\maketitle


\section{Introduction}

The localization of optical waves in non-periodic scattering media with open boundaries is a central problem in both quantum and classical physics with numerous technological applications ranging from the engineering of energy-efficient semiconductor devices with disordered potentials to novel photonic and phononic nanostructures, to metamaterials with tailored spectral singularities \cite{arnold2019computing,arnold2016effective,filoche2012universal,filoche2017landscape,lagendijk1996resonant,shubitidze2024localization,skipetrov2018finite,skipetrov2024higher}. In fact, the rich behavior of open scattering systems described by non-Hermitian (NH) and strongly non-normal (NN) operators has only recently begun to show its full potential for the manipulation of resonant wave transport, localization effects, thermalization phenomena, and for the engineering of non-trivial topological phases of matter \cite{ashida2020nonhermitian,balasubramanian2020many,GueryOdelin2026}. All these scenarios require the precise understanding of the shape and spatial confinement of the structural resonances of NH and NN systems, motivating the development of a unified predictive framework that enables the efficient design of open electromagnetic systems with arbitrary spatial correlations. 

The localization landscape theory, originally introduced by Filoche and Mayboroda \cite{filoche2012universal} and further developed in subsequent work \cite{arnold2016effective,arnold2019computing}, established a powerful mathematical framework for the study of the quantum eigenmodes and electron localization phenomena based on the landscape function $u$. In particular, it was shown that for a self-adjoint, positive-definite elliptic operator $H$, the
Filoche--Mayboroda construction~\cite{filoche2012universal} replaces
eigenvalue computations by a single inhomogeneous solution of $Hu=\one$,
and identifies mode location from the valleys of the reciprocal
effective potential $W_\eff=1/u$~\cite{arnold2016effective,
arnold2019computing,filoche2017landscape,piccardo2017landscape,
sridhar1992experimental}. 
This theory offers an exceptionally efficient geometric perspective to investigate eigenfunctions and Anderson localization phenomena in wave problems described by stationary elliptic operators that are positive and Hermitian, such as Schr\"{o}dinger operator $L=-\nabla^{2}+V$ with disordered potential $V$, which naturally produce a unique and positive landscape \cite{filoche2012universal,arnold2016effective,skipetrov2024higher}. 

More recently, the localization landscape theory has been extended to magnetic Schr\"{o}dinger operators with singular potentials \cite{Poggi2024,Mugnolo2024}, interacting many-body systems \cite{balasubramanian2020many}, optical waves in complex media \cite{shubitidze2024localization}, and attempts have also been made to capture quantum systems beyond static, elliptic, and Hermitian settings \cite{GueryOdelin2026}. 
Unfortunately, these constructions cannot be applied to the electromagnetic vector problems that are described, to all orders of multiple scattering, by the
coupled-dipole Green's matrix $G$ that is directly related to the effective Hamiltonian of the electromagnetic system \cite{goetschy2011lumiere} and enables the efficient design of large-size, non-periodic, open scattering
arrays for photonics and metamaterial technologies~\cite{lagendijk1996resonant,rusek1996localization,Rusek2000,
skipetrov2014absence,skipetrov2018finite,sgrignuoli2019localization,sgrignuoli2020subdiffusive,sgrignuoli2021hyperuniformity,sgrignuoli2022subdiffusive,monsarrat2022pseudogap}. In fact, in these situations, a localization landscape directly introduced through the Green's matrix as $Gu=\one$ would result in a complex
solution, left and right eigenvectors of $G$
will differ~\cite{chalker1998eigenvector}, and the pseudospectrum of $G$
would matter as much as its spectrum~\cite{trefethen2005spectra}. 
Moreover, an electromagnetic landscape that predicts the spatial locations of localized optical waves in a system of coupled resonators is normally introduced based on the Hamiltonian operator of the system, which in fact corresponds to a suitably regularized form of $G^{-1}$.
Therefore, this scenario is naturally conducive to the Feinberg--Zee hermitization
of a quantum Green's function embedded into a $2N{\times}2N$ Hermitian-structured matrix $B^{-1}(z,\eta)$
whose Schur blocks fully encode the pseudospectral response of $G-zI$ \cite{feinberg1997nonhermitian,metz2019spectral,
ashida2020nonhermitian,goetschy2013euclidean}.

In this paper, we build on this powerful approach by considering the non-Hermitian Green's matrix $G$ of planar dipolar arrays and we use the Schur algebra of $B^{-1}(z,\eta)$ to construct real-valued positive
landscapes that inherit the FM interpretation. In particular, we consider the Tikhonov-regularized Gramian inverse of the $G$ matrix (referred to as ``$\Bop$''),
the diagonal Schur block of the Hermitized inverse of the $G$ matrix (referred to as ``$\SY$''), which
carries the Metz--Neri--Rogers spectral density, and finally the
full Feinberg-Zee hermitized inverse (referred to as ``$\SBf$'') contracted against a polarization-isotropic
source. When the electric dipoles are oriented perpendicular to the plane of the array, i.e., for the transverse magnetic (TM) polarization, the three landscapes that we introduced form a progression from computationally inexpensive and
spectrum-blind to spectrally aware. Moreover, each
successive landscape resolves finer structure of the 
long-lived modes. When the dipoles are oriented in the plane of the array, i.e., for the transverse electric (TE) case, the source-driven $\Bop$ and $\SBf$ landscapes acquire the
polarization-isotropic variants $\BBop,\SBBf$, while the spectrally aware
$\SY$ is polarization-balanced by construction. The rest of the paper
discusses the three landscapes precisely, demonstrates the improvements, and defers the algebraic survey, the diamagnetic
bound, the TE adjustment, and the numerical recipe to the appendices. 

Our approach provides a rigorous strategy to predict the spatial localization of electromagnetic waves in open scattering (non-Hermitian) dipole arrays with arbitrary structural complexity solely based on the spectral properties of operators constructed directly from their electromagnetic Green’s
matrix. The proposed constructions extend directly to the dyadic Green's function of three-dimensional resonator systems with Lorentzian response, opening avenues for the engineering and optimization of classical
and quantum structures with desired scattering resonances.

\section{Three landscapes for the non-Hermitian Green's matrix}
\label{sec:three}
We consider $N$ identical in-plane point scatterers at positions
$\{\rr_j\}_{j=1}^N$ with isotropic frequency-dependent polarizability
$\alpha(\omega)$. In TE polarization, the electric field
$\mathbf E(\rr)=(E_x,E_y)^{\!\top}$ in the array plane obeys the
vector wave equation
\begin{equation}
  \nabla\!\times\!\nabla\!\times\!\mathbf E(\rr)
  \;-\;k_0^2\Bigl[1+\alpha(\omega)\!\!\sum_{j=1}^{N}\!\delta(\rr-\rr_j)\Bigr]\mathbf E(\rr)=0,
  \qquad k_0=\omega/c,
  \label{eq:vecwave}
\end{equation}
in the presence of an incident field
$\mathbf E^{\mathrm{inc}}(\rr)$.
We use the time dependence $e^{-i\omega t}$ throughout, and
$\mathbf G_0$ denotes the outgoing (retarded) Green's tensor, so that
the scattered field $\mathbf E-\mathbf E^{\rm inc}$ satisfies the
radiation condition.
Writing~\eqref{eq:vecwave} as a Lippmann--Schwinger integral equation with the
vacuum dyadic Green's tensor
$\mathbf G_0(\rr,\rr^\prime)$~\cite{lagendijk1996resonant,rusek1996localization} and absorbing the singular
self-term into the on-shell single-scatterer $t$-matrix, which is fixed by
the optical theorem $\mathrm{Im}\,t=|t|^2$ to
$1/t(\omega)=-\mathrm{i}+z(\omega)$ with $z\in\mathbb R$, it reduces
the result to the compact Foldy--Lax
form~\cite{foldy1945multiple,lax1951multiple}
\begin{equation}
  \bigl(G-zI\bigr)\,\mathbf E \;=\; -\tfrac{1}{t(\omega)}\,
  \mathbf E^{\mathrm{inc}},
  \label{eq:cde}
\end{equation}
for the $2N$-vector of on-site (i.e., locally exciting) fields
$\mathbf E=(\mathbf E_1,\dots,\mathbf E_N)$, where $G$ is the
$2N\times 2N$ TE Green's matrix of the array, with the dyadic
propagator between distinct scatterers in its off-diagonal blocks
and the radiative damping $\mathrm{i}\,I_2$ on its diagonal
(derivation in Appendix~\ref{app:setup}). For a Lorentzian single scatterer of
resonance frequency $\omega_0$ and radiative linewidth $\Gamma_0$,
$z(\omega)=-2(\omega-\omega_0)/\Gamma_0$ is a real detuning, and
$W(z)=G-zI$ is the system matrix at that detuning; up to the factor
$2/\Gamma_0$. Here $W(z)^{-1}$ is the resolvent of the open-system
effective Hamiltonian
$H_{\mathrm{eff}}=\omega_0 I-(\Gamma_0/2)\,G$
\cite{goetschy2011lumiere,Rusek2000}. The scalar TM case follows the
same construction with the scalar Hankel Green's function replacing
the dyadic tensor~\cite{skipetrov2014absence}, and yields an
$N\times N$ Green's matrix $G^{\mathrm{TM}}$. In both polarizations
$G$ is complex symmetric by
reciprocity but neither Hermitian nor
normal, so its eigenvalues
$G|v_n\rangle=\lambda_n|v_n\rangle$ are complex and its left and right
eigenvectors differ. Through $H_{\mathrm{eff}}$ the eigenvalue
$\lambda_n$ carries both the mode frequency and its radiative
lifetime~\cite{skipetrov2014absence,skipetrov2018finite},
$\omega_n=\omega_0-(\Gamma_0/2)\mathrm{Re}\,\lambda_n$ and
$\Gamma_n=\Gamma_0\,\mathrm{Im}\,\lambda_n\ge0$: long-lived
(subradiant) modes are the eigenvalues closest to the real axis. We present additional details in Appendix~\ref{app:setup}.

\paragraph{The landscape problem for an open scattering system.}
The Filoche--Mayboroda construction for an Hermitian positive-definite
Hamiltonian solves $H u=\mathbf 1$ for a positive envelope that
predicts where the low-energy eigenmodes
live~\cite{filoche2012universal,arnold2016effective}. When loking for modes around a fixed frequency, the direct
transcription to the open scattering system of Eq.~\eqref{eq:cde} is to solve
$W(z)\,u=\mathbf 1$: the response of the array to a uniform drive at
the complex frequency labeled by $z$. However, three obstacles make the direct solve
unsuitable. First, $W(z)^{-1}$ is not a positive matrix, so $u$ is
complex and oscillatory rather than a positive confinement envelope.
Second, $W(z)$ is singular whenever $z\in\sigma(G)$ and nearly
singular close to each $\lambda_n$, which is exactly the physically
interesting regime. To this point, it is well-known that standard techniques from random matrix theory, like the classic Stieltjes transform, fail for non-Hermitian ensembles because the eigenvalues are complex, rendering the resolvent
operator unbounded within the support of the spectrum.
Third, $G$ is non-normal, so $\|W(z)^{-1}\|$ is
not controlled by the distance from $z$ to the
spectrum~\cite{trefethen2005spectra}. The three difficulties are resolved by
working with the singular values of $W(z)$ instead of the eigenvalues
of $G$, and by replacing the non-normal resolvent of $G$ with the
Hermitian resolvent on a doubled space introduced by Feinberg and
Zee~\cite{feinberg1997nonhermitian}. The singular values of $W$ are
real and non-negative, and they are the eigenvalues of the
hermitization $H_0(z)=\bigl(\begin{smallmatrix}0&W\\W^\dagger&0\end{smallmatrix}\bigr)$;
a Tikhonov regulator $\eta>0$ then makes
$B(z,\eta)=H_0(z)+\mathrm{i}\eta I$ invertible for every $z$ with
$\|B^{-1}\|_2\le 1/\eta$~\cite{metz2019spectral}, and its blocks carry
both the regularized resolvent of $G$ and the right-eigenvector LDOS
(Appendix~\ref{app:rigorous}). The three landscapes introduced below
are the real-valued, positive per-site contractions of these blocks
against a uniform or polarization-isotropic source.

In all our computations we take an optical density
$\mathrm{OD}=\rho\lambda^{2}=10$, where $\lambda$ is the wavelength
and $\rho$ is the site density; our conclusions do not depend on the
choice of $\mathrm{OD}$ provided it is large enough to produce
localized modes. Explicitly, the regularized Feinberg--Zee matrix and
its block inverse read
\begin{equation}
  B(z,\eta) =
  \begin{pmatrix}
    \mathrm{i}\eta\,I & W \\ W^\dg & \mathrm{i}\eta\,I
  \end{pmatrix},
  \qquad
  B^{-1} =
  \begin{pmatrix}
    -\mathrm{i}\eta\,X & WY \\ YW^\dg & -\mathrm{i}\eta\,Y
  \end{pmatrix},
  \label{eq:B}
\end{equation}
where $X=(WW^\dg+\eta^2I)^{-1}$ and $Y=(W^\dg W+\eta^2I)^{-1}$ are
the regularized Gram inverses of $W=G-zI$.

\paragraph{Gramian landscape ($\Bop$).}
We introduce here the FM-continuous, spectrum-blind $\Bop$ reference landscape as:
\begin{equation}
  u_\Bop(j) = \big|\big[(G^\dg G+\eta^2 I)^{-1}\one\big]_j\big|^2 .
  \label{eq:gramdef}
\end{equation}
By singular value decomposition (SVD), $\Bop$ amplifies the small-$\sigma$ right singular vectors of
$G$ contracted uniformly against the array. It is the $z=0$ limit of
$\SY$ (introduced below) and admits a rigorous diamagnetic confinement bound in the
$\eta\to 0$ limit (Appendix~\ref{app:diamag}). Our approach builds on the fundamental connection between the scattering matrix and the Green matrix that has enabled control of light transport in random media via the analysis of singular vectors and the statistical properties of orthogonal transmission channels \cite{cao2022shaping,rotter2017light}.   

\paragraph{Spectral landscape ($\SY$).}
The Metz--Neri--Rogers-aware, $z$-dependent Schur block, is here used to define the $\SY$ landscape as:
\begin{equation}
  u_\SY(j) = \big|[Y(z,\eta)\,\one]_j\big|^2 .
  \label{eq:specdef}
\end{equation}
Since the diagonal elements of $Y$ appear under the trace in the empirical
spectral density $\rho_G(z)$~\cite[Eq.~(33)]{metz2019spectral}, 
$\SY$ inherits a first-principles connection to the local pseudomode
content of $G$ at frequency $z$. We span $z$ across the longest-lived
eigenvalues of $G$ (the eigenvalues with the smallest imaginary part in
the complex-eigenvalue coordinates
~\cite{skipetrov2014absence,skipetrov2018finite}), that correspond to the
modes with the longest radiative lifetimes. These are the most spatially
localized modes and therefore $z$ is the natural spectral parameter at which
to probe the local pseudomode content of $G$ with the generalized
landscapes.

\paragraph{Hermitized-inverse landscape ($\SBf$).}
The full $B^{-1}$ applied to the symmetric doubled source 
$[\one;\one]$ is used to introduce the $\SBf$ landscape, defined by:
\begin{equation}
  \begin{pmatrix} \mathbf{a}\\\mathbf{b} \end{pmatrix}
    = B^{-1}\!\begin{pmatrix} \one\\ \one \end{pmatrix},
  \qquad
  u_\SBf(j) = |a_j|^2 + |b_j|^2 .
  \label{eq:hermdef}
\end{equation}
$\SBf$ combines the regularized direct solve with $\SY$ through the
exact identity \eqref{eq:herm-tik-decomp}; all three landscapes are
defined for every $z$ when $\eta>0$~\cite{feinberg1997nonhermitian},
and differ in the spectral filter applied to the right singular
vectors of $W$.


\paragraph{Why the three landscapes probe eigenmodes of $G$.} 
A short derivation makes it clear that
Eqs.~\eqref{eq:gramdef}--\eqref{eq:hermdef} are not three independent
choices of Schur blocks but three readouts of the \emph{same}
eigenmode-recovery identity for $G$. Appendix~\ref{app:rigorous}  supplies the full
derivation and we outline the two key ideas here. First, for any invertible matrix $M$ the identity
$\|M^{-1}\|_2 = 1/\sigma_{\min}(M)$ holds by direct substitution in the
operator-norm maximum. Applied to $M = G-zI$, we obtain that the
resolvent norm is \emph{exactly} the inverse of the smallest singular
value probed by the regularized Gram inverses $X$ and $Y$. Second, if $G$ is
diagonalisable with biorthogonal eigenvectors
$G\lvert v_n\rangle = \lambda_n\lvert v_n\rangle$,
$\langle u_n\rvert G = \lambda_n\langle u_n\rvert$, the Laurent
expansion of the resolvent at a simple eigenvalue is
$R(z) = P_n/(z-\lambda_n) + R_{\mathrm{reg}}(z)$ with
$P_n = \lvert v_n\rangle\langle u_n\rvert/\langle u_n|v_n\rangle$.
Combining the two,
$\|R(z)\|_2\to\kappa_n/|z-\lambda_n|$ as $z\to\lambda_n$, with the
eigenvalue condition
number
$\kappa_n=\|v_n\|\,\|u_n\|/|\langle u_n|v_n\rangle|$ ~\cite{wilkinson1965algebraic,bauer1960norms,trefethen2005spectra}. The square
$\kappa_n^2$ is the Petermann excess-noise factor from laser
physics~\cite{petermann1979calculated,patra1999excess}, and it
provides a quantitative signature of non-normality: $\kappa_n$ becomes
large precisely when left and right eigenvectors nearly decouple.

The Metz--Neri--Rogers (MNR) framework~\cite{metz2019spectral} supplies
a second, projector-level identity that does not require explicit
knowledge of $(\lvert u_n\rangle,\lvert v_n\rangle)$. The $2N\times 2N$
hermitization $B(z,\eta)$ of \eqref{eq:B} has, at
$z=\lambda_n\in\sigma(G)$, the two-fold eigenvalue $\mathrm{i}\eta$
with eigenvectors $(\lvert\hat u_n\rangle;0)^{\!\top}$ and
$(0;\lvert\hat v_n\rangle)^{\!\top}$, where
$\lvert\hat u_n\rangle=\lvert u_n\rangle/\|u_n\|$ and
$\lvert\hat v_n\rangle=\lvert v_n\rangle/\|v_n\|$ are the left and
right eigenvectors of $G$ each normalized to unit length, and the
small-$\eta$ projector reconstruction
\begin{equation}
\mathrm{i}\eta\,B^{-1}(\lambda_n,\eta)
=
\begin{pmatrix}
\lvert\hat u_n\rangle\langle\hat u_n\rvert & 0 \\
0 & \lvert\hat v_n\rangle\langle\hat v_n\rvert
\end{pmatrix}
+ O\!\left(\frac{\eta}{\sigma_{N-1}}\right),
\label{eq:MNRreconstruction}
\end{equation}
where $\sigma_{N-1}>0$ is the second-smallest singular value of
$W(\lambda_n)$. Reading \eqref{eq:MNRreconstruction} block by block
in the inverse of \eqref{eq:B}, whose diagonal blocks are
$-\mathrm{i}\eta X$ and $-\mathrm{i}\eta Y$, yields the pointwise
identities
\begin{equation}
\eta^{2}\,Y_{jj}(\lambda_n,\eta)\xrightarrow[\eta\to 0^{+}]{}
|\hat v_n(j)|^{2},
\qquad
\eta^{2}\,X_{jj}(\lambda_n,\eta)\xrightarrow[\eta\to 0^{+}]{}
|\hat u_n(j)|^{2},
\label{eq:LDOSidentity}
\end{equation}
valid for $\eta\ll\sigma_{N-1}$, so that the diagonal of $Y$ is, up
to the regulator, the local density of states (LDOS) of the right
eigenvector of $G$ at frequency $z$. The landscapes of
\eqref{eq:specdef}--\eqref{eq:hermdef} inherit this interpretation
through a Cauchy--Schwarz envelope: $u_\SY(j)$ is the source-driven
contraction of $Y$ against $\one$, so sites at which the
right-eigenvector amplitude $|\hat v_n(j)|^2$ vanishes also vanish in
$u_\SY$; $u_\SBf(j)$ is the analogous contraction of the full
$B^{-1}$. For reciprocal $G$ the two LDOS branches coincide at every
site ($X_{jj}=Y_{jj}$, a direct consequence of $G=G^{\!\top}$; see
Appendix~\ref{app:rigorous}), so what $\SBf$ adds over $\SY$ is the
off-diagonal blocks $WY$ and $YW^\dg$ of $B^{-1}$, not a separate
left-eigenvector contribution.
The scaling
$\|R(z)\|_2\sim\kappa_n/|z-\lambda_n|$  shows that these pointwise
envelopes are amplified both by small spectral distance and by the
non-normality of the mode, so the landscapes become sharp at modes
that are long-lived \emph{and} spectrally sensitive (see Appendix~\ref{app:rigorous} for additional details).

The final step, developed as a four-link chain in
Appendix~\ref{app:rigorous:chain}, explains why these long-lived,
sensitive modes are also \emph{spatially localized}: a small
$\sigma_{\min}(W)$ drives a small $\Imag\lambda_n$, and a small
$\Imag\lambda_n$ forces the biorthogonal pair to lie approximately in
the kernel of the radiation operator $\Imag G$. This near-kernel
consists of configurations that radiate no coherent far field, and in
aperiodic arrays such configurations can only be realized by tightly
packed sub-wavelength dipolar clusters. The Herm, Spec, and Gram
landscapes therefore render the same object, basins at the spatially
compact subradiant eigenmodes of $G$, consistent with their origin
as different source-driven contractions of \eqref{eq:B}.


The three landscapes coincide at $z=0$ only in the sense that $\SY|_{z=0}=\Bop$, by $(W^\dg W+\eta^2 I)^{-1}|_{z=0}=(G^\dg G+\eta^2 I)^{-1}$. $\SBf$ does \emph{not} reduce to $\Bop$ at $z=0$: the diagonal $\mathrm{i}\eta$ blocks of $B^{-1}$ carry a Gram-like contribution, but the off-diagonal blocks $WY$ and $YW^\dg$ apply a different filter. For reciprocal $G$ this gives the exact pointwise identity
\begin{equation}
  u_\SBf(j) \;=\; 2\,\bigl|[YW^\dg\one]_j\bigr|^2 \;+\; 2\eta^2\,u_\SY(j),
  \label{eq:herm-tik-decomp}
\end{equation}
in which $YW^\dg\one$ is the Tikhonov-regularized solution of $Wb=\one$: the stabilized, $z$-dependent form of the direct landscape equation (cf.\ Appendix~\ref{app:survey}). In spectral-filter terms, $\SBf$ weights a singular pair $(\sigma_k,v_k)$ in intensity by $1/(\sigma_k^2+\eta^2)$ while $\SY$ weights it by $1/(\sigma_k^2+\eta^2)^2$: $\SBf$ therefore admits \emph{more} pseudomodes at larger $\sigma_k$, not fewer, and has the broader filter of the two. Appendix~\ref{app:survey} rules out the other possible Schur-block
candidates (off-diagonal blocks, $X$-block, direct FM transcription,
biorthogonal) on algebraic grounds. Appendix~\ref{app:tepol}
introduces the TE polarization-isotropic replacements
$\BBop,\SBBf$ and the source-free cross-check $\Ydiag$.

\section{Operator comparison across aperiodic geometries}
\label{sec:results}

We compare the three landscapes on a number of aperiodic, both random
and deterministic, two-dimensional (2D) point patterns of electric
dipoles. In particular, we consider here binary Bernoulli random arrays, the three primitive classes of deterministic arrays from deterministic aperiodic substitutions (i.e., Fibonacci, Thue--Morse, and Rudin-Shapiro), sub-random arrays constructed from the quasi-random Halton sequence, deterministic aperiodic Pinwheel patterns with isotropic response, as well as additional aperiodic structures from number theoretic constructs (co-prime arrays and Eisenstein primes) \cite{dal2012deterministic,lee2010spatial,sgrignuoli2020subdiffusive,sgrignuoli2022subdiffusive,
sgrignuoli2021hyperuniformity,wang2018spectral,dal2021wave,sgrignuoli2020multifractality,baake2013aperiodic,schroeder2006number,dalnegro2022waves}.
Our selection covers the three spectral classes of the Lebesgue decomposition theorem --- pure point, singular continuous, and absolutely continuous --- which exhausts the possible diffraction measures of aperiodic point sets \cite{baake2013aperiodic,dal2012deterministic,macia2006role,macia2020quasicrystals}.   
In the supplementary material, we consider additional examples of aperiodic
structures.  

Each pattern is normalized to a common aperture and truncated to
approximately $N=3500$ interacting dipole sites for the scalar TM
calculations throughout the main text. Vector TE calculations use the
same aperture; the operator doubles to $2N\times 2N$ and the dense
eigendecomposition and solve costs scale as $N^3$, so some TE figures
are produced at a reduced $N$ when runtime requires. Every figure
states its $N$ explicitly in the caption:
Figure~\ref{fig:spec_grid}--\ref{fig:herm_grid} use $N\le 1500$ to
make the $z$-sweep across eight probe values tractable, and the
Poisson figure~\ref{fig:grid} reports the realized count ($N=3394$
TM, $N=2803$ TE) of a Poisson point process whose cardinality
fluctuates by construction. For every
pattern--operator pair, we compute the FM watershed segmentation of
the landscape into cells and overlay the top $K=100$ long-lived
eigenmodes of the Green's matrix, which provides the spatial structure of the electric field eigenmodes \cite{sgrignuoli2019localization,sgrignuoli2021hyperuniformity}, on the resulting basins to assess their spatial
confinement visually.  Explicit forms for $G^{\mathrm{TM}}$ and
$G^{\mathrm{TE}}$ and the complete numerical recipe are given in
Appendix~\ref{app:setup}.
\begin{figure}[!ht]
    \centering
    \begin{subfigure}{0.39\textwidth}
    \centering
    \includegraphics[width=\linewidth]{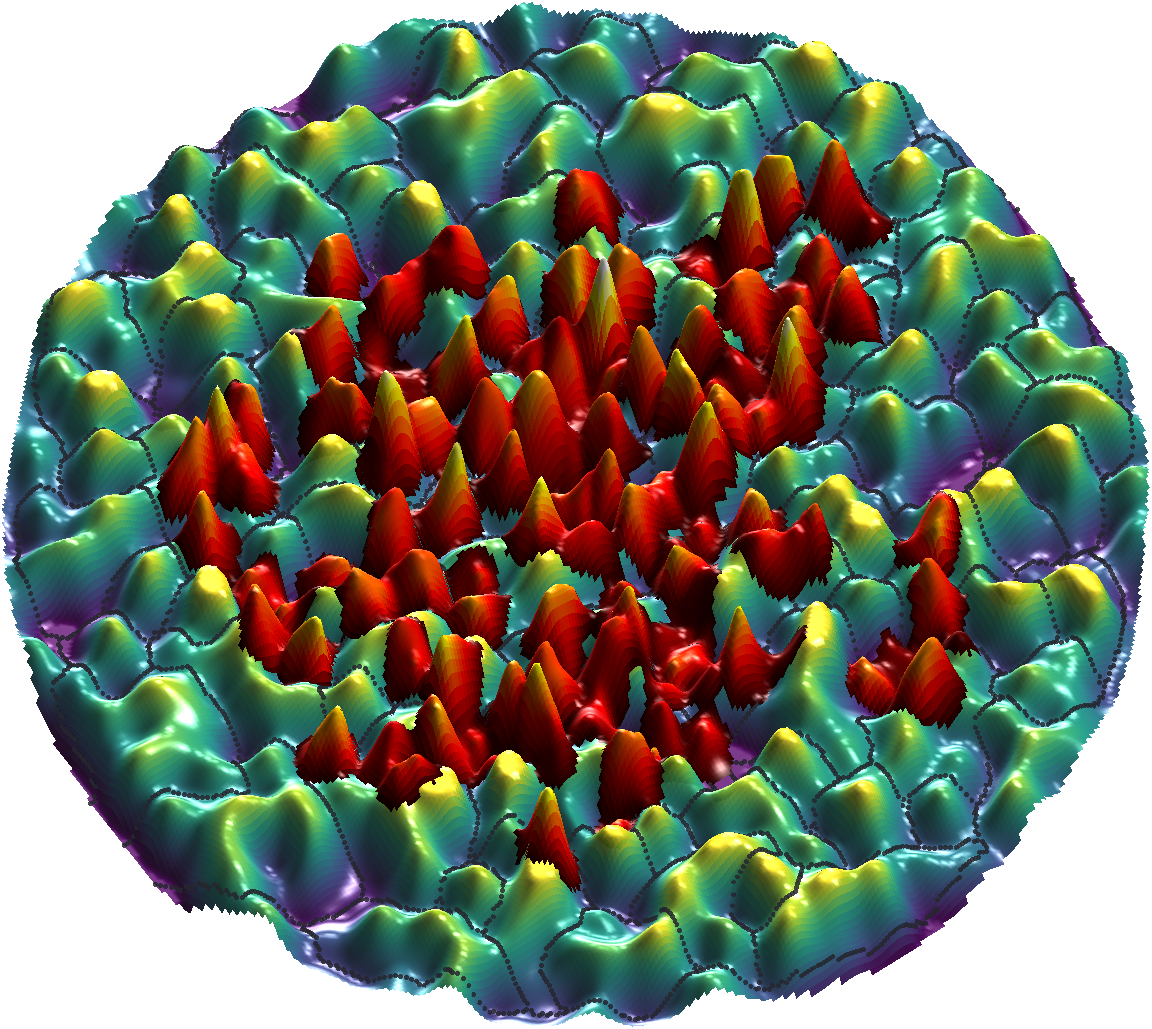}
    \caption{}
    \end{subfigure}
    \begin{subfigure}{0.39\textwidth}
    \centering
    \includegraphics[width=\linewidth]{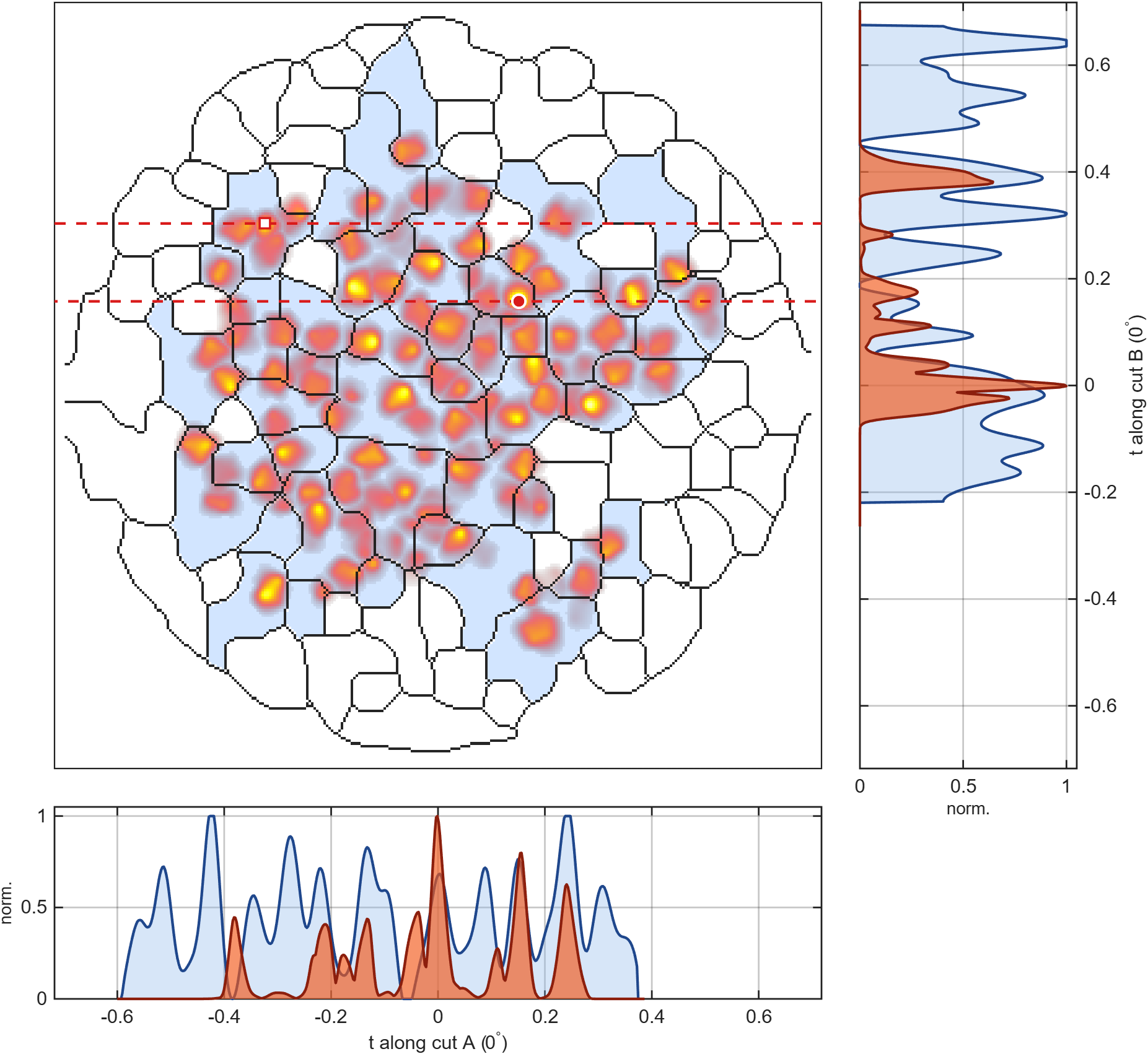} 
    \caption{}
    \end{subfigure}
    \caption{Bernoulli lattice ($p=0.05$), TM, $N=3500$, top 100 modes. (a) The 3D visualization of the Herm landscape function showing the top 100 localized modes. (b) 2D top-down with the watershed ridges black, host-cell union in blue , top-100 modes' max-projection in hot. Two 1D cross-sections are shown to highlight localization.}
    \label{fig:bernoulli-tm}
\end{figure}

\begin{figure}[!ht]
    \centering
    \begin{subfigure}{0.39\textwidth}
    \centering
    \includegraphics[width=\linewidth]{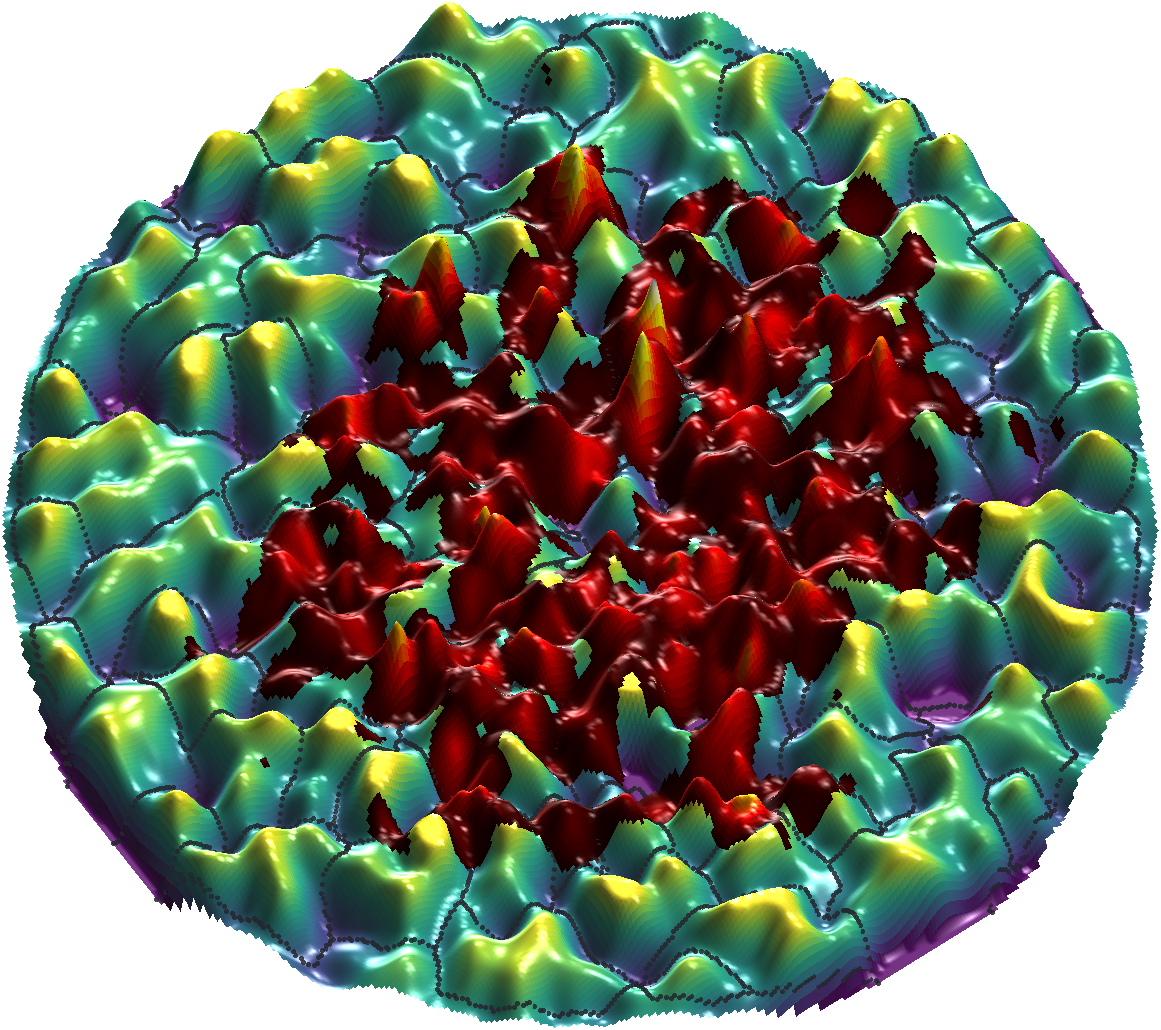}
    \caption{}
    \end{subfigure}
    \begin{subfigure}{0.39\textwidth}
    \centering
    \includegraphics[width=\linewidth]{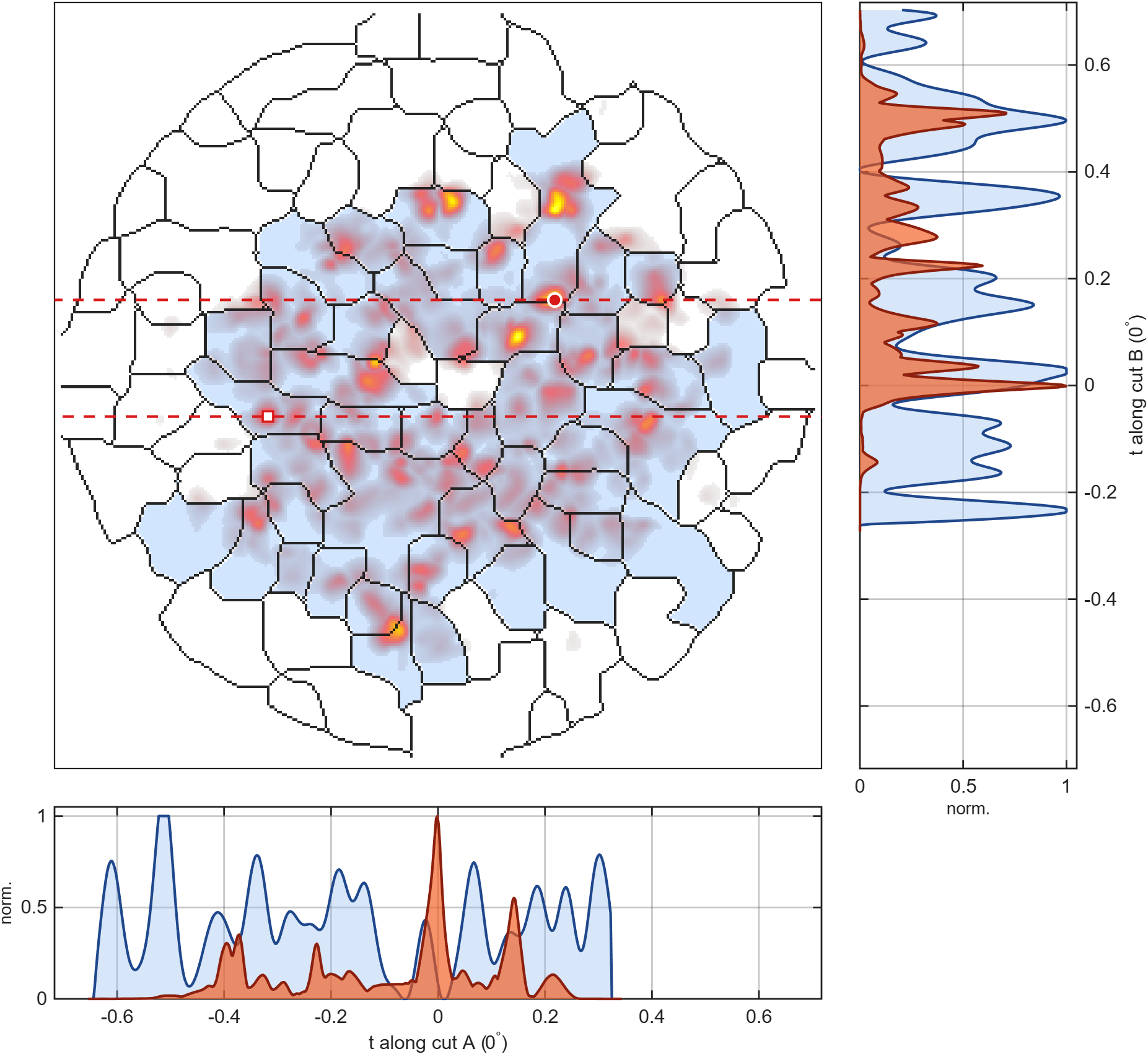} 
    \caption{}
    \end{subfigure}
    \caption{Bernoulli lattice ($p=0.05$), TE, $N=3500$, top 100 modes. (a) The 3D visualization of the landscape function showing the top 100 localized modes. (b) 2D top-down with the watershed ridges black, host-cell union in blue , top-100 modes' max-projection in hot. Two 1D cross-sections are shown to highlight localization.}
    \label{fig:bernoulli-te}
\end{figure}
Our approach is first demonstrated with respect to canonical random systems in Fig.~\ref{fig:bernoulli-tm} and
Fig.~\ref{fig:bernoulli-te}, where we show the $\SBf$ landscapes overlapped with a few representative modes of a random
Bernoulli ($p=0.05$) binary array of electric dipoles under both TM and TE polarizations,
respectively.  Specifically, in Fig.~\ref{fig:bernoulli-tm}(a) we visualize in
three spatial dimensions (3D) the landscape function showing the top 100 localized
modes, each fully contained within the predicted regions by the $\SBf$ landscape.  In Fig.~\ref{fig:bernoulli-tm}(b), we show the two-dimensional (2D) top-down view
of the Green's matrix $\SBf$ landscape with the watershed ridges black,
host-cell union in blue, and the top-100 modes' max-projection in hot colors.
Two one-dimensional (1D) cross-sectional cuts corresponding to the two horizontal dashed lines in red are also shown to highlight the
coincidence of the Green's matrix eigenmode profiles and the landscape localization regions.  Figure~\ref{fig:bernoulli-te} shows the same analysis for TE polarization: the modes again sit inside the predicted basins, with the modified landscape of Appendix~\ref{app:tepol} applied to the vector problem.

\begin{figure}[!ht]
    \centering
    \begin{subfigure}{0.39\textwidth}
    \centering
    \includegraphics[width=\linewidth]{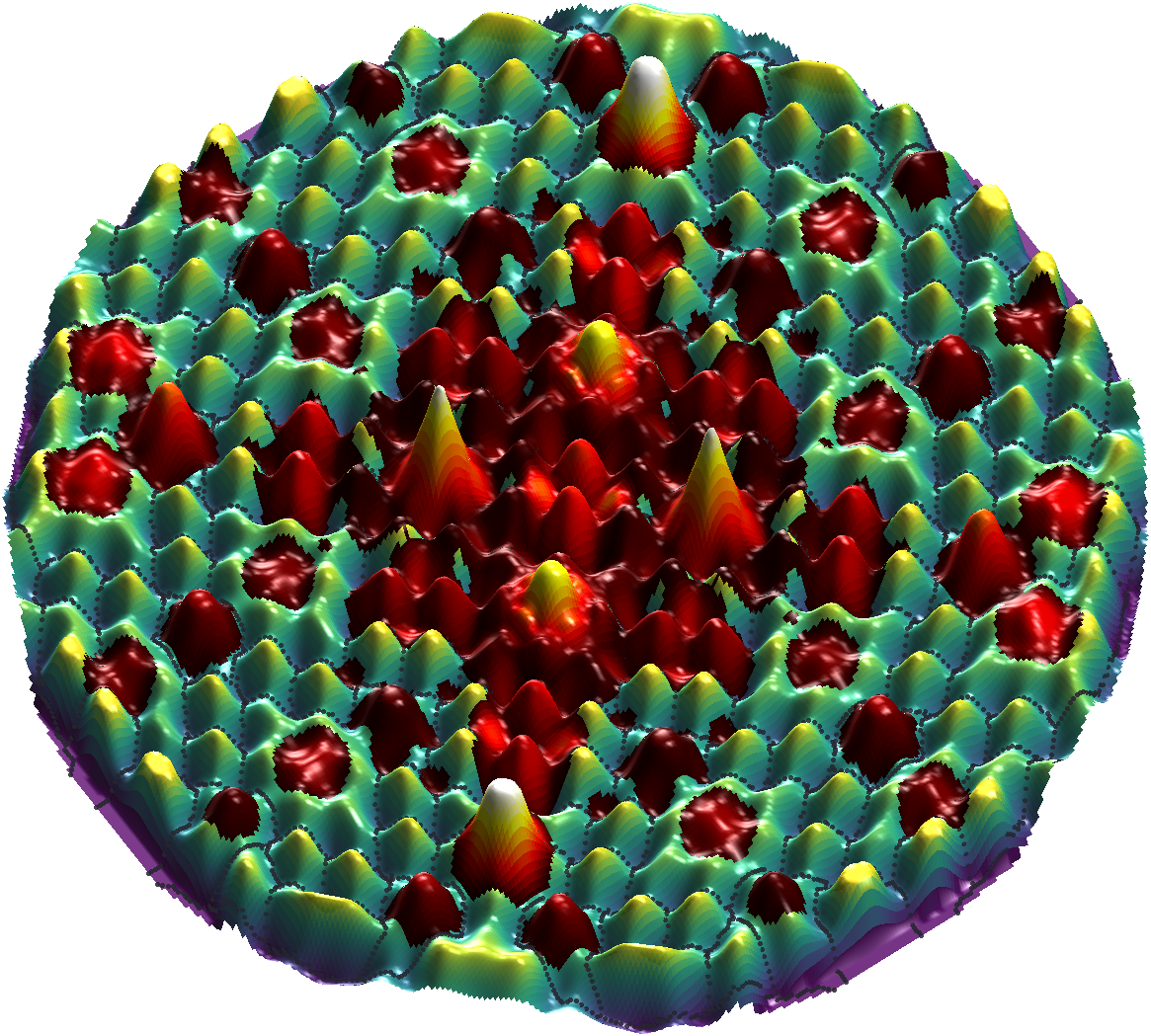}
    \caption{}
    \end{subfigure}
    \begin{subfigure}{0.39\textwidth}
    \centering
    \includegraphics[width=\linewidth]{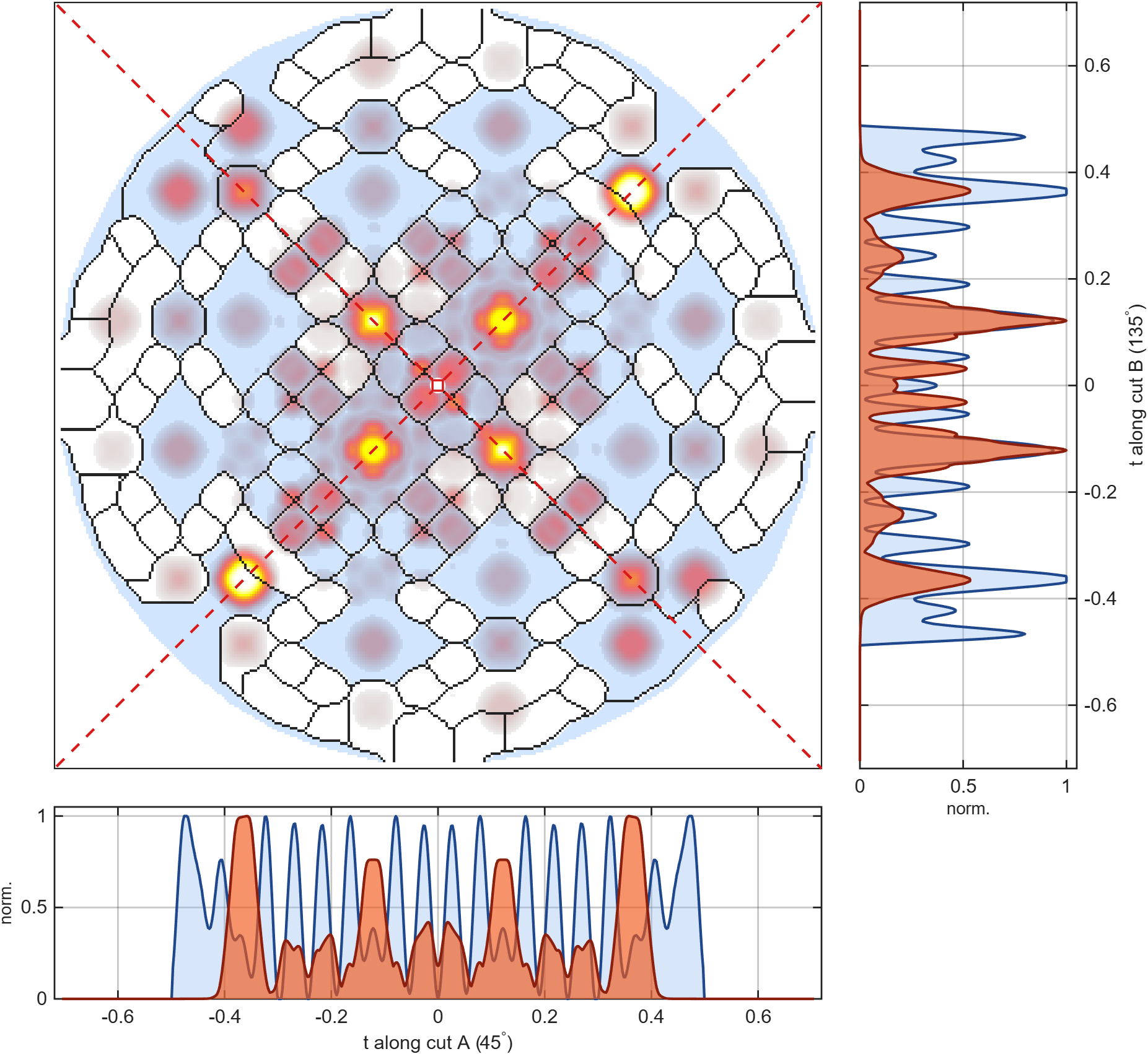} 
    \caption{}
    \end{subfigure}
    \caption{Thue-Morse, TM, $N=3500$, top 100 modes. (a) The 3D visualization of the landscape function showing the top 100 localized modes. (b) 2D top-down with the watershed ridges black, host-cell union in blue , top-100 modes' max-projection in hot. Two 1D cross-sections are shown to highlight localization.}
    \label{fig:thuemorse-tm}
\end{figure}

\begin{figure}[!ht]
    \centering
    \begin{subfigure}{0.39\textwidth}
    \centering
    \includegraphics[width=\linewidth]{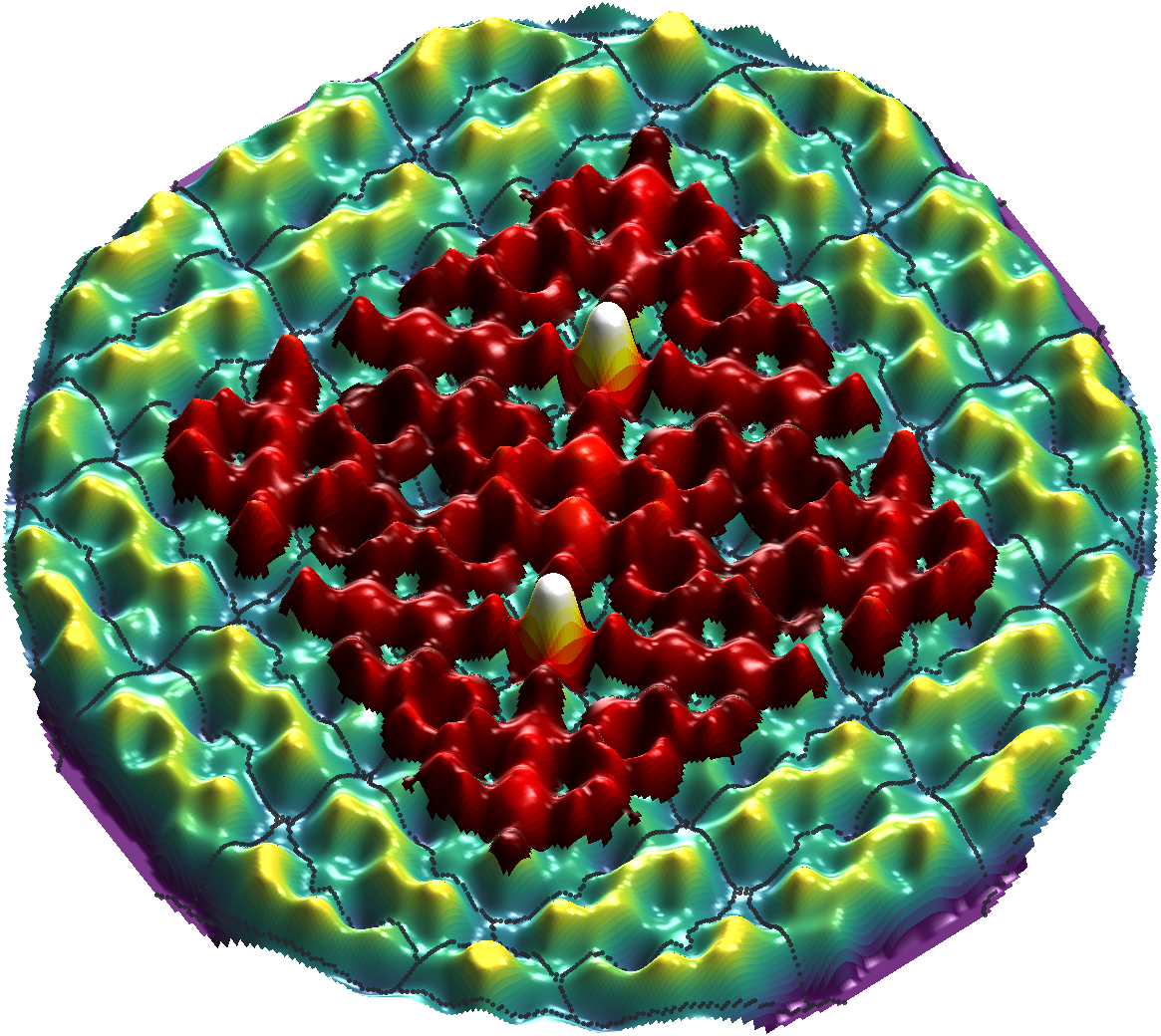}
    \caption{}
    \end{subfigure}
    \begin{subfigure}{0.39\textwidth}
    \centering
    \includegraphics[width=\linewidth]{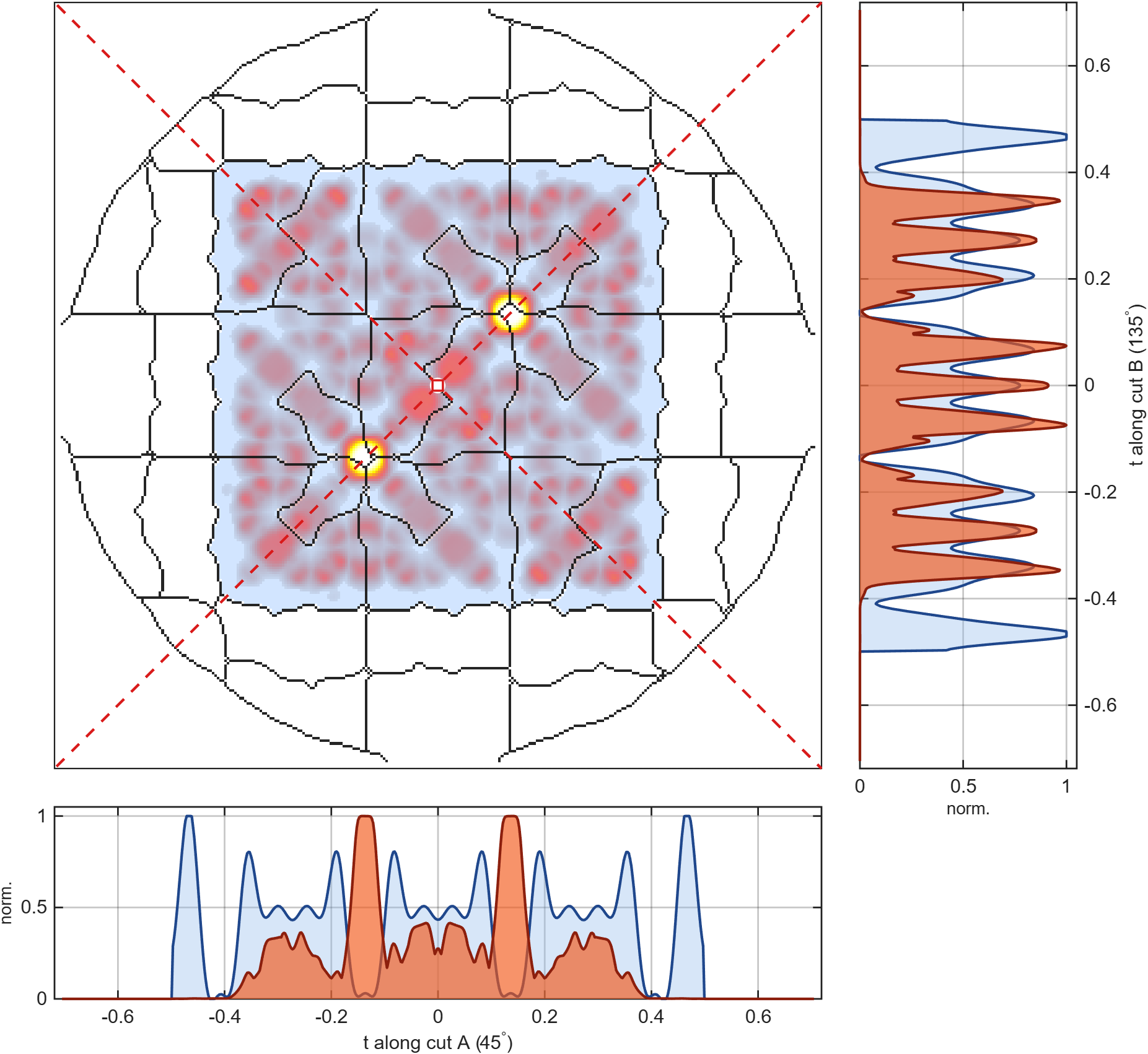} 
    \caption{}
    \end{subfigure}
    \caption{Thue-Morse, TE, $N=3500$, top 100 modes. (a) The 3D visualization of the landscape function showing the top 100 localized modes. (b) 2D top-down with the watershed ridges black, host-cell union in blue , top-100 modes' max-projection in hot. Two 1D cross-sections are shown to highlight localization.}
    \label{fig:thuemorse-te}
\end{figure}

We address the case of a Thue--Morse deterministic aperiodic structure in Fig.~\ref{fig:thuemorse-tm} and
Fig.~\ref{fig:thuemorse-te} for arrays of
electric dipoles under TM and TE polarizations, respectively. The Thue--Morse structure is representative of aperiodic systems with singular continuous spectral correlation properties and gives rise to distinctive multifractal mode patterns, called critical modes, that fluctuate at multiple scales across the array \cite{dal2012deterministic,baake2013aperiodic,noh2011lasing,dal2004photon,macia2006role,macia2020quasicrystals}. 
However, also in this more complex case we  notice in Fig.~\ref{fig:thuemorse-tm} and
Fig.~\ref{fig:thuemorse-te} that the mode envelopes feature characteristic peaks that sit inside the union of host cells
identified by the $\SBf$ landscape without any adjustments. The watershed walls
never truncate a mode peak: mode maxima coincide with maxima of the
landscape $u$ (equivalently, minima of the effective potential $1/u$),
and the watershed lines of $-u$ run along the valley lines of $u$ that
separate neighbouring basins.
The $\SBf$ landscape therefore predicts the spatial regions of localization even for the amplitude fluctuations of the critical modes of the Thue--Morse structure, extending the approach beyond uncorrelated random media.

\begin{figure}[!ht]
    \centering

    \noindent
\begin{tabular}{@{}c@{\hspace{2pt}}cccc@{}}
& \small point cloud & \small Gram & \small Spec & \small Herm \\
\rlbl{TM}      & \pnl{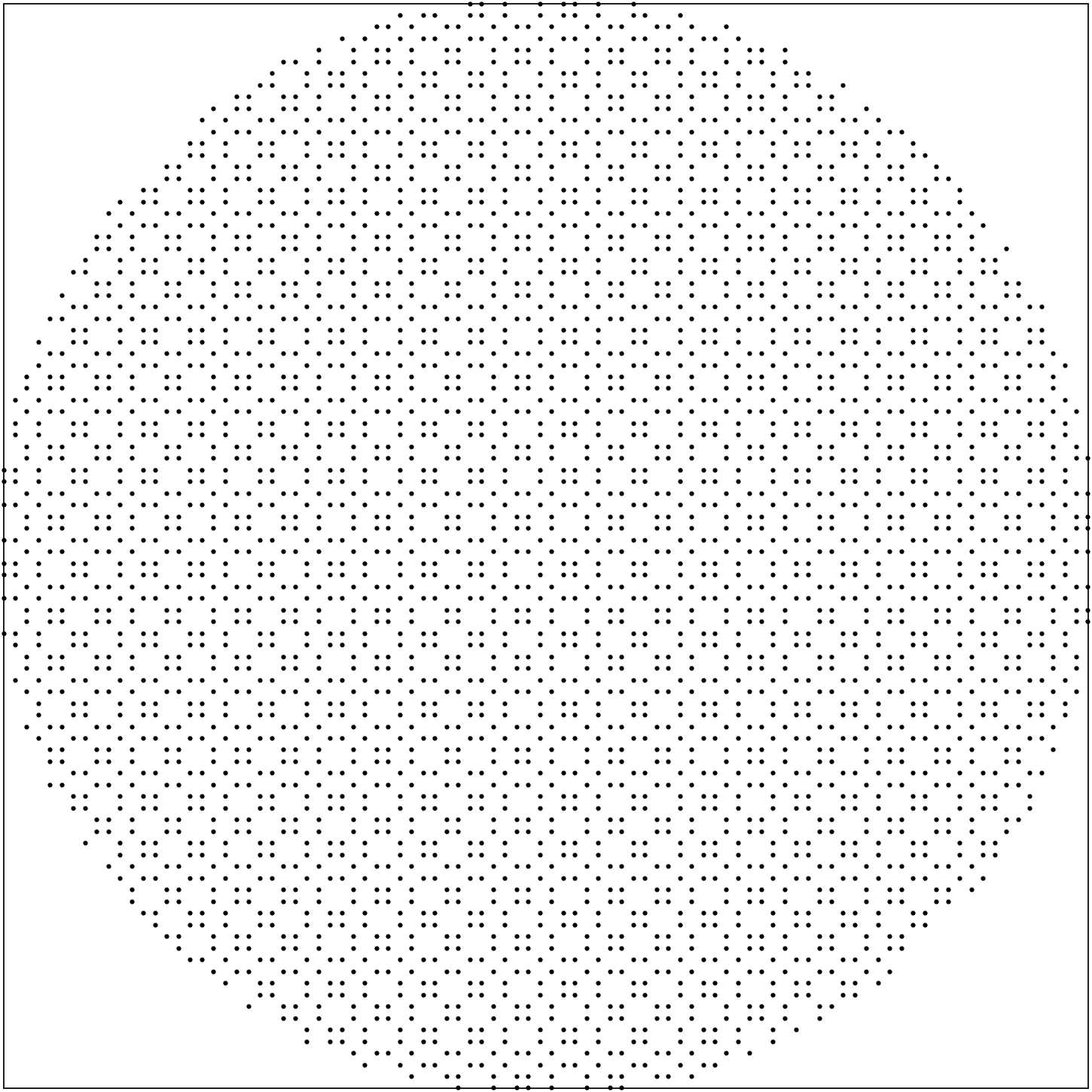} & \pnl{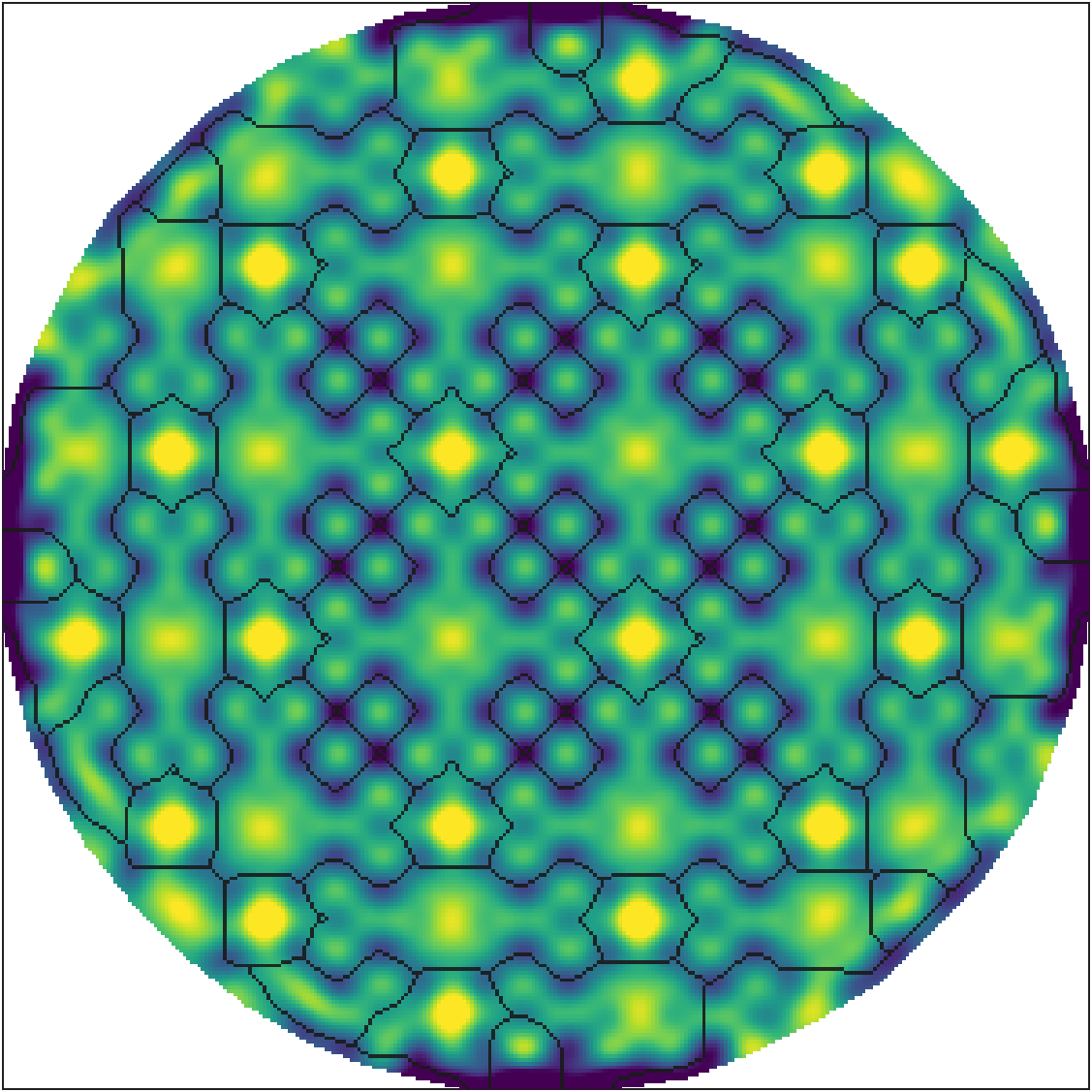} & \pnl{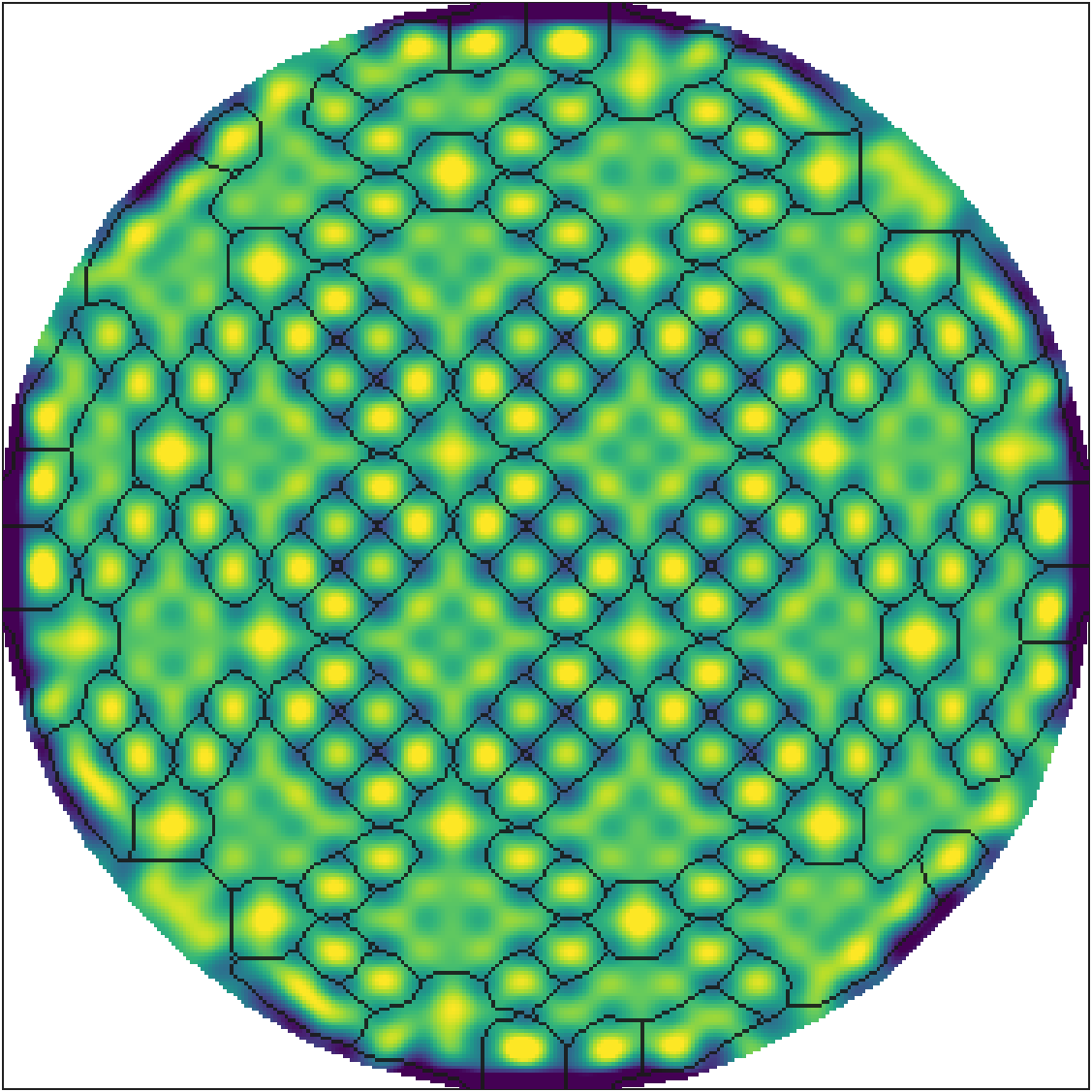} & \pnl{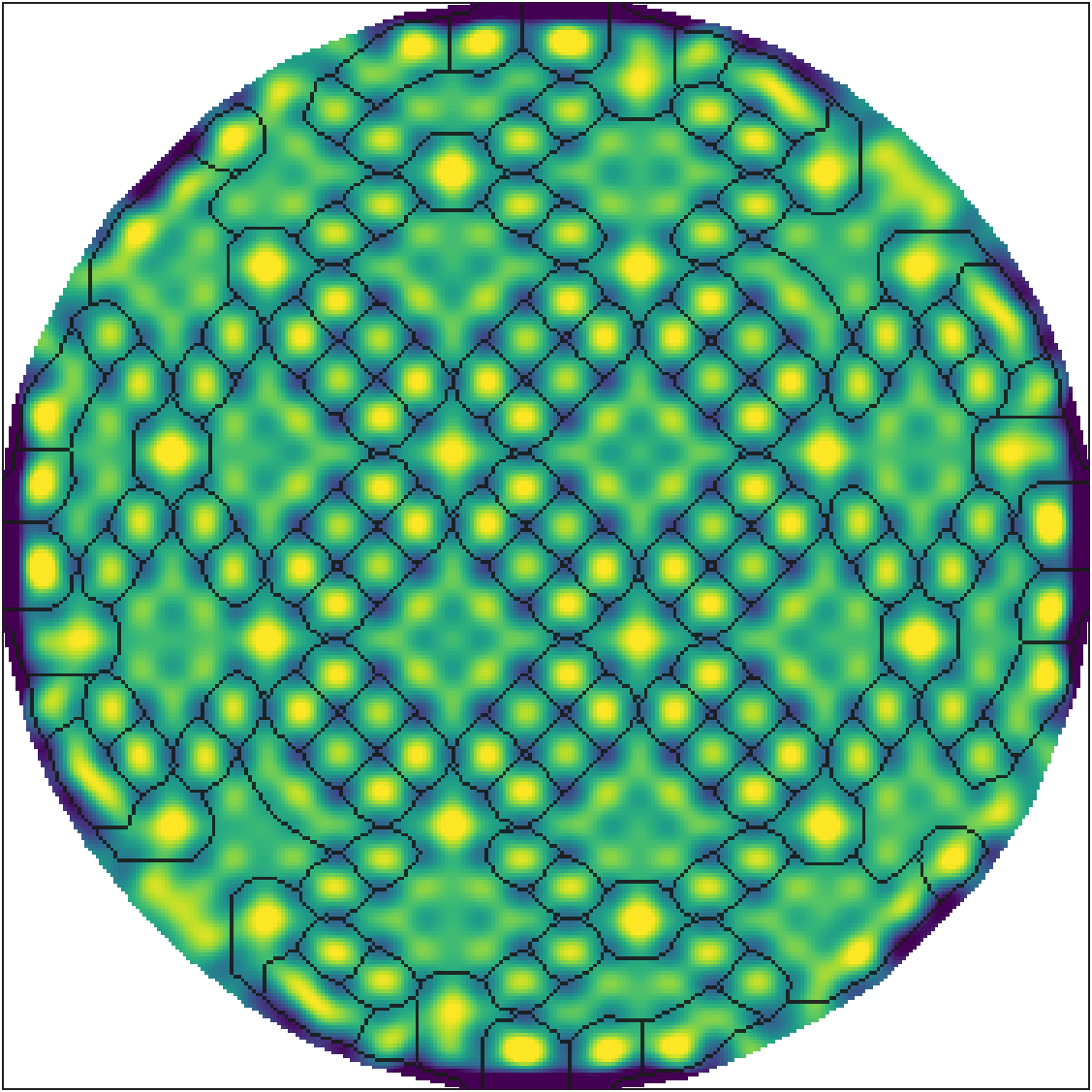} \\
\rlbl{TE-iso}  & \pnl{images/ThueMorse/pointcloud.png} & \pnl{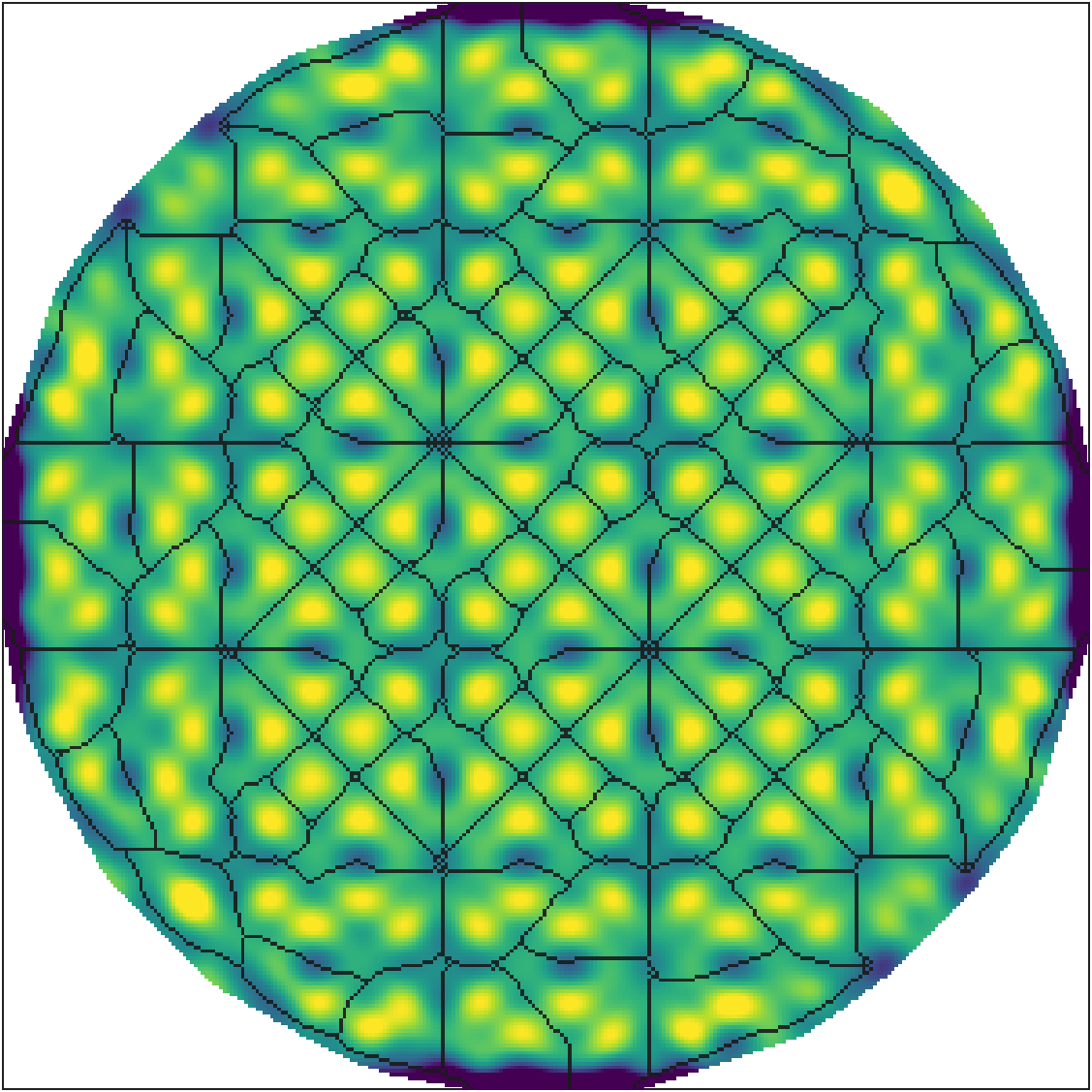} & \pnl{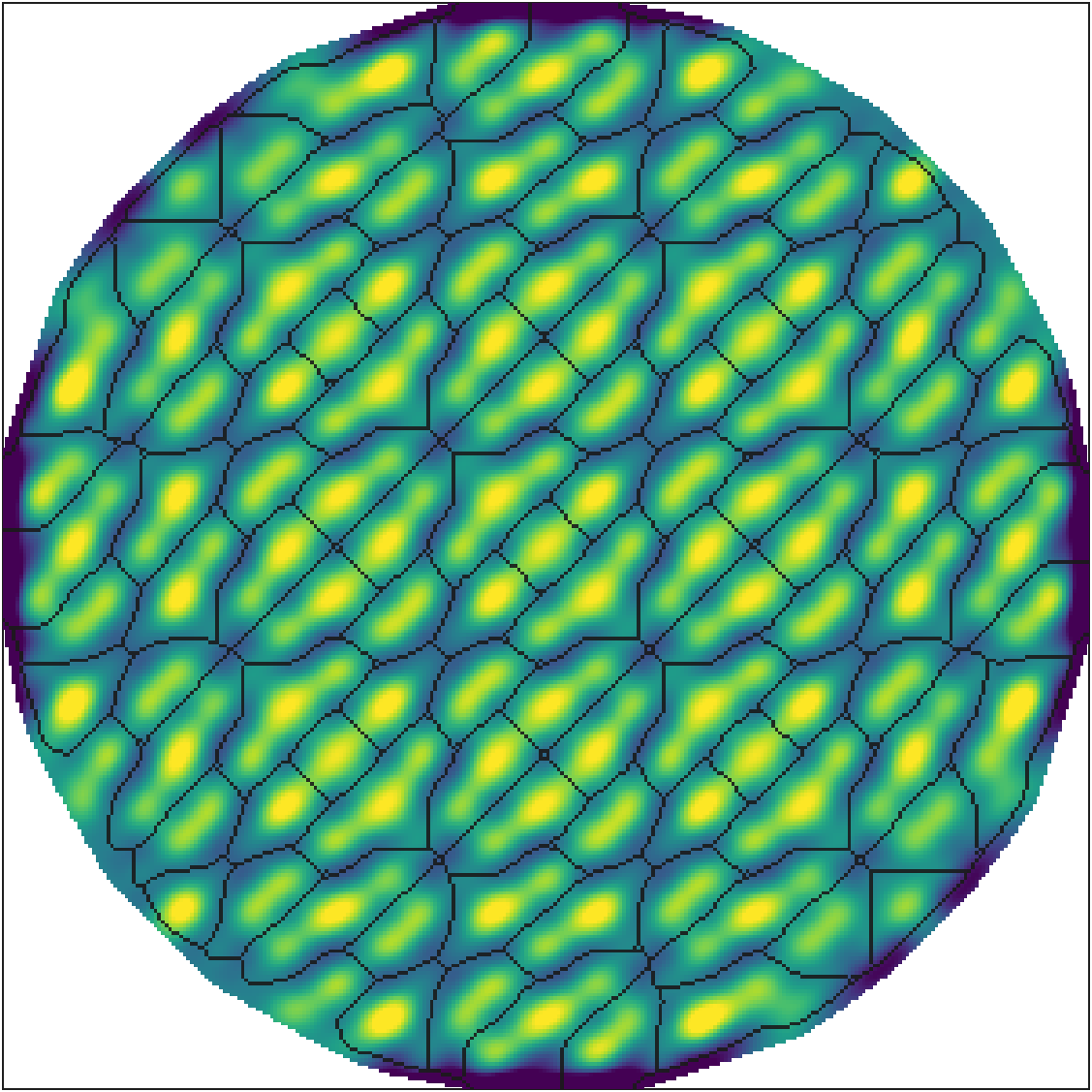} & \pnl{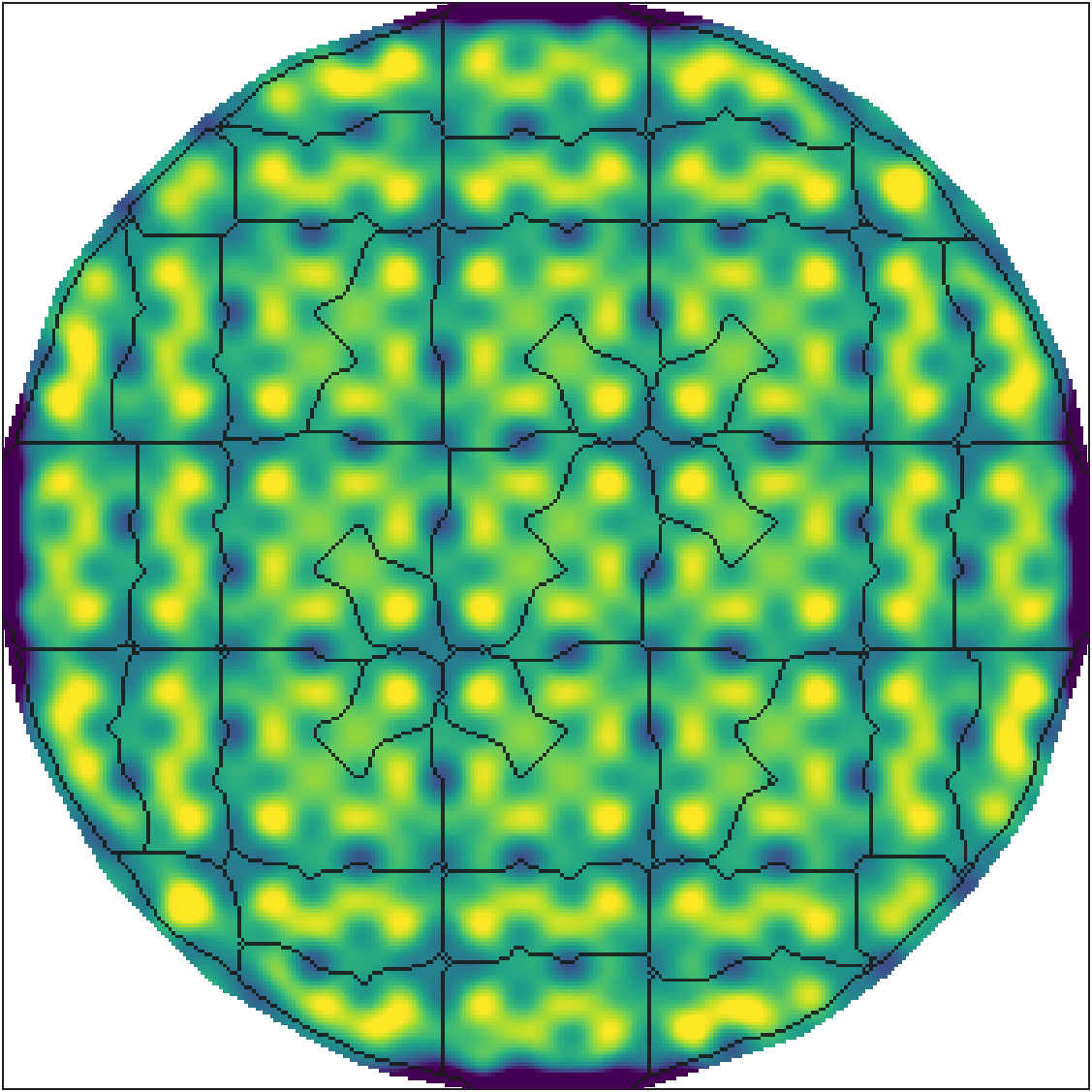} \\
\end{tabular}
    
    \caption{Thue-Morse array, $N{=}3500$ ~-- ~TM vs TE-iso landscapes with watershed. Both rows use exactly the same $N{=}3500$ Thue--Morse points; TM
is the scalar Hankel Green's matrix, TE-iso is the vector Green's matrix
with isotropic (sum of x- and y-polarized) source in the landscape. }
    \label{fig:thuemorse-tmte}
\end{figure}

The three landscapes defined in Section~\ref{sec:three} are compared
side by side in Fig.~\ref{fig:thuemorse-tmte} for the Thue--Morse
array under both TM (top row) and TE-iso (bottom row) polarizations.
Each row displays the point pattern together with the Gram $\Bop$,
Spec $\SY$, and Herm $\SBf$ landscapes rendered as viridis heatmaps
with the watershed ridges in black.  Moving rightward along a row,
the watershed partitions the array into progressively finer basins:
Gram, which operates at $z=0$, yields the coarsest segmentation;
Spec, anchored at the longest-lived eigenvalue, resolves
pseudospectral structure that Gram collapses; and Herm, retaining all
four Schur blocks in coherent superposition, exposes additional
sub-basin structure that Spec itself merges.  In both polarizations
the eight-fold symmetry of the Thue--Morse pattern is clearly
reproduced by the basin topology, and the TM and TE-iso rows produce
visually congruent segmentations despite the TE case using the
vectorial $2N{\times}2N$ Green's matrix and a polarization-isotropic
source (Appendix~\ref{app:tepol}), confirming that the progression
Gram\,$\to$\,Spec\,$\to$\,Herm is not an artefact of the scalar TM
Green's matrix.

\paragraph{Choosing the appropriate landscape.}
The three landscapes introduced in the paper are not
interchangeable proxies of a single ``true'' object; they trace a
principled progression from spectrum-blind to fully hermitization-faithful,
and each is the natural first tool for a distinct question about the
array.  \emph{Gram} $B_2 = |(G^{\dagger}G+\eta^2 I)^{-1}\mathbf{1}|^2$
is the cheapest to compute (one dense solve at $z=0$), is
spectrum-blind by construction, and is the one variant with a rigorous
$\eta\to 0$ diamagnetic confinement bound (Appendix~\ref{app:diamag});
it is the safe default and the right first choice for a previously
unstudied pattern.  \emph{Spec} $S_Y = |Y(z,\eta)\mathbf{1}|^2$ makes
the spectral shift $z$ explicit -- its diagonal enters the trace of the
Metz--Neri--Rogers empirical spectral density $\rho_G(z)$ -- so it
inherits a first-principles connection to the local pseudomode content
at frequency $z$; anchored at the longest-lived eigenvalue, Spec is the
right tool for mode-selective work inside the pseudospectrum, and is
the object of first algebraic resort because it carries a spectral
density directly. 

\emph{Herm} combines the regularized direct solve $YW^\dg\one$
with the Spec contraction $Y\one$ through the pointwise identity
\eqref{eq:herm-tik-decomp}. All three landscapes are defined for
every $z\in\C$ when $\eta>0$, since $W^\dg W+\eta^2 I$,
$G^\dg G+\eta^2 I$, and $B(z,\eta)$ are all invertible with operator
norm bounded by $1/\eta$~\cite{feinberg1997nonhermitian}; what
distinguishes them is numerical conditioning, not existence. Spec and
Gram form the normal equations, so
$\mathrm{cond}(W^\dg W+\eta^2 I)=\mathrm{cond}(B)^2$, while Herm
solves the equivalent augmented system
$B[\mathbf a;\mathbf b]=[\one;\one]$ with condition number
$\mathrm{cond}(B)$. At the working value $\eta=5\times10^{-3}\|G\|_F$
both are modest (of order $10^{3}$ and $30$ respectively at
$N=1500$), so the distinction matters only for much smaller $\eta$
or for iterative solvers; the price Herm pays for the better
conditioning is a $2N\times 2N$ solve rather than the $N\times N$
solves of Gram and Spec. At $z=0$, Spec coincides with Gram, but
Herm does not reduce to Gram because its off-diagonal blocks apply a
different filter (see \eqref{eq:herm-tik-decomp}). We therefore run all three on any new array: disagreement flags a
fragile mode identification, and agreement (as observed on Fibonacci,
Eisenstein prime, and Thue--Morse) corroborates the mode localization
independently of the block choice.

\begin{figure}[!ht]
    \centering
    {\setlength{\panelW}{32mm}%
\begin{tabular}{@{}ccccc@{}}
\lbl{images/ThueMorse/pointcloud.png}{point cloud} &
\lbl{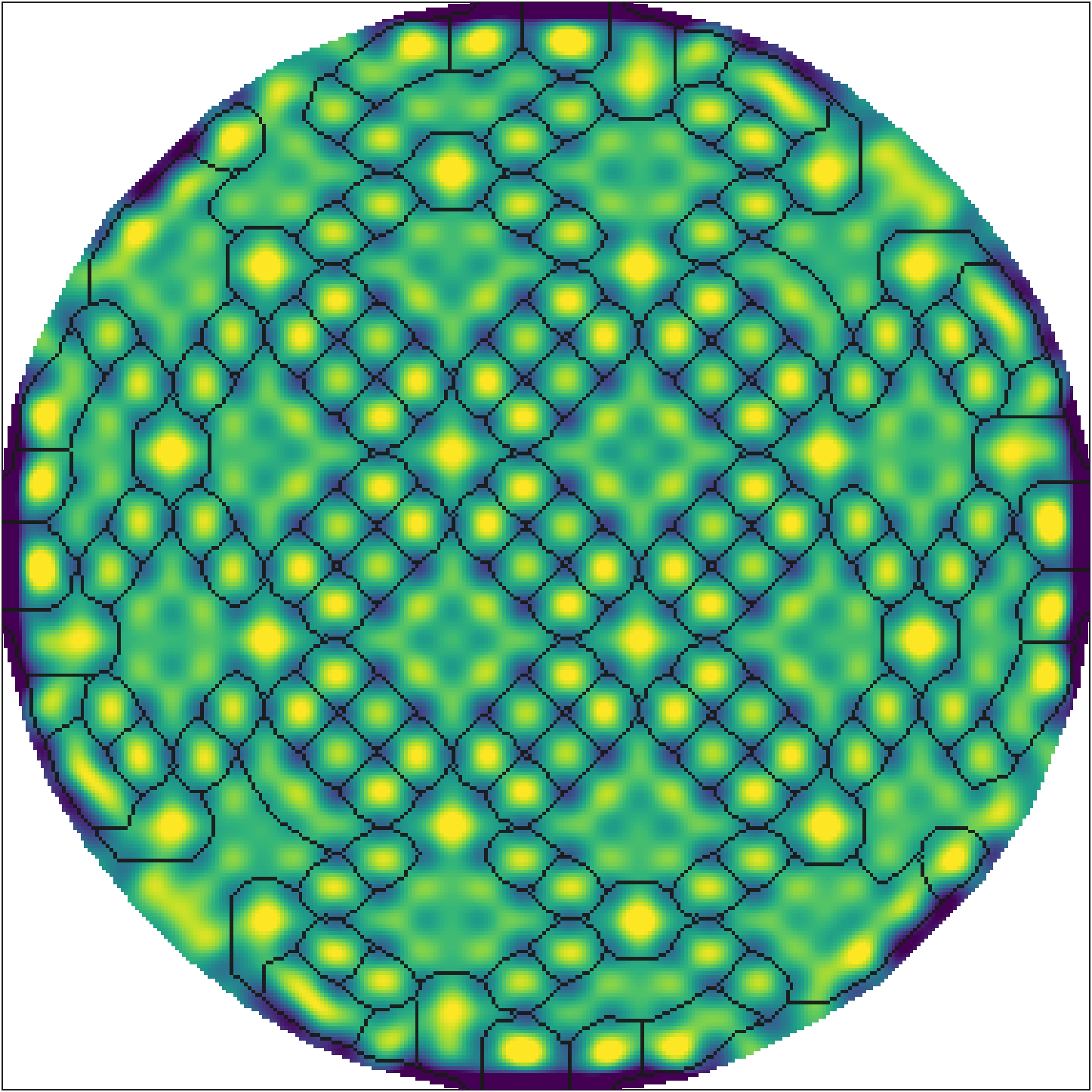}{Herm + watershed} &
\lbl{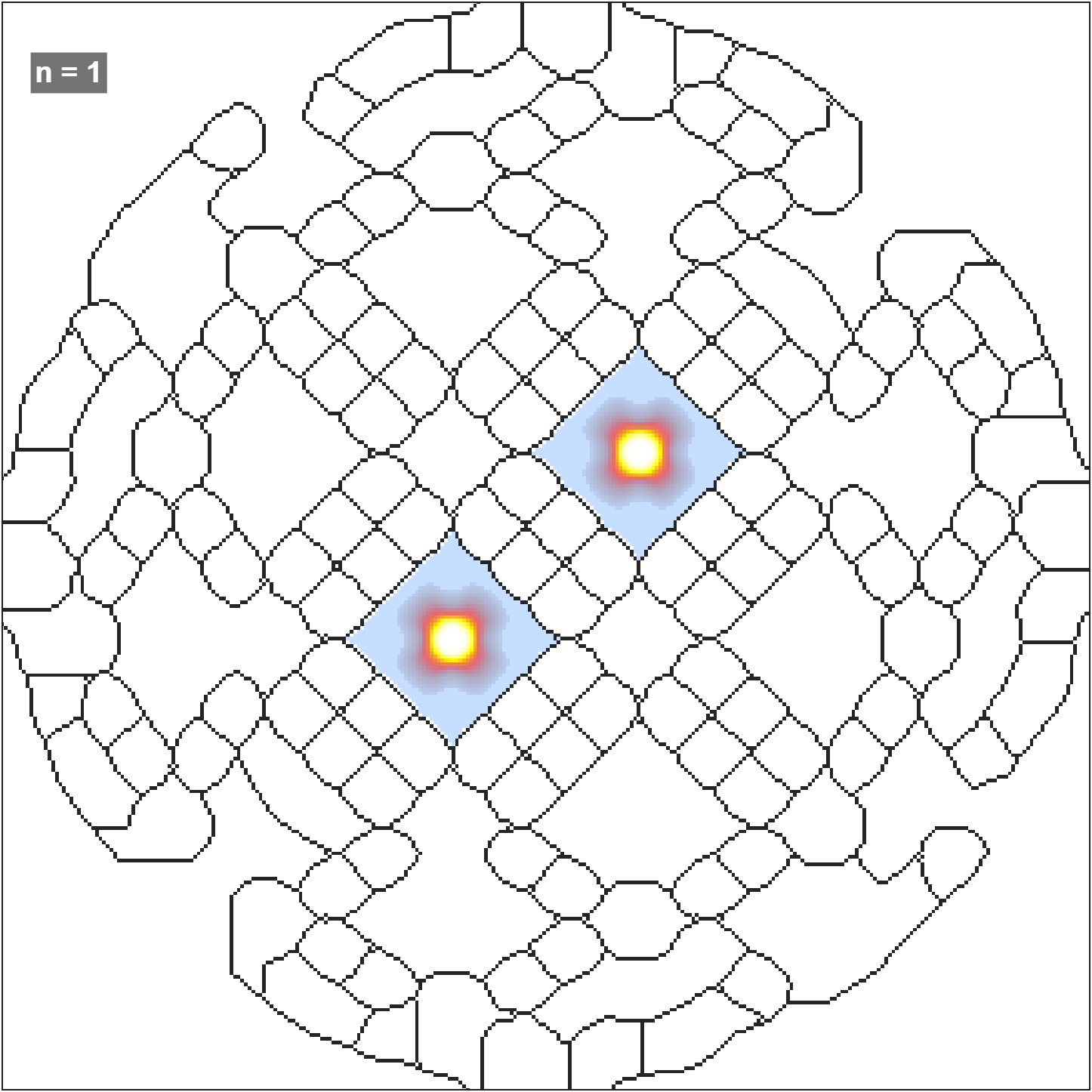}{$n=1$} &
\lbl{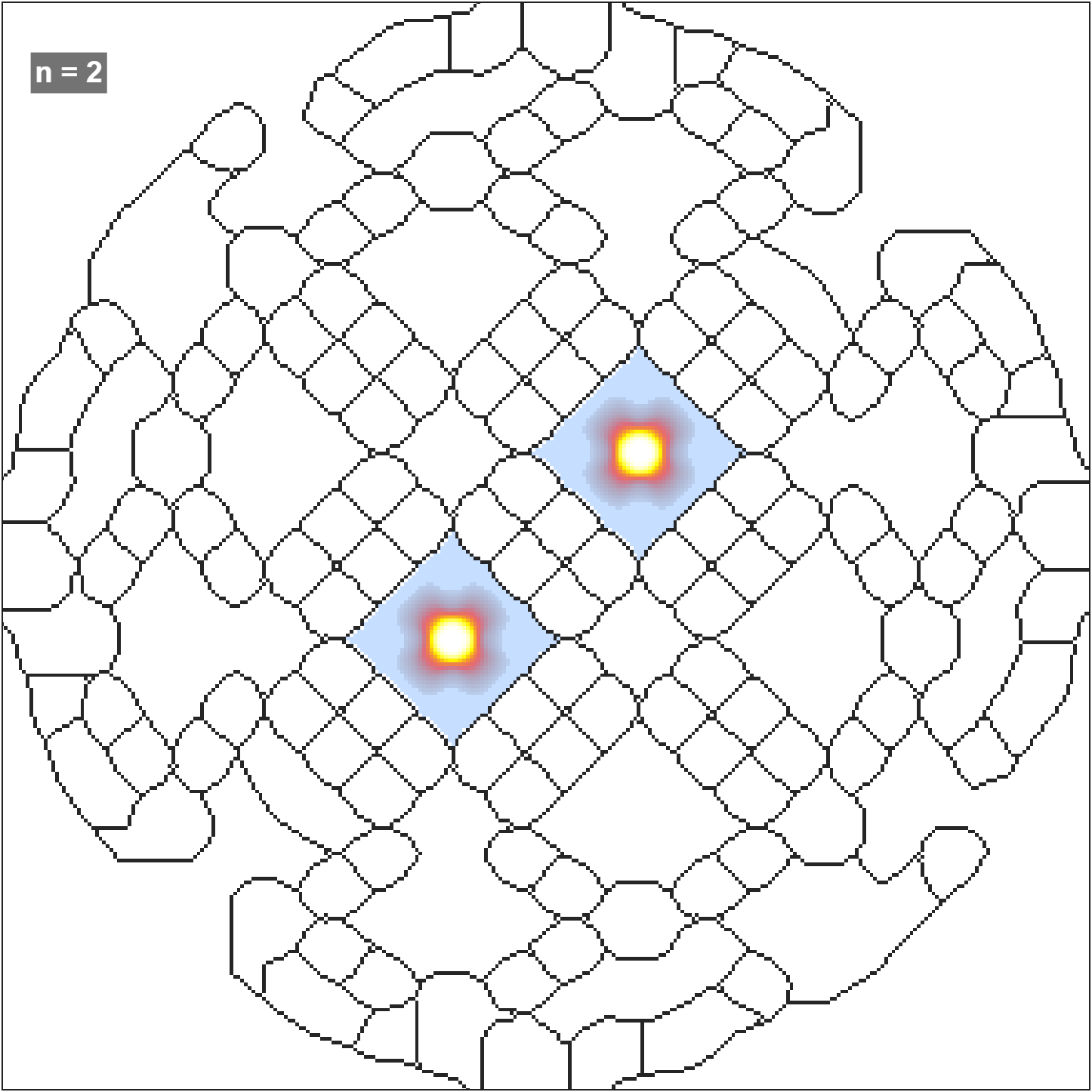}{$n=2$} &
\lbl{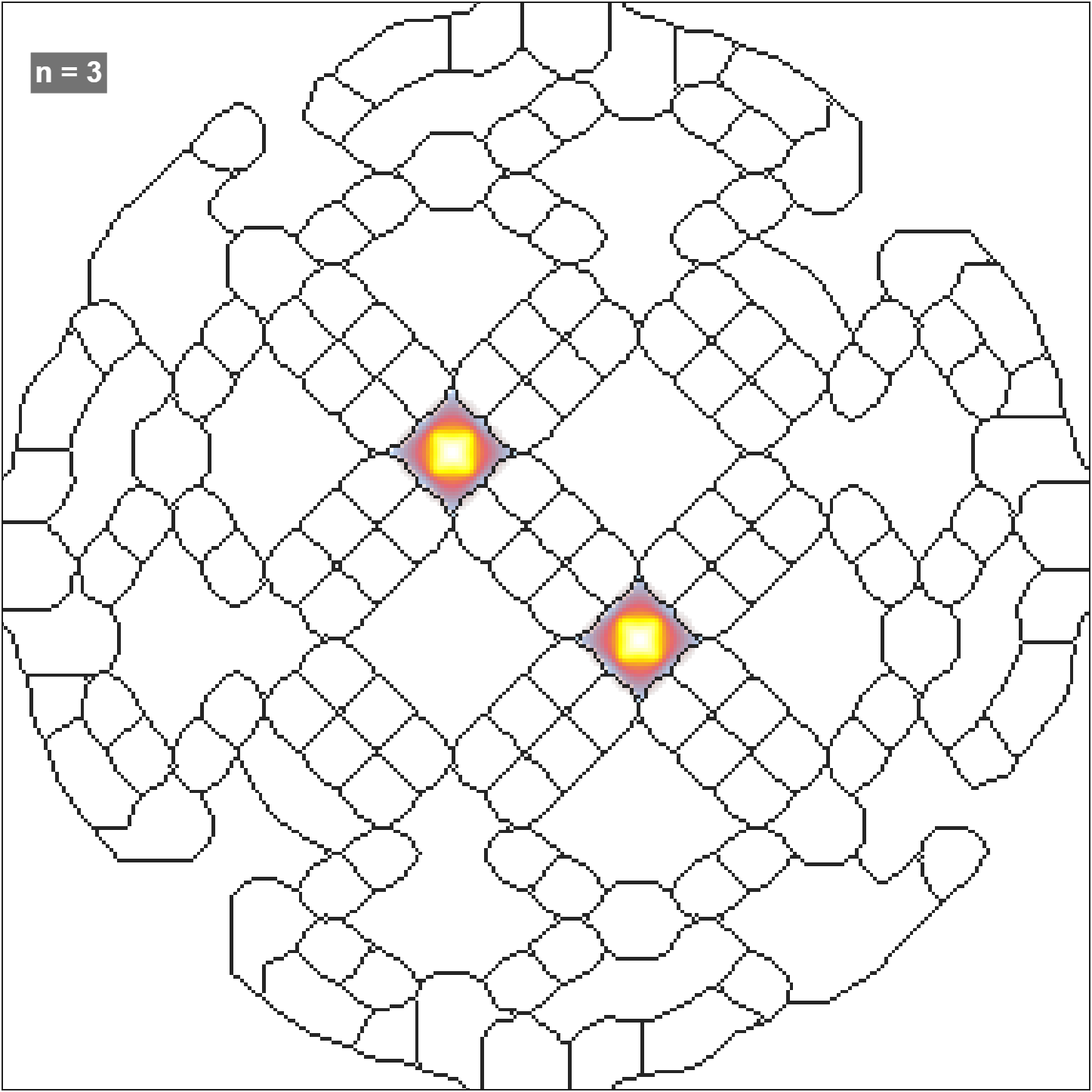}{$n=3$} \\
\lbl{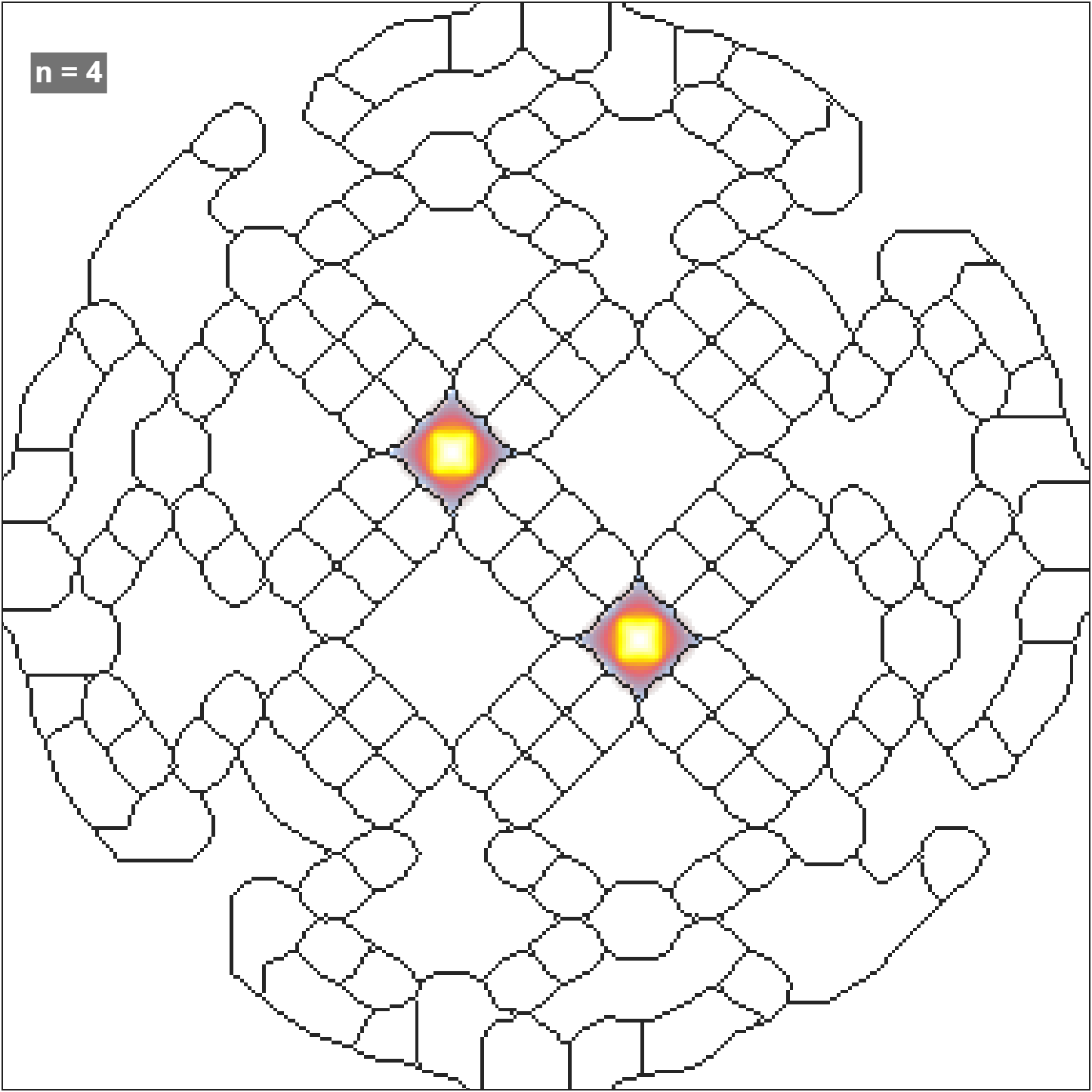}{$n=4$} &
\lbl{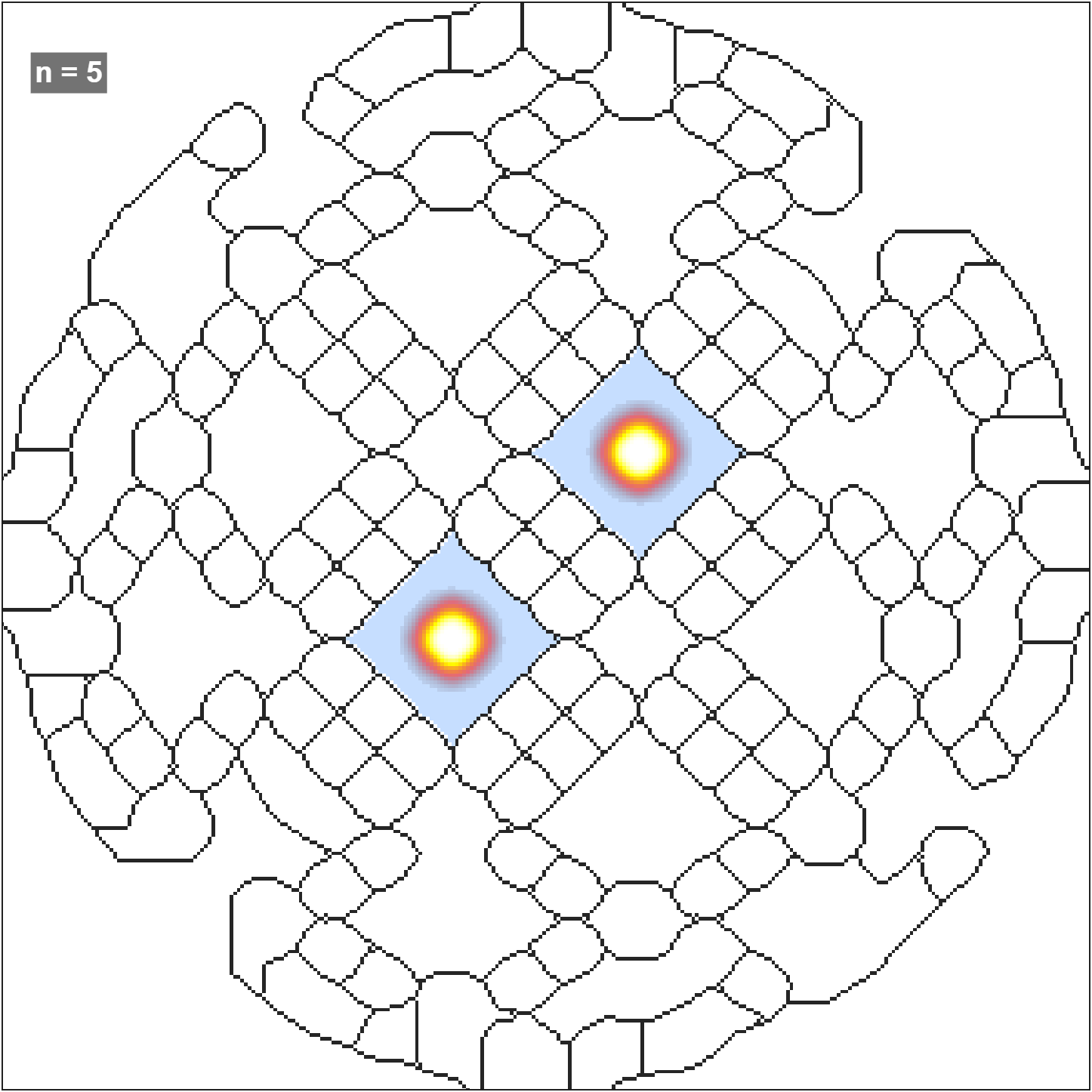}{$n=5$} &
\lbl{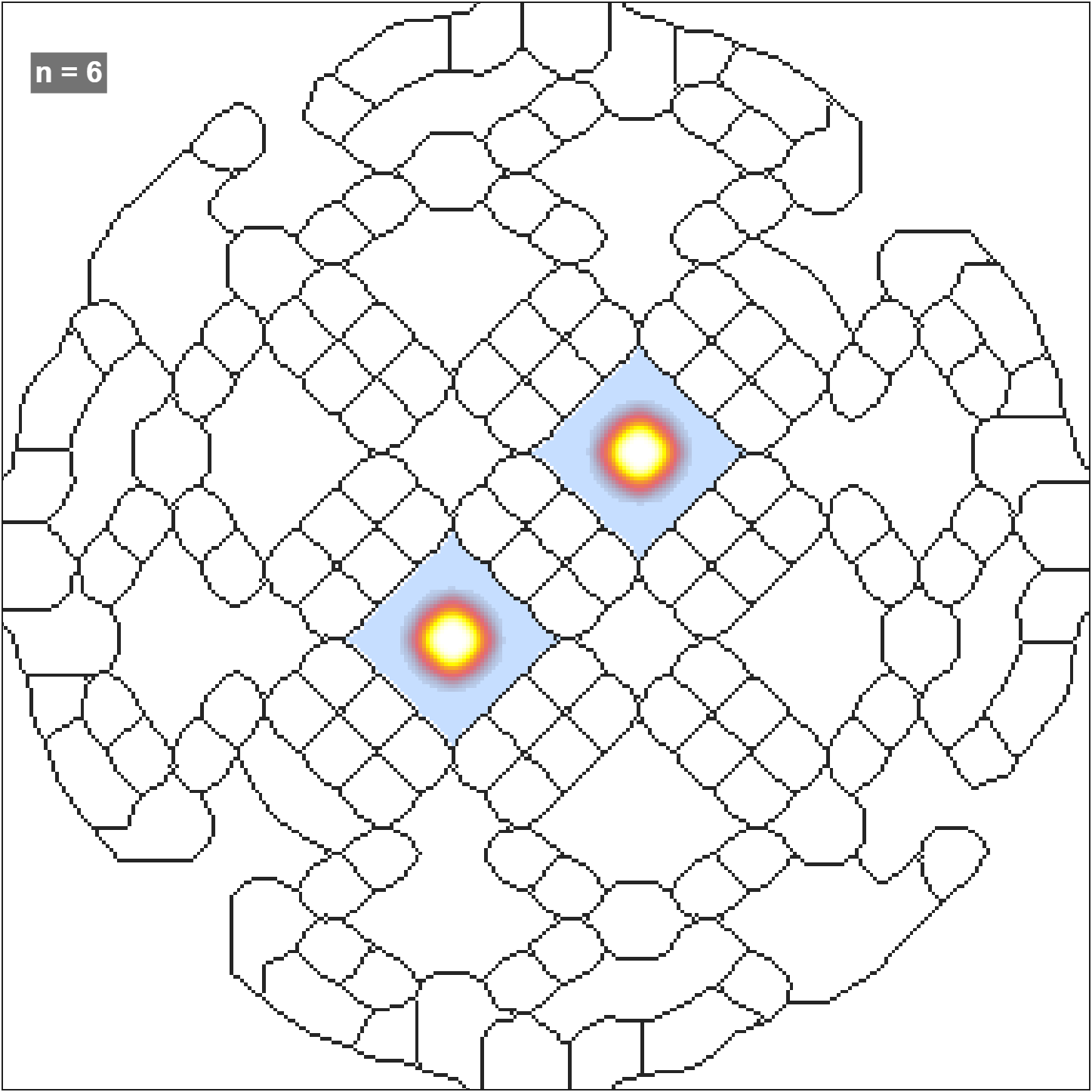}{$n=6$} &
\lbl{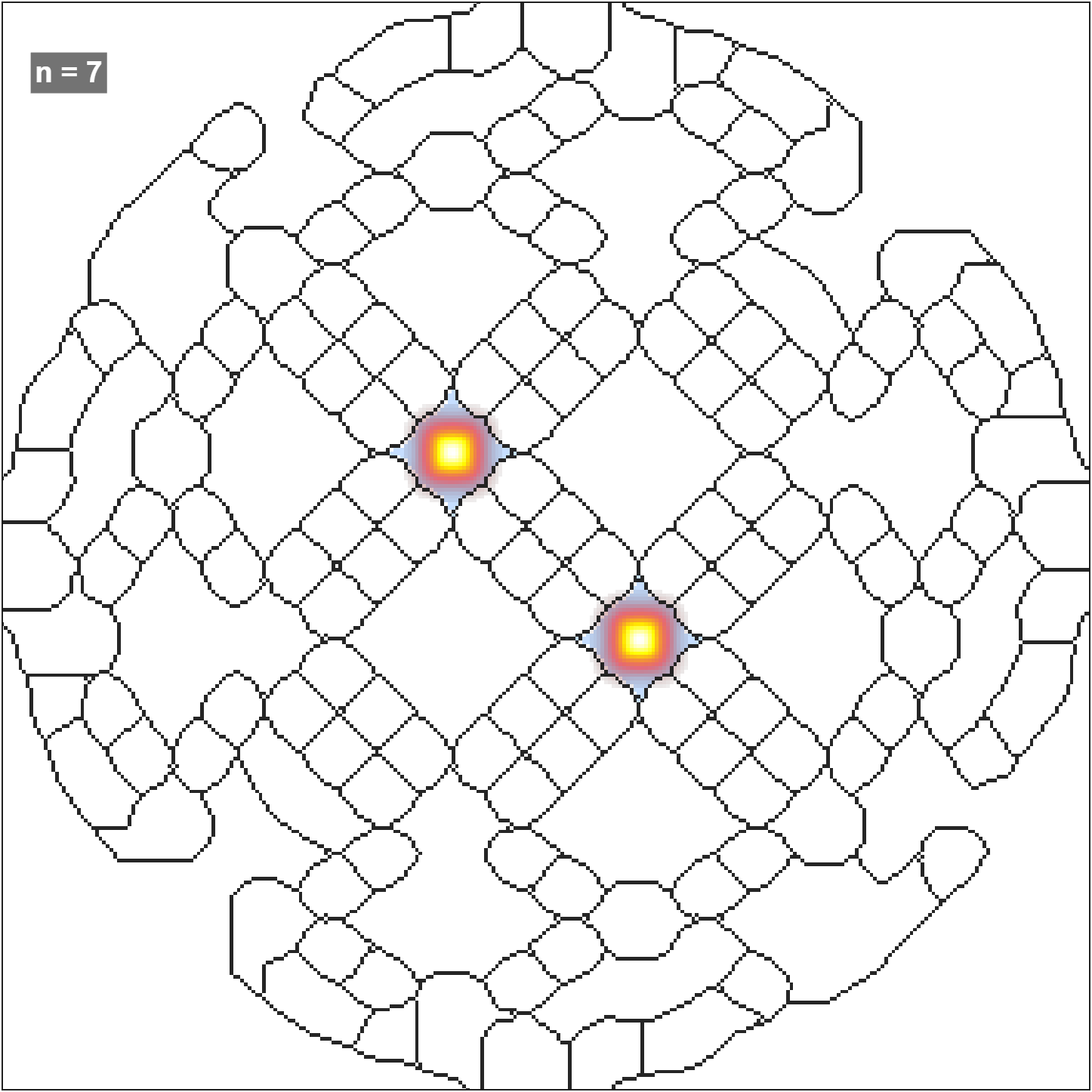}{$n=7$} &
\lbl{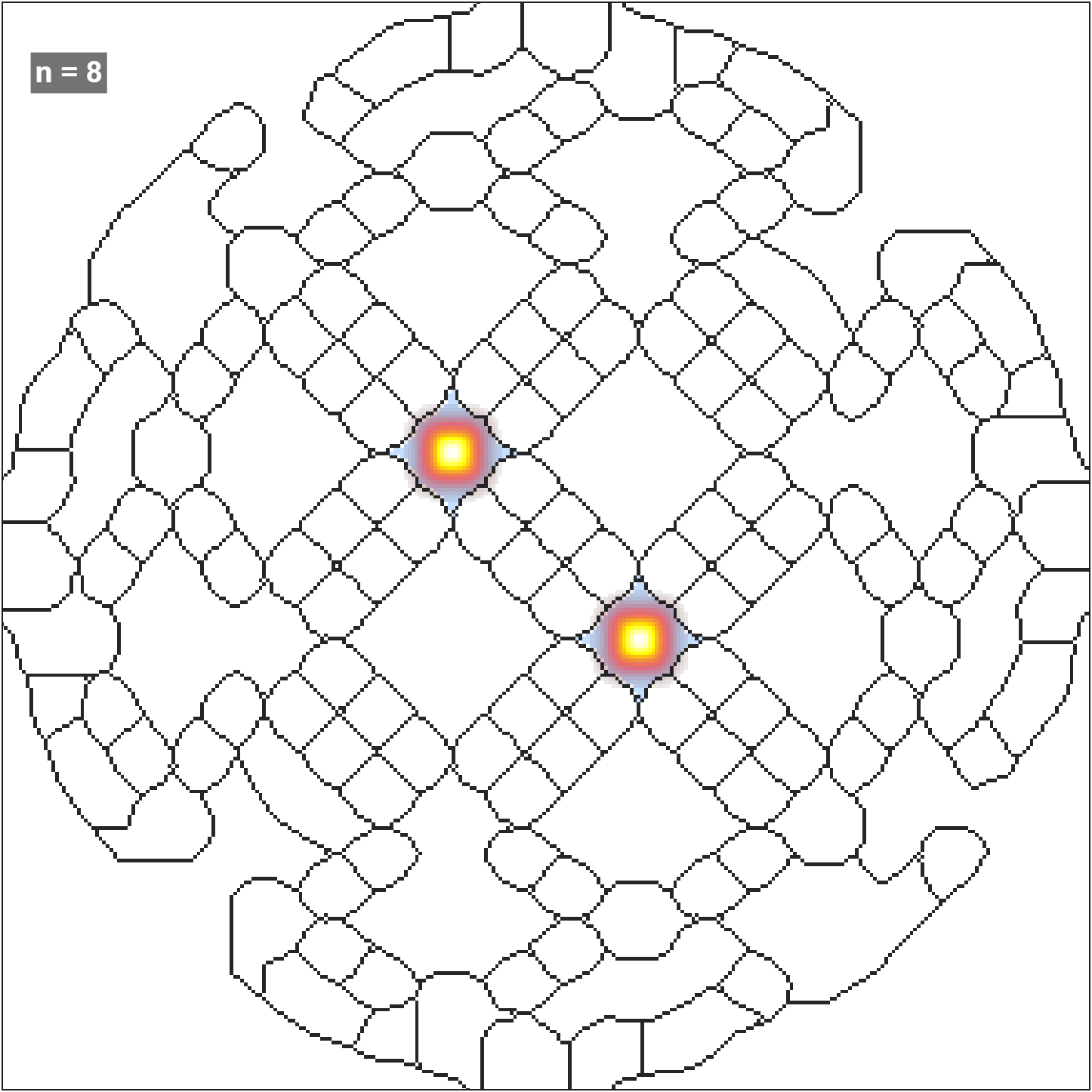}{$n=8$} \\
\lbl{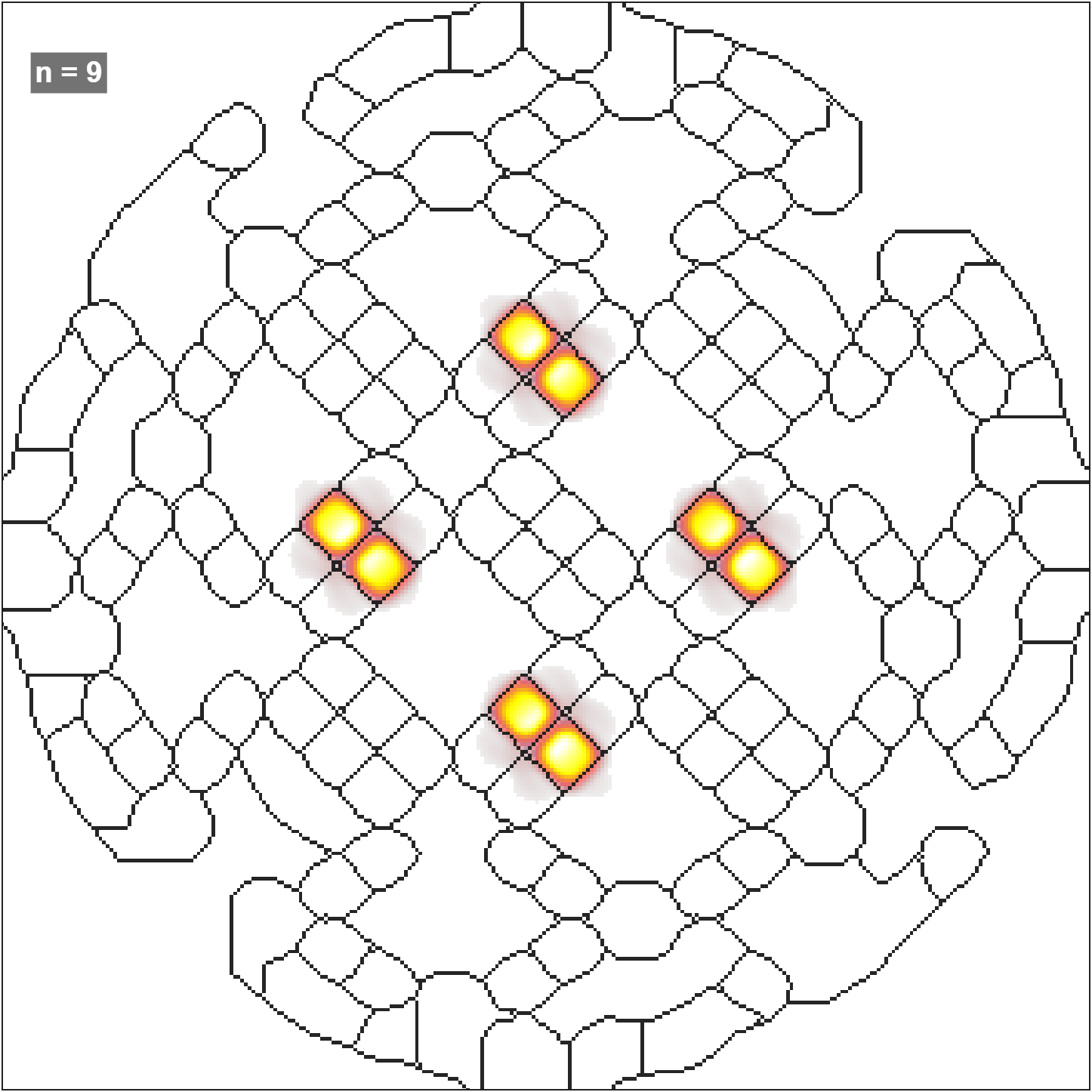}{$n=9$} &
\lbl{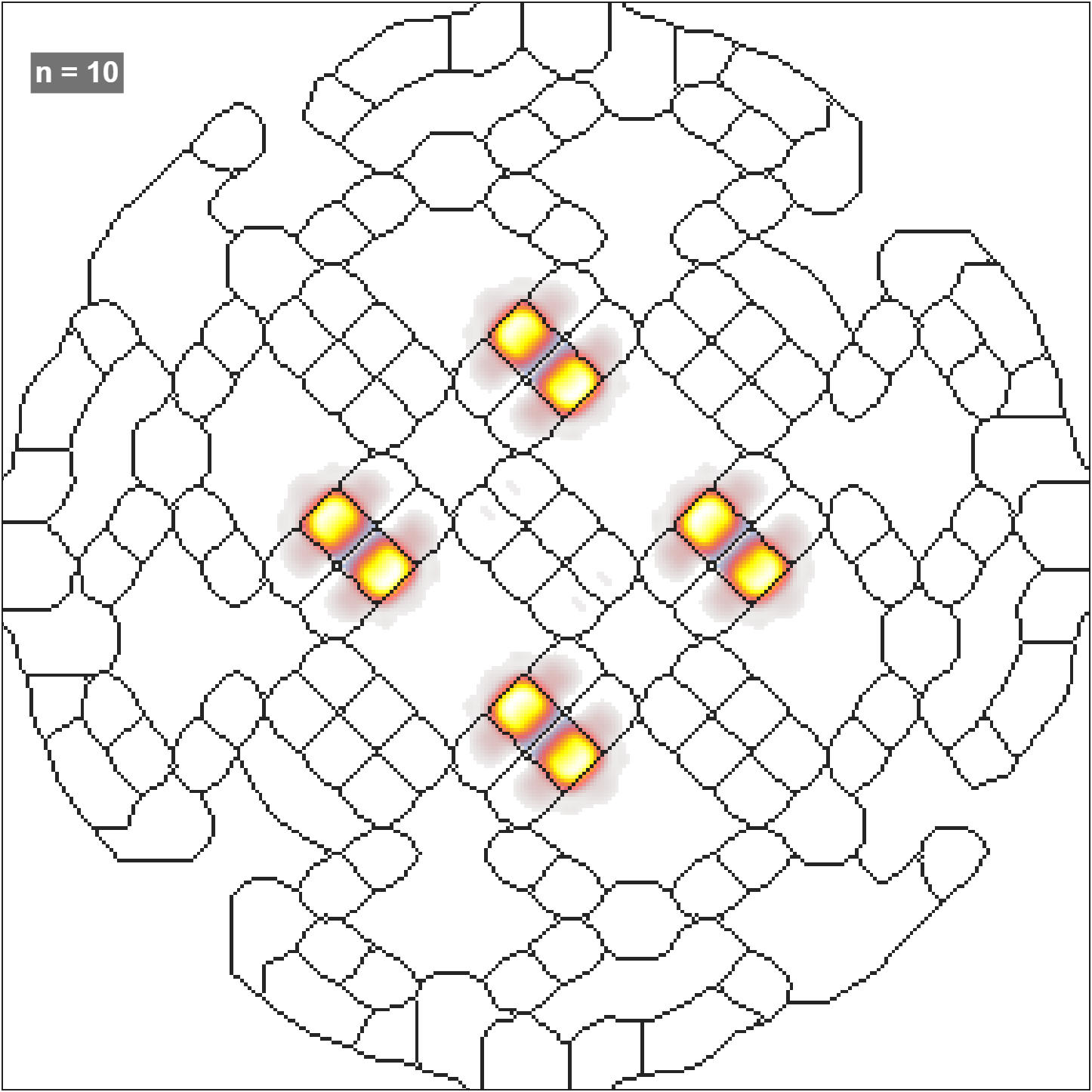}{$n=10$} &
\lbl{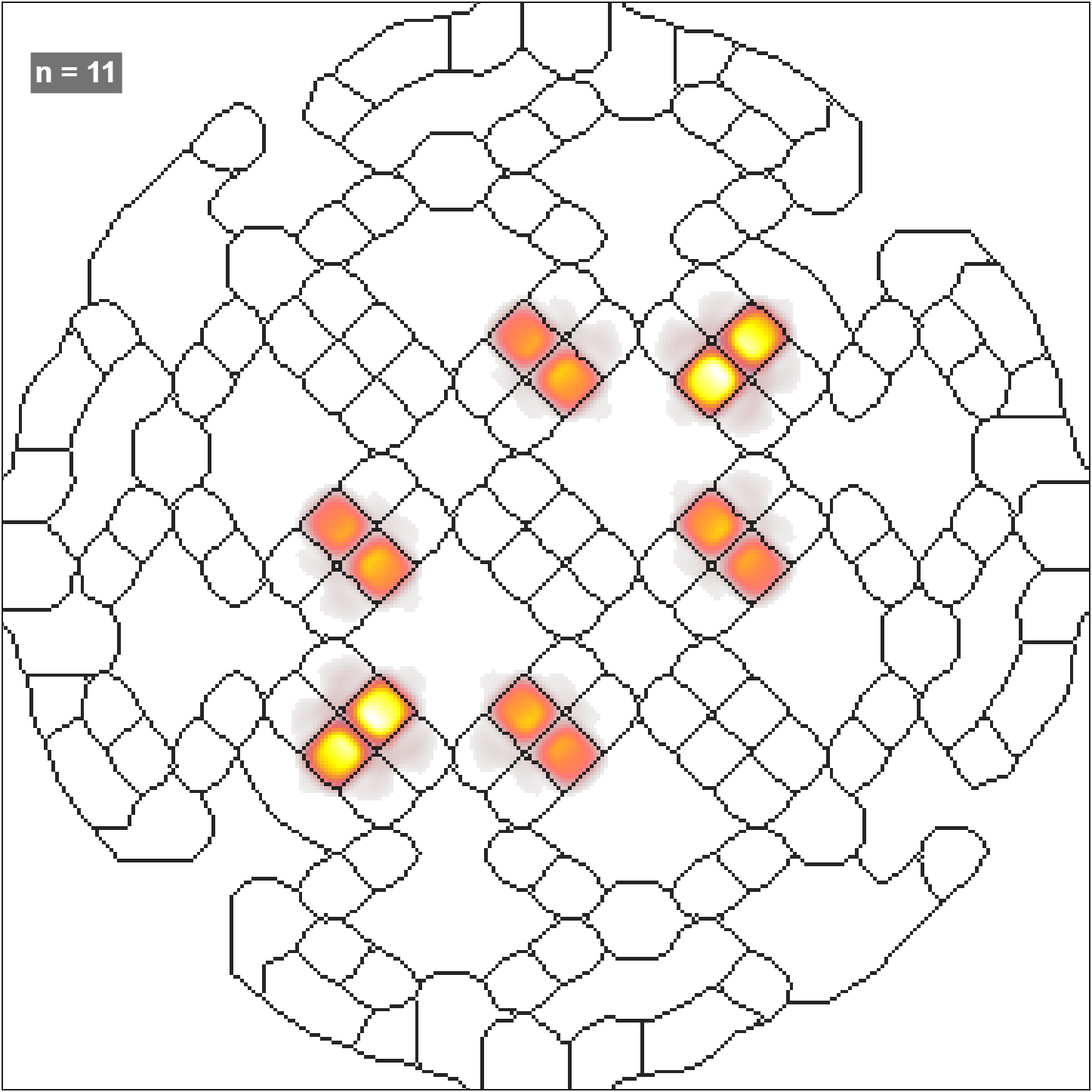}{$n=11$} &
\lbl{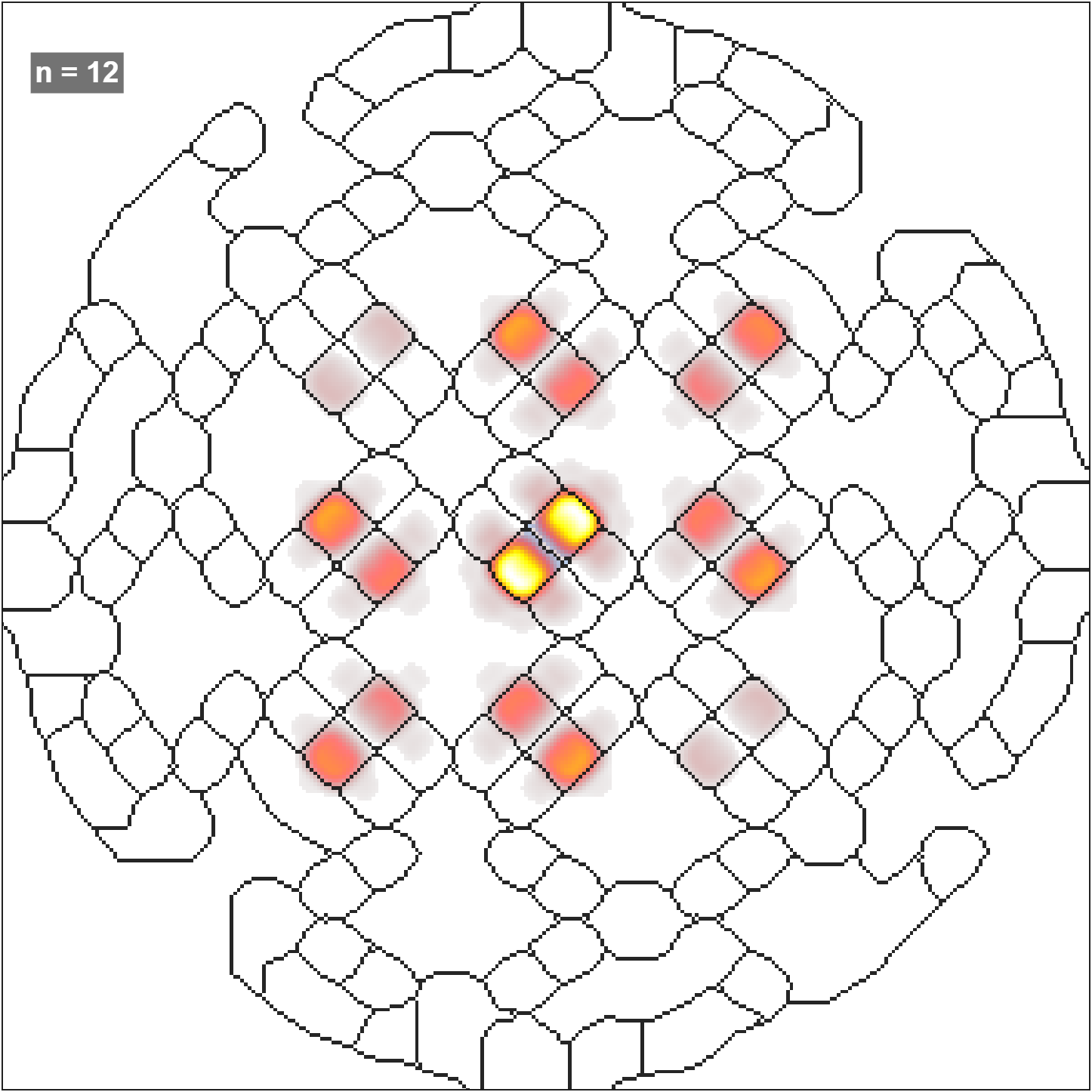}{$n=12$} &
\lbl{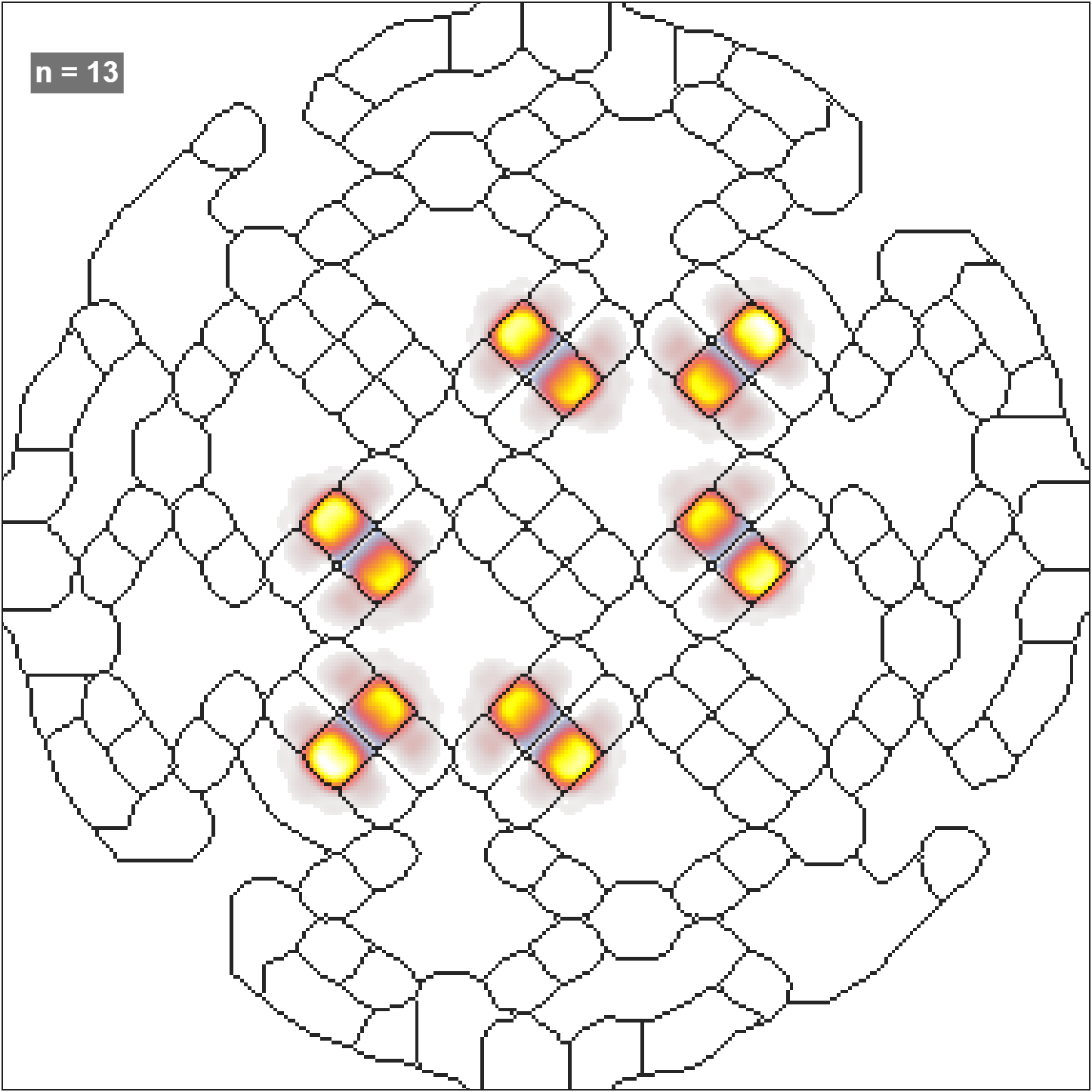}{$n=13$} \\
\end{tabular}}
    \caption{Thue--Morse array, TM, $N{=}3500$.  Starting from the
    point cloud and the Herm landscape with its watershed
    segmentation (top row, left two panels), the remaining 13 panels
    show the first 13 long-lived eigenmodes individually, overlaid on
    the same Herm watershed.  The index $n$ labels the eigenmode by
    decreasing lifetime, so $n=1$ is the longest-lived mode of $G$ ---
    the mode with the smallest imaginary part of its complex
    eigenvalue --- and $n=13$ the thirteenth longest-lived mode.  For
    every mode the host cells predicted by the landscape are tinted in
    blue and the mode intensity is shown in hot; the peaks of each
    eigenmode sit inside the predicted basins without per-mode
    adjustment, and the $C_8$ symmetry of the pattern is reflected in
    the correspondence between symmetric mode pairs and symmetric
    basin pairs.}
    \label{fig:thuemorse-modes}
\end{figure}

\paragraph{Mode-by-mode identification on the Thue--Morse array.}
Figure~\ref{fig:thuemorse-modes} makes the mode-by-mode picture
explicit for the deterministic Thue--Morse array under TM polarization.
The 3\,$\times$\,5 panel first shows the point-dipole cloud and the Herm
landscape with its watershed segmentation, and then follows the first
13 long-lived eigenmodes individually with the corresponding host cells
tinted in blue: the peaks of each eigenmode sit inside the very basins
predicted by the Herm landscape, without any per-mode re-tuning.  
\begin{figure}[!ht]
    \centering
    \noindent

    \begin{tabular}{@{}c@{\hspace{2pt}}ccc@{}}
& \small point cloud & \small Spec landscape + watershed & \small top-100 modes on watershed \\
\rlbl{Pinwheel}          & \pnl{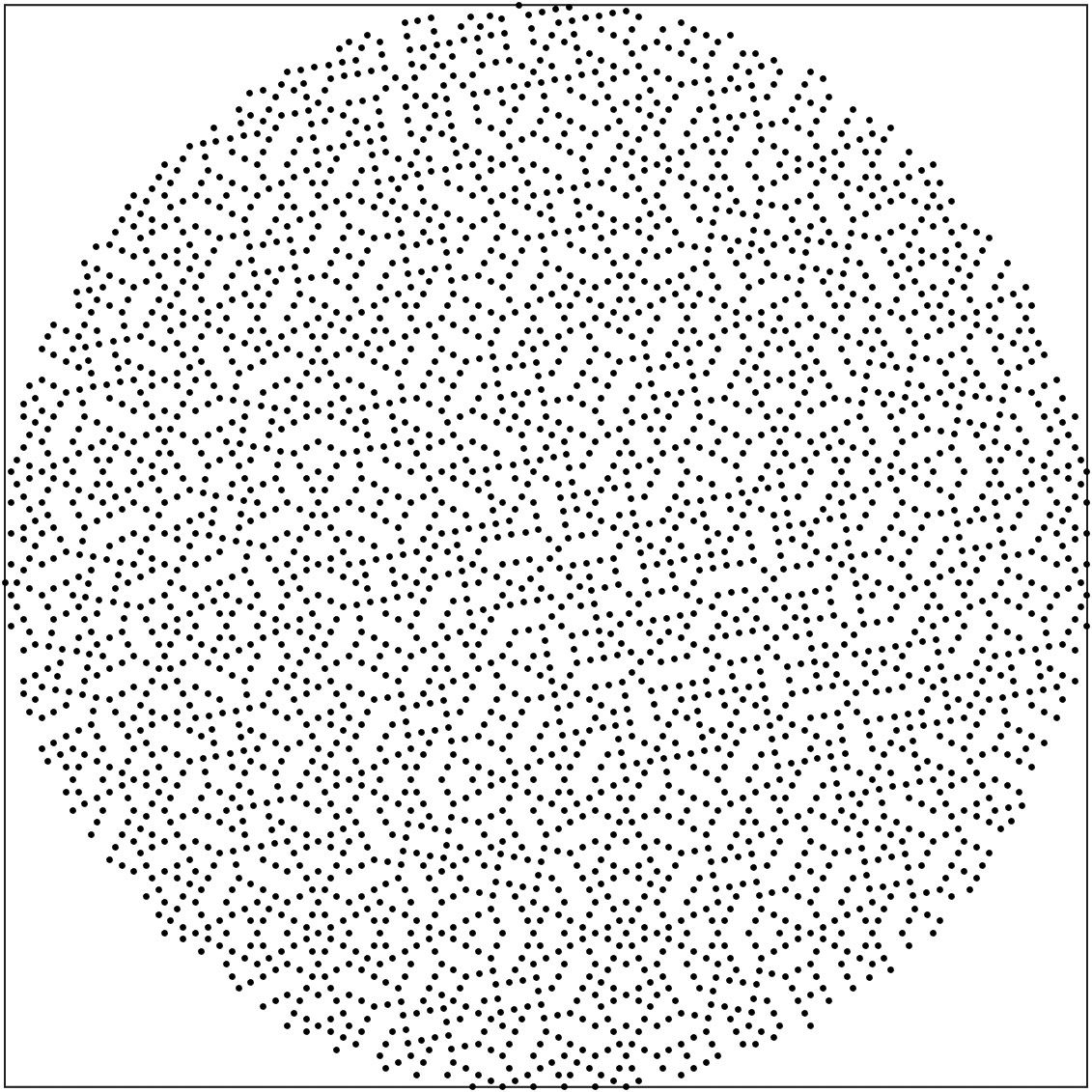}       & \pnl{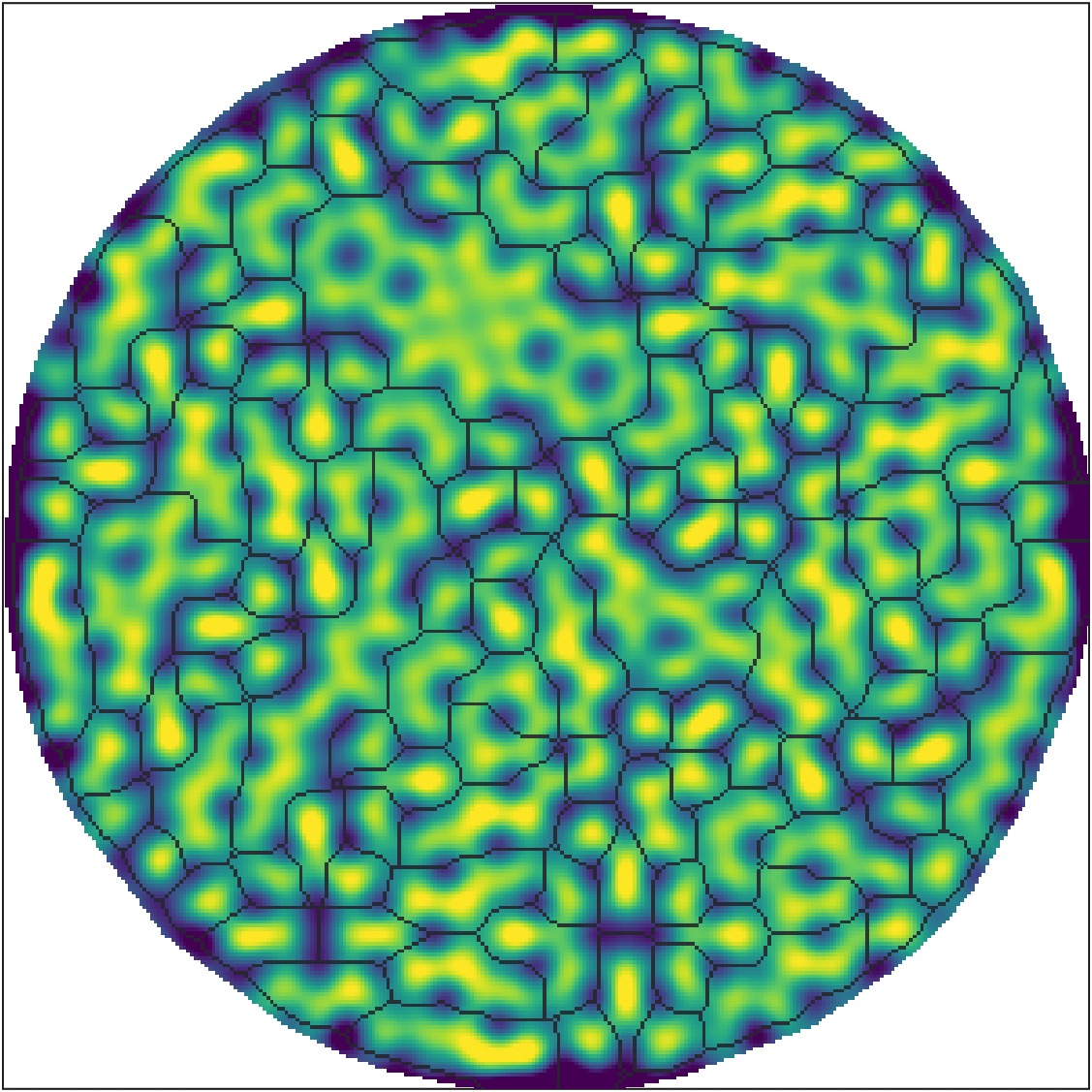}       & \pnl{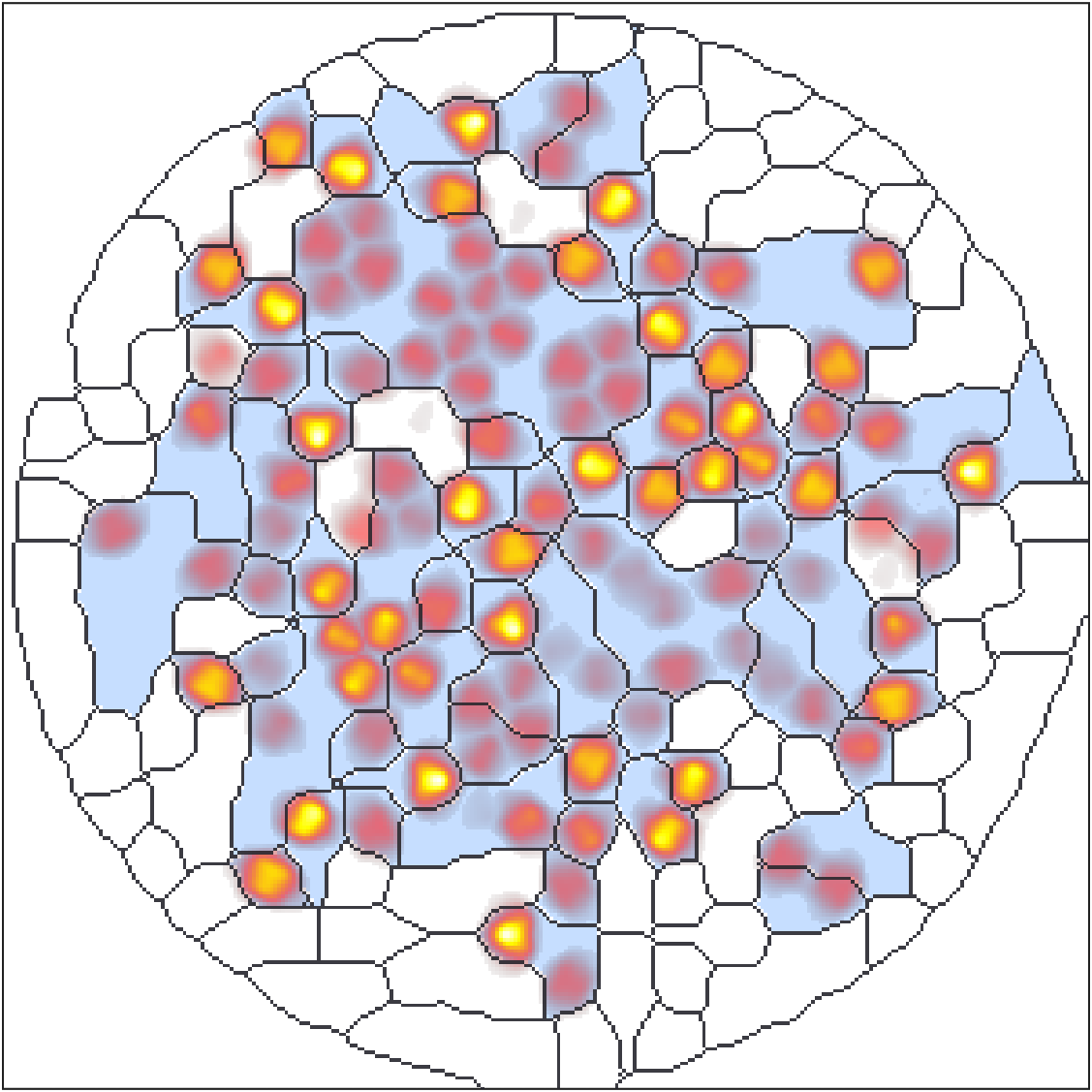}       \\
\rlbl{Halton}             & \pnl{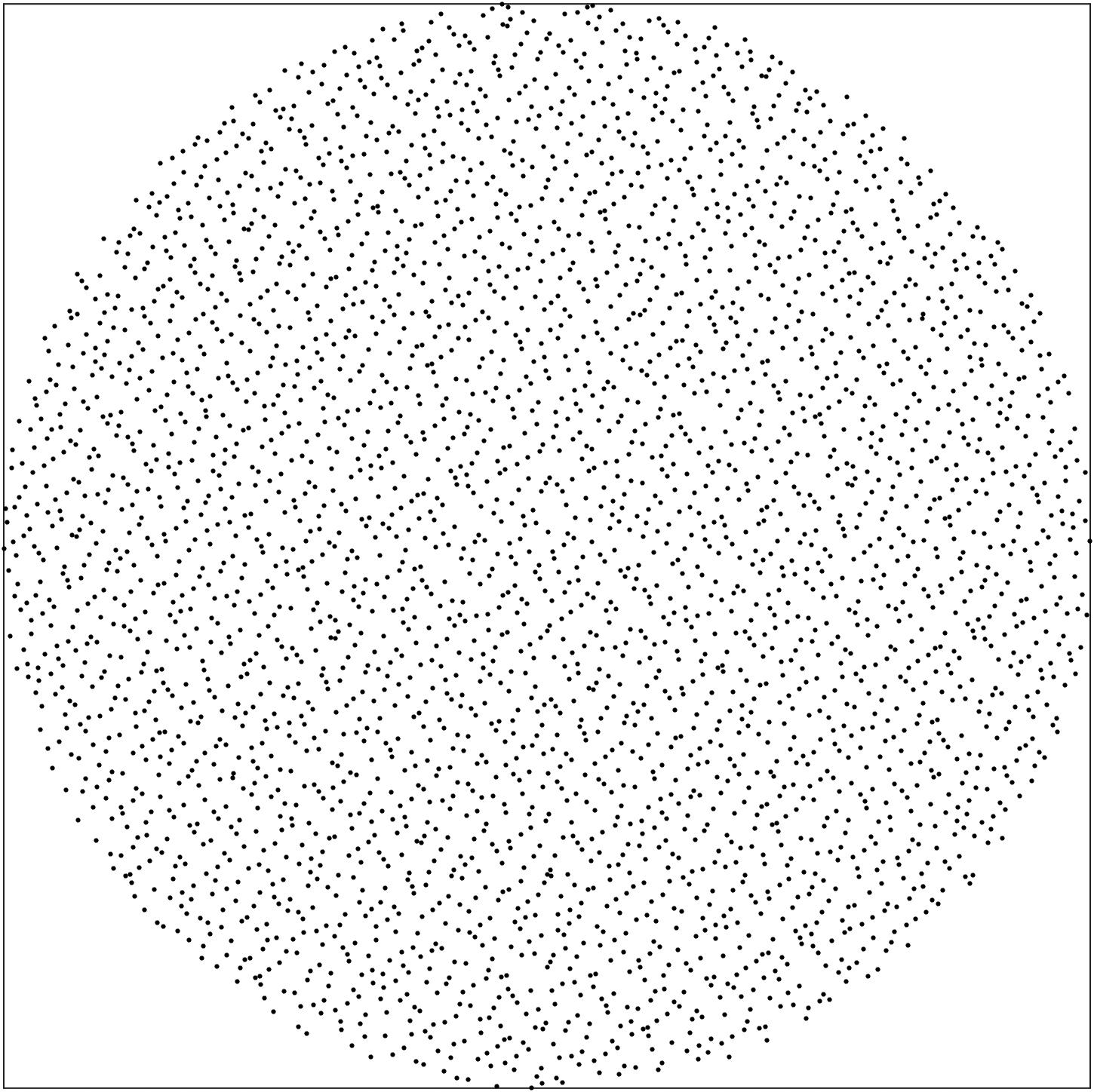}          & \pnl{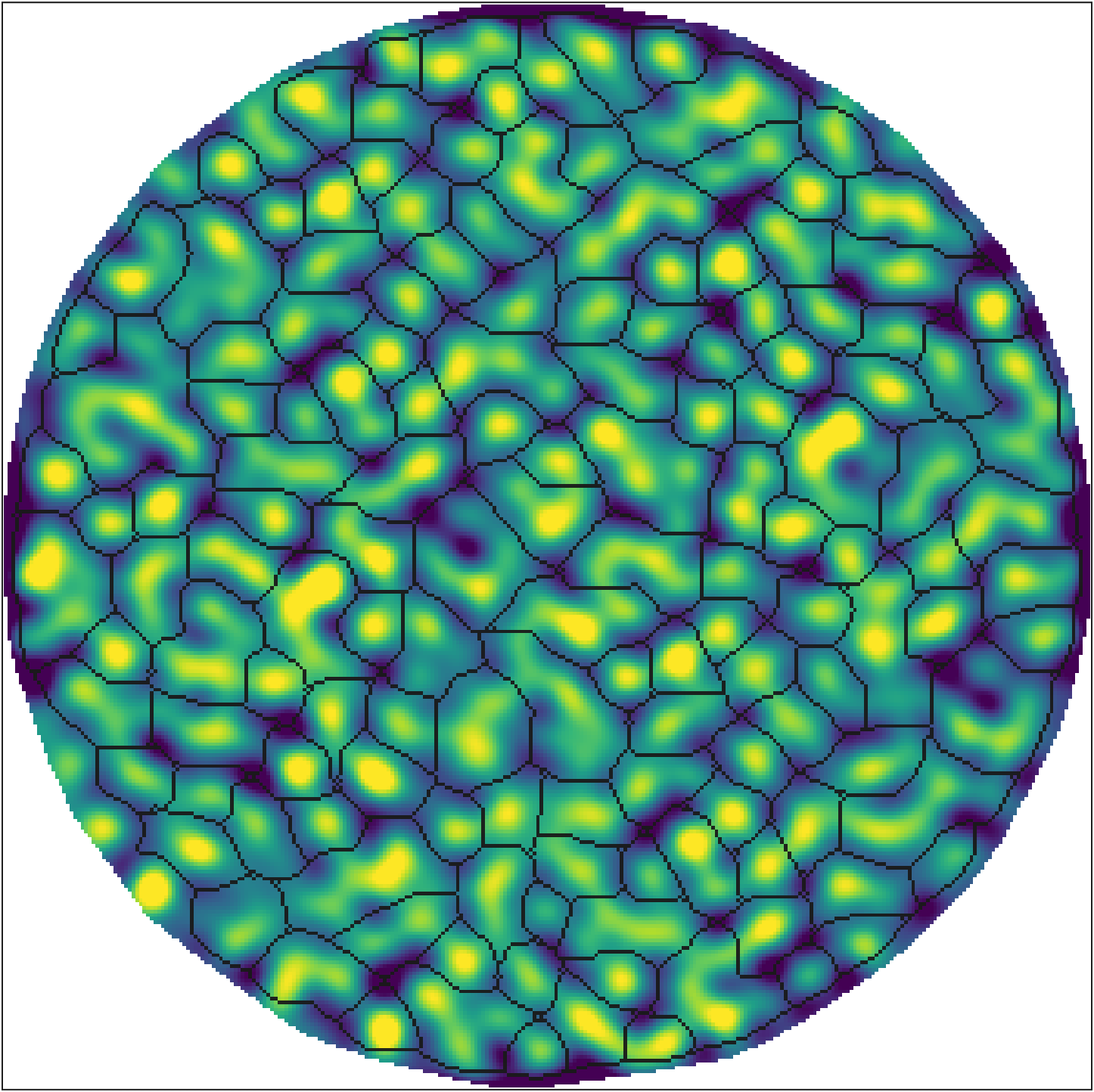}          & \pnl{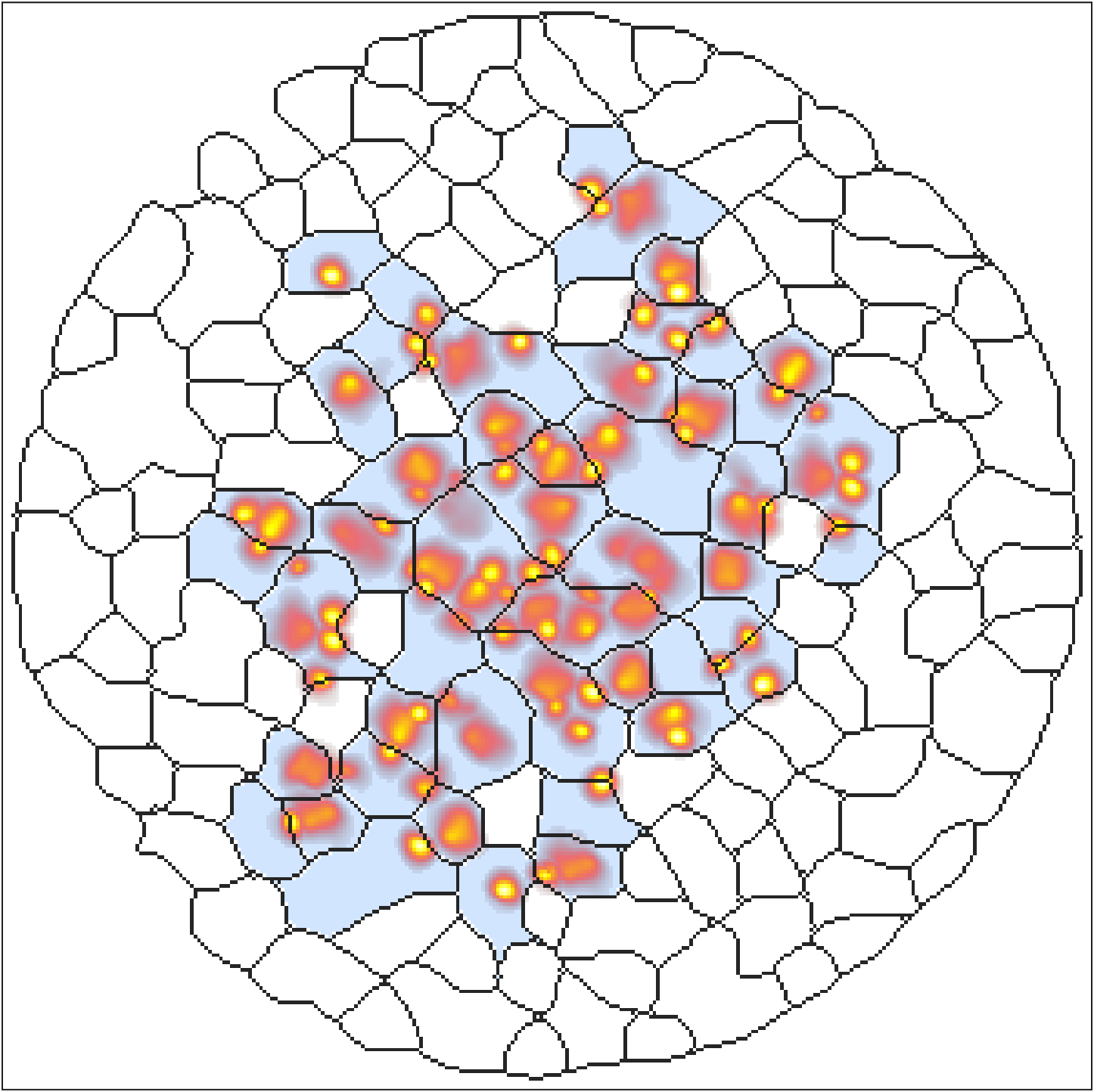}          \\
\rlbl{Eisenstein Prime}   & \pnl{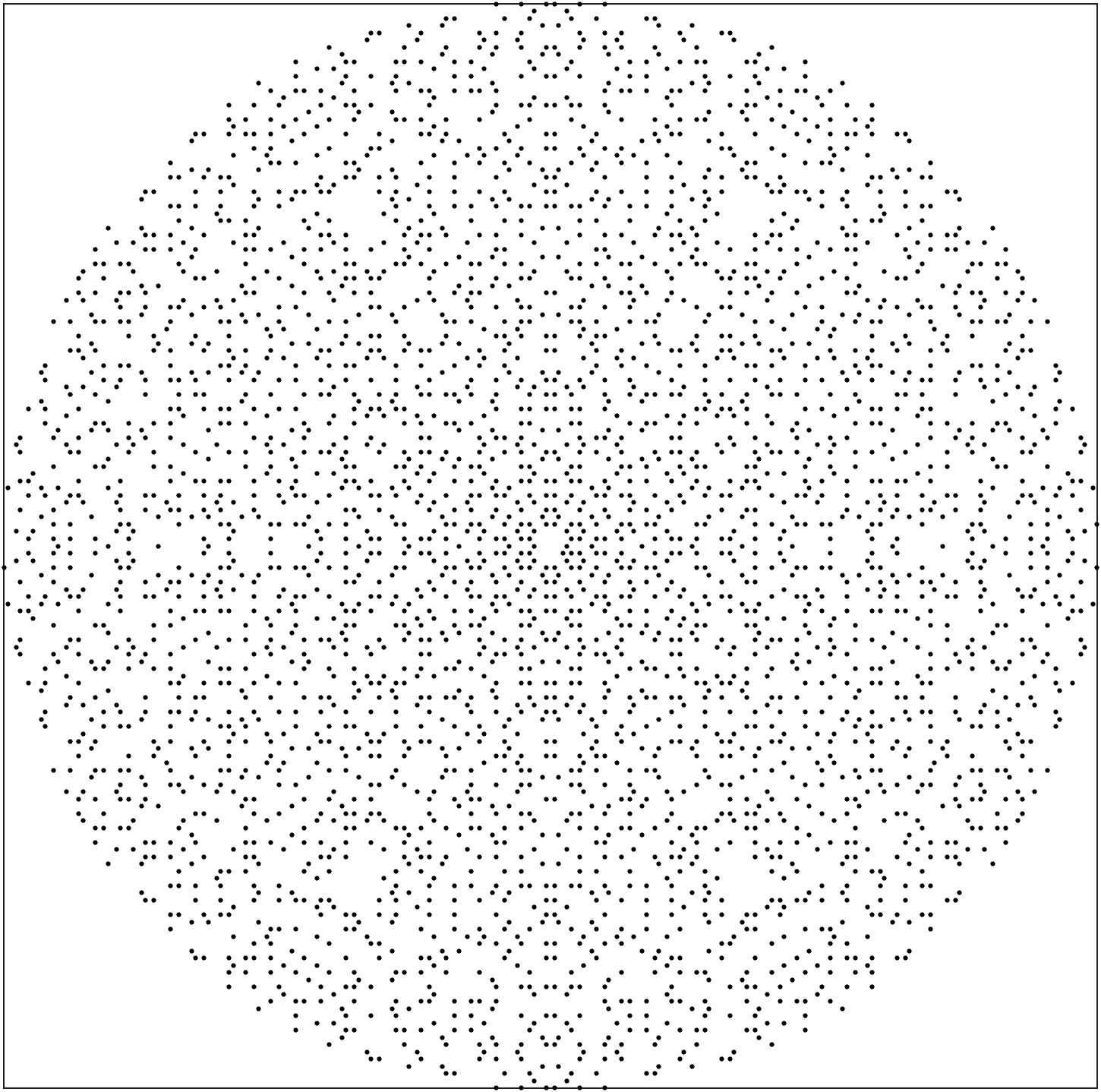} & \pnl{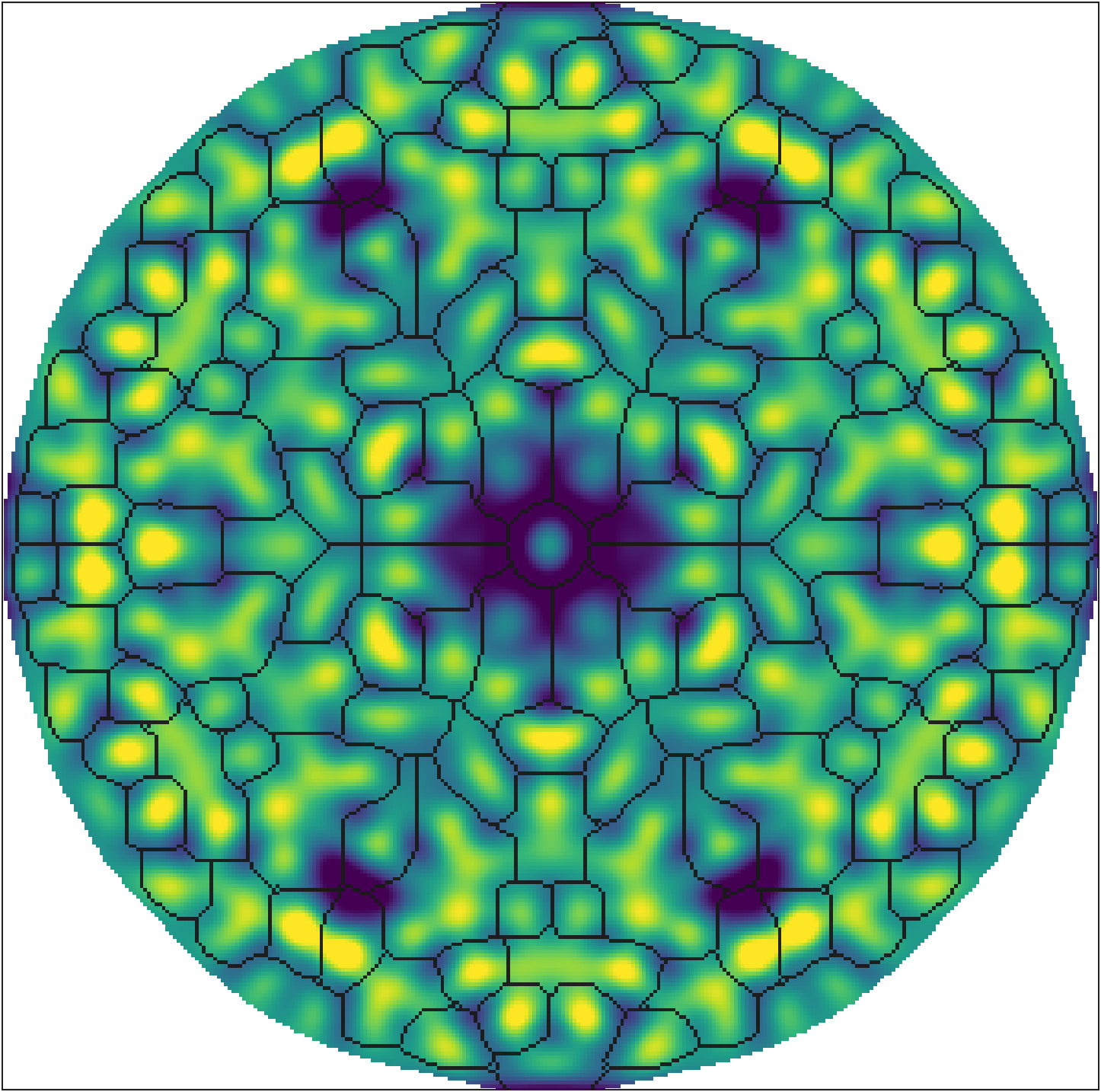} & \pnl{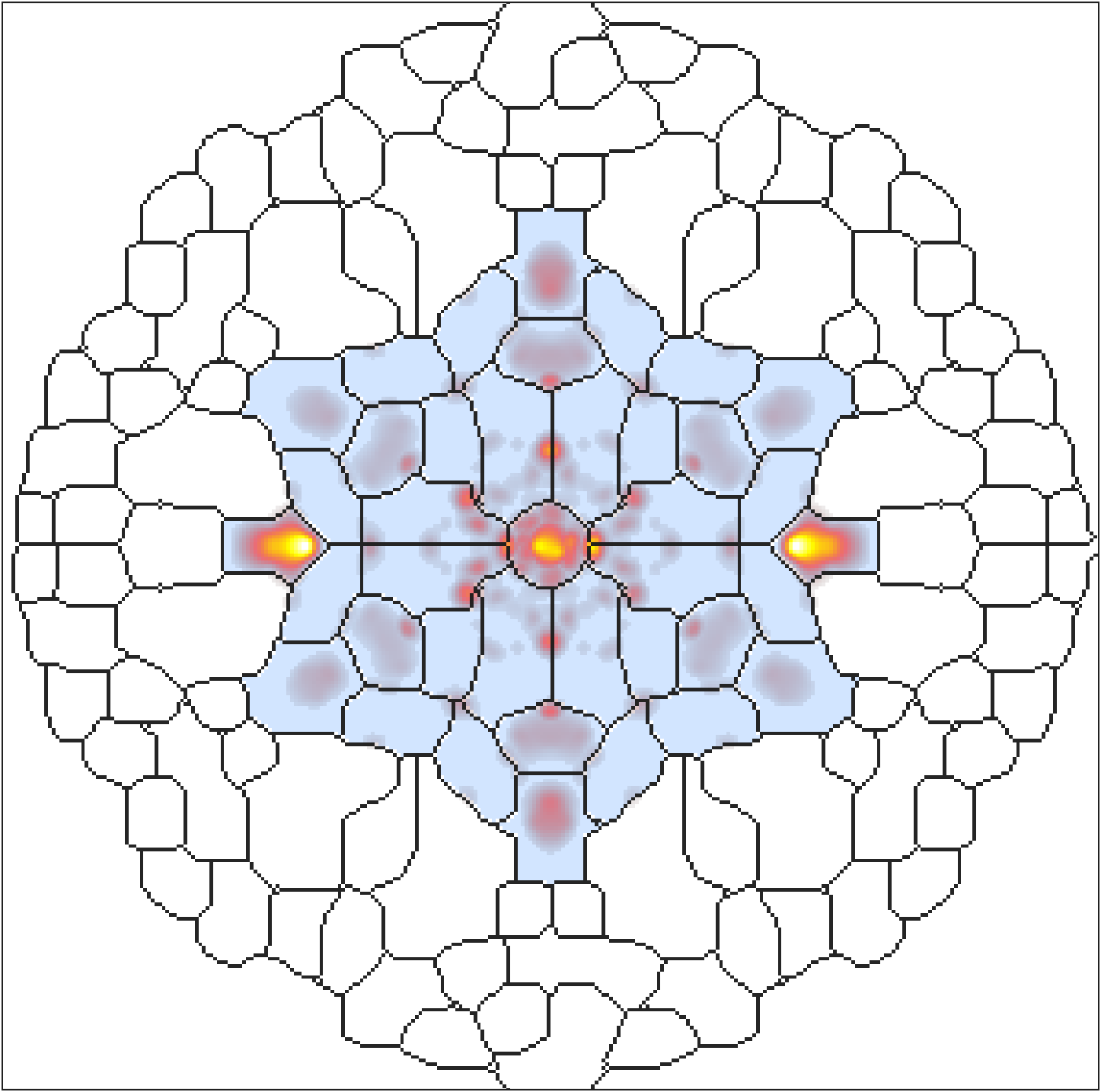} \\
\rlbl{Fibonacci}          & \pnl{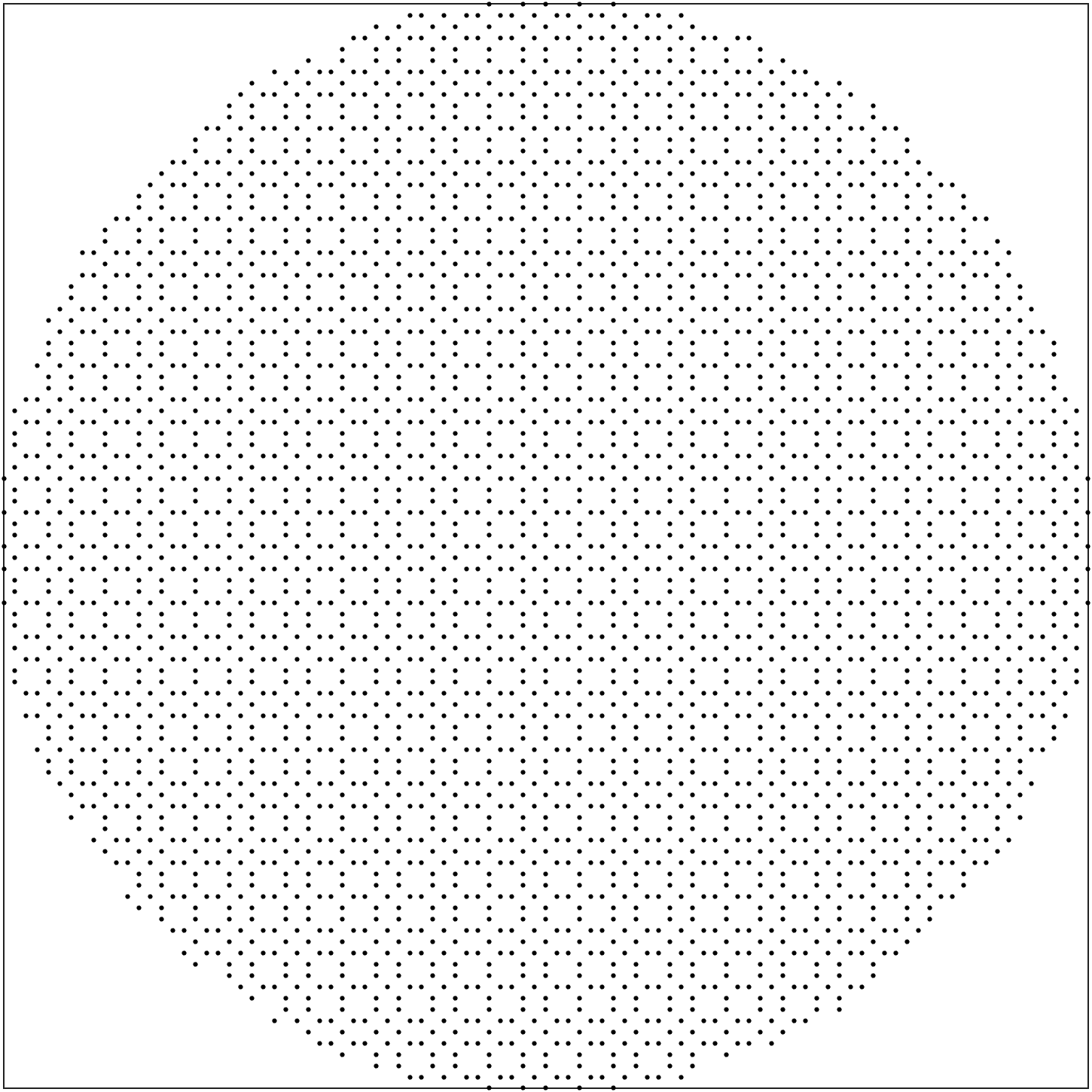}       & \pnl{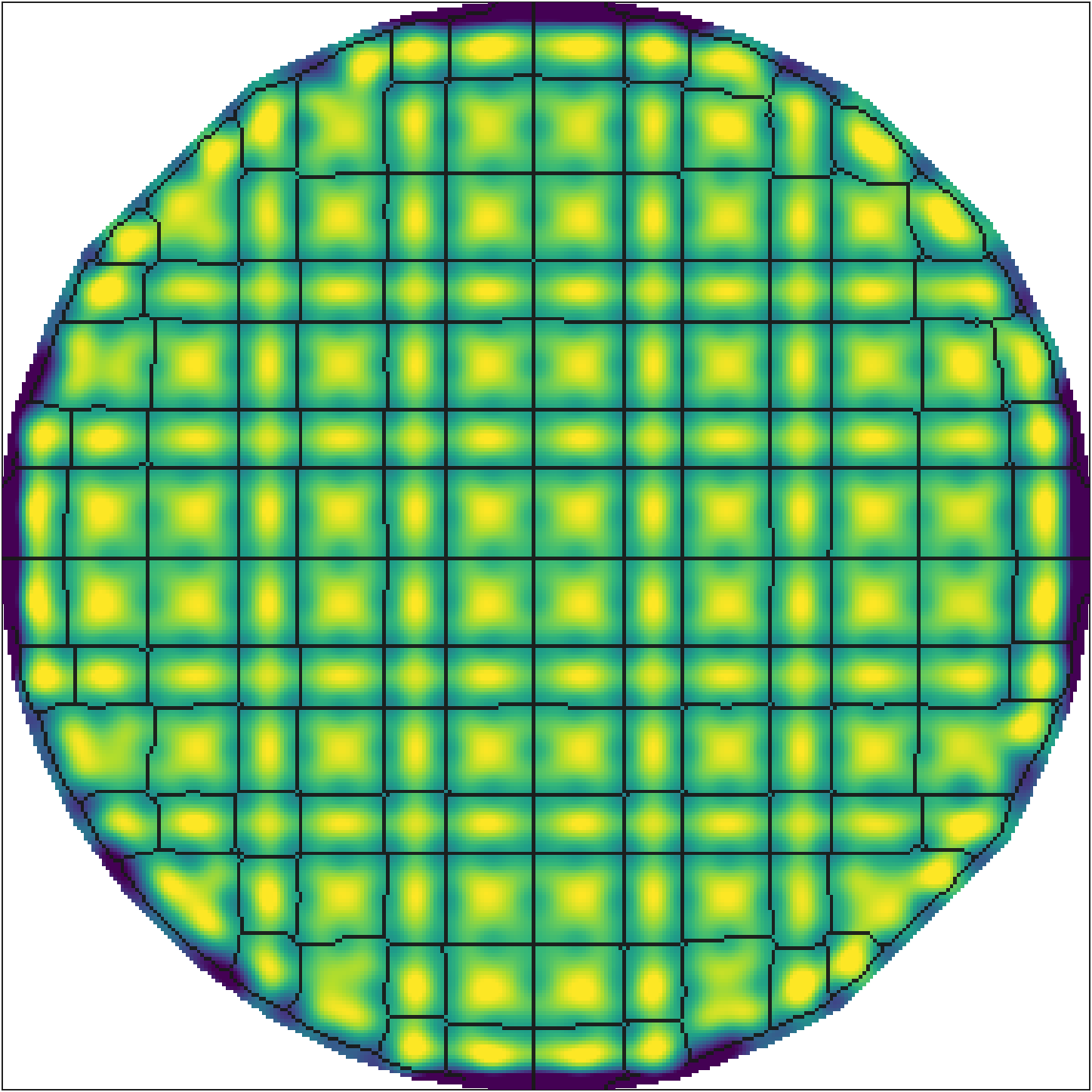}       & \pnl{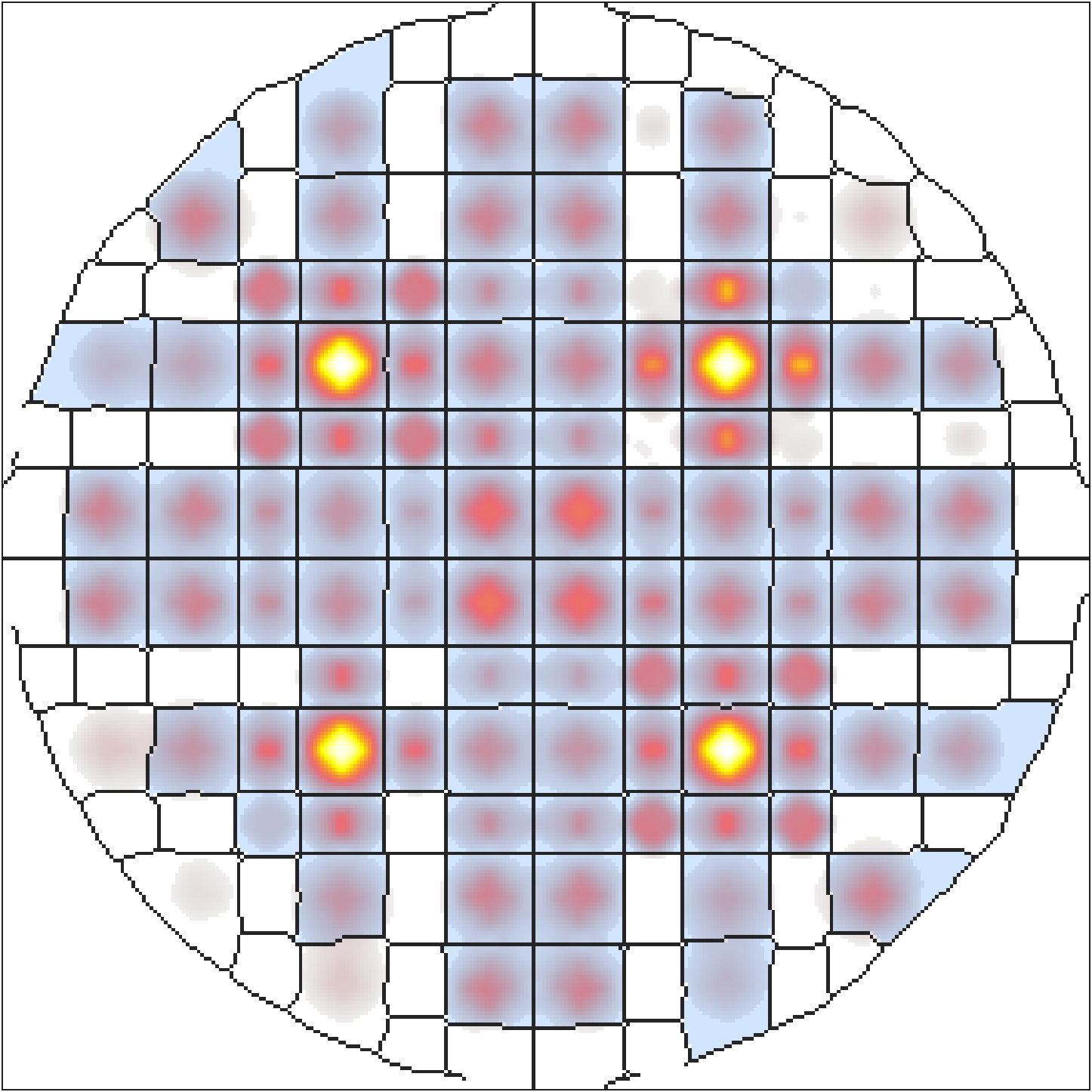}       \\
\rlbl{Rudin-Shapiro}          & \pnl{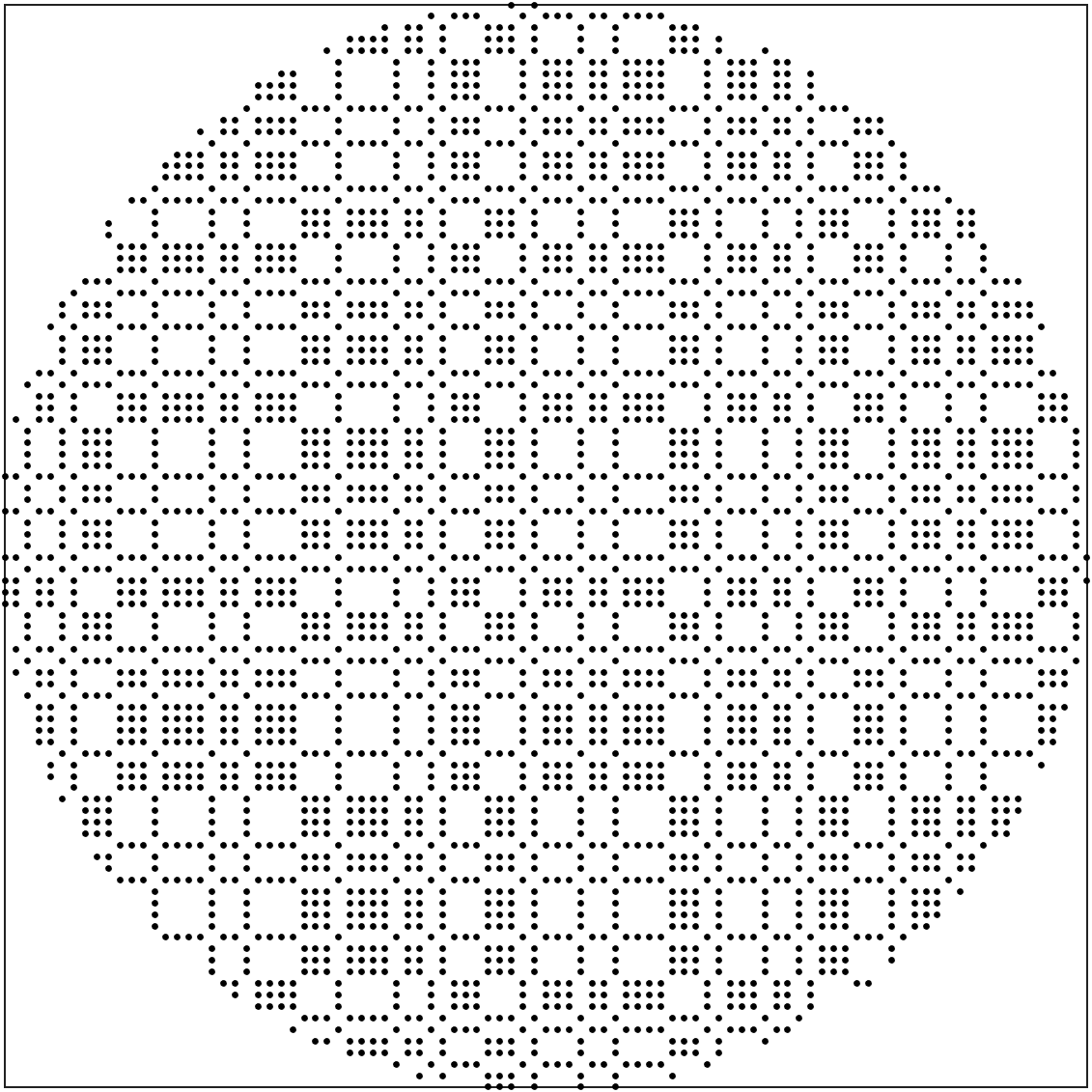}       & \pnl{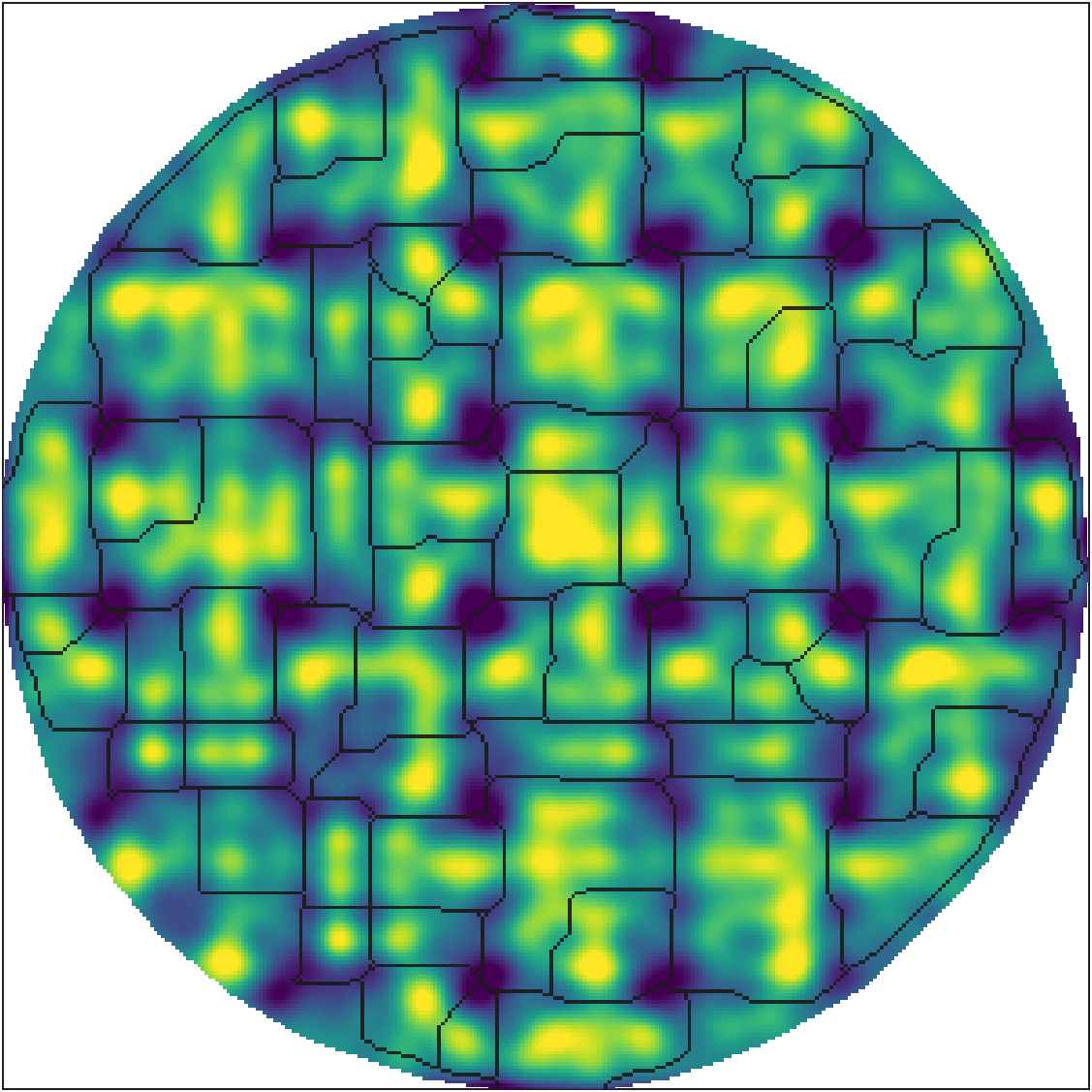}       & \pnl{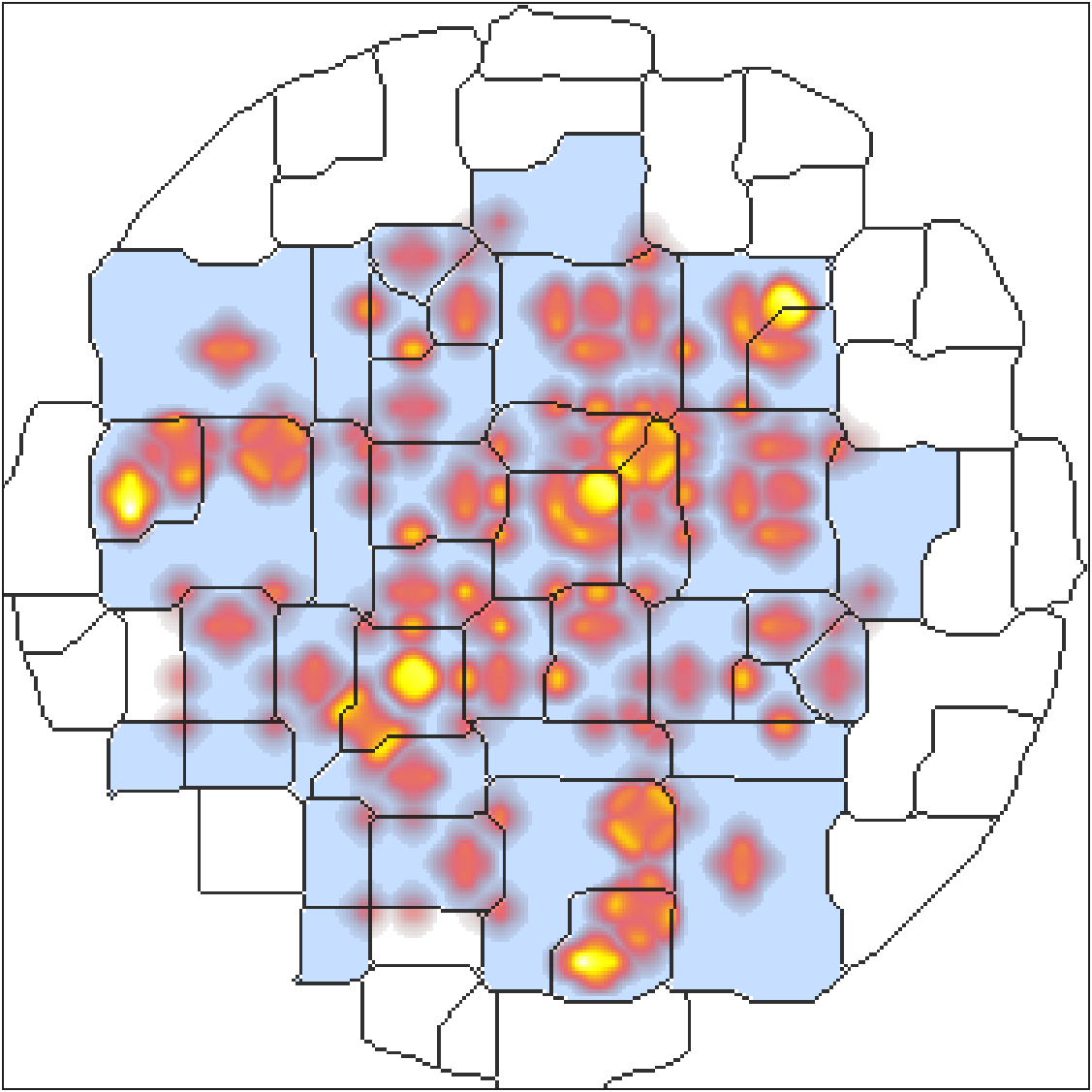}       \\
\end{tabular}

    \caption{ Point patterns vs Spec landscape and top-100 mode localization ~-- ~TM, $N{=}3500$. }
    \label{fig:validation-grid}
\end{figure}
The
same construction was carried out to $K=100$ long-lived modes as an
independent stress test of the landscape's predictive value beyond a
few low-order modes; the extended per-mode collection is deferred to
the supplementary material.  The mode envelopes continue to track the
Herm basins through the full $K=100$ window.  Modes related by the
$C_8$ symmetry of the Thue--Morse pattern distribute their intensity
across a corresponding pair of symmetric basins, so each eigenmode is
contained within the union of its predicted host cells, as expected
from the point-group symmetry of the array.  In
both TM and TE (Figs.~\ref{fig:thuemorse-tm} and
\ref{fig:thuemorse-te}) the top-100 mode max-projection is contained
within the Herm basins, and the 1D cross-sections show that individual
peaks reach the watershed walls but do not cross them: the Herm
landscape provides a genuinely predictive map of eigenmode support
across a range of mode indices much larger than the first 13.

The results obtained for Thue--Morse extend to the other aperiodic
families in our set.  On pattern classes that carry strong geometric
symmetries --- Fibonacci and the Eisenstein-prime lattices --- the
Herm watershed reproduces the pattern's symmetries in its basin
structure and the top-100 mode envelope is contained within the union
of basins.  On stochastic and low-discrepancy patterns (Bernoulli,
Halton) the basin shapes are irregular but the same coverage property
holds, with Herm resolving finer basins than Spec.  The validation
gallery shown in Fig.~\ref{fig:validation-grid} (Pinwheel, Halton, Eisenstein
prime, Fibonacci, and Rudin--Shapiro at the Spec landscape)
demonstrates the same qualitative picture across the full TM pattern
set, and the corresponding TE results are collected in Appendix~\ref{app:validation}. 
Finally, in Figure \ref{fig:grid} we also consider the landscapes for Poisson point patterns of electric dipoles under both TM and TE polarizations. 
Together, Figs.~\ref{fig:bernoulli-tm}
through~\ref{fig:validation-grid} and the supplementary galleries
show that the Gram/Spec/Herm working set, algebraically justified
in Appendix~\ref{app:rigorous}, is consistent with the eigenmodes
across every pattern class we have tested and are robust predictors of wave localization phenomena.

\section{Conclusions}
\label{sec:conclusions}
We introduced three real-valued, positive landscapes for the vector electromagnetic open problem of scattering dipoles that build on the Feinberg--Zee hermitization of the vector Green's matrix $G$. We considered in particular the regularized Gramian inverse ($\Bop$), the
diagonal Schur block that carries the Metz--Neri--Rogers spectral
density ($\SY$), and the full hermitized inverse, or the system's Green's function, contracted against a uniform source (TM) or an isotropic polarization pair (TE) ($\SBf$). The three landscapes form a
progression from spectrum-blind to fully hermitization-faithful and agree
in the limit $z\to 0$. Moreover, by applying this approach to a large number of random and deterministic aperiodic arrays with varying degrees of structural correlations we found that at nonzero
spectral shift $\SY$ resolves modes that $\Bop$ collapses, and $\SBf$
resolves modes that $\SY$ collapses.
Across all patterns tested, the three landscapes predict the
regions that host the long-lived eigenmodes of $G$ in both TM and TE
polarizations, so that a watershed of the landscape partitions the
array into basins that contain the eigenmodes without
diagonalization. The constructions extend directly to the dyadic
Green's function of three-dimensional resonators with
Lorentzian response, where the same Feinberg--Zee hermitization and
the same choice of source contraction apply; we expect this to be
useful for the transport and imaging of light in complex media, as well as the inverse design and optimization of photonic structures
with enhanced light-matter interaction with prescribed patterns of resonant modes.

\section*{Data Availability}
The materials and methods developed in this paper can be obtained directly from the authors upon reasonable request.

\begin{acknowledgments}
L.D.N. acknowledges partial support
from the U.S. Army Research Office under Award No.~W911NF-2510284 and the Shared Computing
Cluster, which is administered by Boston University’s
Research Computing Services. In addition, L.D.N. would like to thank Marcel Filoche and Arthur Goetschy for insightful discussions and suggestions.
\end{acknowledgments}

\section*{Author Contributions}
L.D.N. and H.S. contributed equally to this work.

\appendix
\renewcommand\thefigure{\thesection.\arabic{figure}}
\setcounter{figure}{0}
\section*{Appendices}
\section{Coupled-dipole Green's matrix and its singular structure}
\label{app:setup}
We consider $N$ point scatterers at positions $\{\rr_j\}_{j=1}^{N}$ with vacuum wavenumber $k_0$. The standard 2D coupled-dipole Green's matrix~\cite{lagendijk1996resonant,rusek1996localization,skipetrov2014absence} assumes scalar form for TM polarization and vectorial $2N\times 2N$ form for TE. We use the time dependence $e^{-i\omega t}$ throughout, and
$\mathbf G_0$ denotes the outgoing (retarded) Green's tensor, so that
the scattered field $\mathbf E-\mathbf E^{\rm inc}$ satisfies the
radiation condition.

Treating the scatterer contribution in~\eqref{eq:vecwave} as a source,
$[\nabla\!\times\!\nabla\!\times\!-\,k_0^2]\mathbf E
= k_0^2\alpha(\omega)\sum_{j}\delta(\rr-\rr_j)\mathbf E(\rr_j)$,
and inverting the free operator with the vacuum dyadic Green's tensor
$\mathbf G_0(\rr,\rr^\prime)$, which satisfies
$[\nabla\!\times\!\nabla\!\times\!-\,k_0^2]\mathbf G_0(\rr,\rr^\prime)
=I_2\,\delta(\rr-\rr^\prime)$, gives the Lippmann--Schwinger
representation
\begin{equation}
  \mathbf E(\rr)
  \;=\;\mathbf E^{\mathrm{inc}}(\rr)
  \;+\;k_0^2\,\alpha(\omega)\sum_{j=1}^{N}\mathbf G_0(\rr,\rr_j)\,\mathbf E(\rr_j),
  \qquad
  \bigl[\nabla\!\times\!\nabla\!\times\!-\,k_0^2\bigr]\,
  \mathbf E^{\mathrm{inc}} \;=\; 0,
  \label{eq:LSapp}
\end{equation}
where $\mathbf E^{\mathrm{inc}}$ is any homogeneous solution --- the
externally imposed drive. Evaluating~\eqref{eq:LSapp} at $\rr=\rr_n$
closes a self-consistent linear system for the on-site fields
$\mathbf E_n\!:=\!\mathbf E(\rr_n)$. The $n\!=\!m$ term carries the
point-coincidence singularity of $\mathbf G_0$; its physical content
is the single-scatterer radiative self-energy, and following the
standard Foldy--Lax
construction~\cite{foldy1945multiple,lax1951multiple,lagendijk1996resonant,rusek1996localization,Rusek2000}
we absorb it into the on-shell $t$-matrix $t(\omega)$. For a
lossless point scatterer the optical theorem forces
$\mathrm{Im}\,t=|t|^2$, equivalently
\begin{equation}
  \tfrac{1}{t(\omega)} \;=\; -\mathrm{i} \;+\; z(\omega),
  \qquad z(\omega)\in\mathbb R,
  \label{eq:optthm}
\end{equation}
and for a Lorentzian resonance at $\omega_0$ with radiative linewidth
$\Gamma_0$,
$z(\omega)=-2(\omega-\omega_0)/\Gamma_0$~\cite{goetschy2011lumiere}.
Grouping the regularized diagonal (the $-\mathrm{i}$ above) with the
off-diagonal propagators into a single array operator produces the
$2N\times 2N$ Green's matrix $G$ of~\eqref{eq:GTE}: $G_{nn}=\mathrm{i}\,I_2$ is the
radiative damping, and $G_{nm}$ for $n\ne m$ is the regular part of
the dyadic kernel between scatterers. 
With this definition, \eqref{eq:LSapp} evaluated at every $\mathbf r_n$ reads
\begin{equation}
  \mathbf E_n=\mathbf E^{\rm inc}_n
  +t(\omega)\sum_{m\neq n}G_{nm}\,\mathbf E_m ,
  \qquad n=1,\dots,N .
  \label{eq:foldylax}
\end{equation}
Dividing by $t(\omega)$ and inserting \eqref{eq:optthm},
$(-i+z)\,\mathbf E_n-\sum_{m\neq n}G_{nm}\mathbf E_m
=\mathbf E^{\rm inc}_n/t(\omega)$.
Moving the term $-i\,\mathbf E_n$ into the sum as the diagonal block
$G_{nn}=i\,I_2$ gives the compact form
\begin{equation*}
  \bigl(G-zI\bigr)\mathbf E=-\frac{1}{t(\omega)}\,\mathbf E^{\rm inc},
\end{equation*}
i.e.\ \eqref{eq:cde}.
The scalar TM derivation is identical, with
the vacuum dyadic $\mathbf G_0$ replaced by the scalar 2D vacuum
Green's function $G_0(\rr,\rr^\prime)=(\mathrm{i}/4)H_0^{(1)}(k_0|\rr-\rr^\prime|)$,
producing $G^{\mathrm{TM}}$ of~\eqref{eq:GTM} with $G^{\mathrm{TM}}_{nn}=\mathrm{i}$.

\paragraph{TM (scalar).}
For TM scalar waves, the matrix elements are defined as
\begin{equation}
  G^{\mathrm{TM}}_{nm} \;=\; \mathrm{i}\,H_0^{(1)}\!\big(k_0\,|\rr_n-\rr_m|\big) \quad (n\ne m),
  \qquad G^{\mathrm{TM}}_{nn} \;=\; \mathrm{i},
  \label{eq:GTM}
\end{equation}
where $H_0^{(1)}=J_0+\mathrm{i}Y_0$ is the zeroth-order Hankel function of the first kind. The diagonal is fixed by the radiative self-term. The matrix dimension is $N\times N$.


\paragraph{TE (vectorial).}
For TE vector waves, each scatterer carries an in-plane two-component dipole and the $(n,m)$ block of $G$ is the in-plane projection of the dyadic Green tensor. With the same normalization as in Eq.~\eqref{eq:GTM}~\cite{skipetrov2018finite,skolnick2026structural}, the $2\times 2$ blocks are defined as
\begin{equation}
  G^{\mathrm{TE}}_{nm} \;=\; 2\mathrm{i}\Big[F_1(k_0 r_{nm})\,I_2 - F_2(k_0 r_{nm})\,\hat{\rr}_{nm}\hat{\rr}_{nm}^{\!\top}\Big] \quad (n\ne m),
  \qquad G^{\mathrm{TE}}_{nn} \;=\; \mathrm{i}\,I_2,
  \label{eq:GTE}
\end{equation}
with
\begin{equation}
  F_1(\xi) \;=\; H_0^{(1)}(\xi) - \frac{H_1^{(1)}(\xi)}{\xi},
  \qquad
  F_2(\xi) \;=\; H_0^{(1)}(\xi) - \frac{2\,H_1^{(1)}(\xi)}{\xi},
  \label{eq:F12}
\end{equation}
where $r_{nm}=|\rr_n-\rr_m|$, $\hat{\rr}_{nm}=(\rr_n-\rr_m)/r_{nm}$ is the in-plane unit displacement vector, $\hat{\rr}_{nm}\hat{\rr}_{nm}^{\!\top}$ its outer product, $I_2$ is the $2\times 2$ polarization identity, and $H_1^{(1)}=J_1+\mathrm{i}Y_1$ is the first-order Hankel function of the first kind. The diagonal is again fixed by the radiative self-term: the regular parts of $F_1$ and $F_2$ tend to $\tfrac12$ and $0$ as $\xi\to 0$, so $\mathrm{i}\,I_2$ is the coincidence limit of the regular part of the off-diagonal blocks, and $\operatorname{Im}G^{\mathrm{TE}}_{nn}=1$ normalizes the single-dipole decay rate to $\Gamma_0$ exactly as in the TM case. 
Matrices \eqref{eq:GTM} and \eqref{eq:GTE} are complex, non-Hermitian ($G\ne G^\dg$), and non-normal ($GG^\dg\ne G^\dg G$).

The Green's matrix $G$ depends on $\omega$ through
$k_0=\omega/c$. If $\Gamma_0\ll\omega_0$, the interesting frequencies
lie within a few $\Gamma_0$ of $\omega_0$ and $k_0$ changes
negligibly over that range. One therefore evaluates $G$ once at
$k_0=\omega_0/c$. This is what makes the eigenvalues of $G$ a
linear eigenvalue problem; without it $G$ would depend on $z$.
The complex eigenvalues $\Lambda_m$ are obtained by diagonalizing the corresponding Green's matrix:
\begin{equation} \label{eq:eigenproblem}
    \hat{G} \mathbf{\psi}_m = \Lambda_m \mathbf{\psi}_m,
\end{equation}
where the total number of modes is $M = N$ for TM polarization and $M = 2N$ for TE polarization. These eigenvalues encode both the resonance frequencies $\omega_m$ and the decay rates $\Gamma_m$ of the system's quasimodes. Following the standard dimensionless notation \cite{skipetrov2018finite}, we define the normalized frequency shift $\nu_m$ and the normalized decay rate $\gamma_m$ as:
\begin{equation} \label{eq:nu_m}
    \nu_m = \frac{\omega_m - \omega_0}{\Gamma_0} = -0.5 \, \text{Re}(\Lambda_m),
\end{equation}
\begin{equation} \label{eq:gamma_m}
    \gamma_m = \frac{\Gamma_m}{\Gamma_0} = \text{Im}(\Lambda_m).
\end{equation}
In this linearized (first-order) formulation, the (generally) wavenumber-dependent Green matrix $G(k)$ is evaluated once at the unperturbed real wavenumber $k_0$, assuming $\lambda(\omega - i\Gamma) \simeq \lambda(\omega_0)$ \cite{rusek1996localization,Rusek2000}. 
Note that the linearized (first-order) approach allows one to accurately approximate the positions of the quasibound resonance states of an open scattering system, i.e., self-sustaining outgoing waves exist in the absence of any incoming driving field, in terms of the spectral properties of the Green's matrix without the need of calculating the poles of the scattering matrix in the complex energy plane. 
This approximation well-describes the resonant properties of electromagnetic scattering arrays under the following validity conditions:
\begin{enumerate}
	\item \textbf{Narrow Resonance Regime:} The single-scatterer linewidth is much smaller than the bare transition frequency ($\Gamma_0 \ll \omega_0$). 
	\item \textbf{Zero-Range Point Scatterers:} The scatterer radius $a$ is much smaller than both the wavelength ($k_0 a \ll 1$) and the inter-particle spacing ($r_{ab} \gg a$).
    In scattering theory, Wigner's causality bound and the centrifugal barrier constrain the bare resonance width of a physical scatterer of radius $a$ to satisfy:
\begin{equation}
	\frac{\Gamma_0}{E_0} \lesssim k_0 a \ll 1.
\end{equation}
Thus, a physically consistent point scatterer cannot possess an arbitrarily broad resonance width.
	\item \textbf{Linear Single-Excitation Dynamics:} The system is weakly driven so that at most one quantum of excitation occupies the ensemble at any time.
\end{enumerate}
Every one of these validity boundaries is directly controlled by the bare width $\Gamma_0$ and the collective widths $\Gamma$.
Moreover, when $\Gamma_0$ is large---or when the sample size $L$ becomes macroscopic---propagation retardation across the medium cannot be neglected. In that regime, $G(k)$ varies substantially across the resonance band, requiring the full nonlinear pole equation to be solved iteratively \cite{rusek1996localization,Rusek2000}.

\paragraph{Modes}
A resonance is a field pattern that sustains
itself with no incident wave: $(G-zI)\textbf{E}=0$ with $\textbf{E}\ne0$. This
requires $z=\lambda_n$, an eigenvalue of $G$, with $\textbf{E}=\ket{v_n}$
the right eigenvector, $G\ket{v_n}=\lambda_n\ket{v_n}$.
Invereting  $z(\omega)\approx{2(\omega-\omega_0)}/{\Gamma_0}$, the corresponding (complex) frequency is
\begin{equation}
  \omega=\omega_0-\frac{\Gamma_0}{2}\lambda_n
  =\underbrace{\omega_0-\frac{\Gamma_0}{2}\mathrm{Re}\,\lambda_n}_{\omega_n}
  -\frac{\mathrm{i}}{2}\underbrace{\Gamma_0\,\mathrm{Im}\,\lambda_n}_{\Gamma_n}.
  \label{eq:modes}
\end{equation}
So $\mathrm{Re}\,\lambda_n$ is the frequency shift and $\mathrm{Im}\,\lambda_n$ is
the decay rate in units of $\Gamma_0$. Long-lived (subradiant) modes
are the eigenvalues closest to the real axis. A real $z$ is a
physical driving frequency; a complex $z$ is an analytic
continuation, needed to place a probe exactly at $\lambda_n$.

\paragraph{Reciprocity.}
Interchanging $\rr_n\leftrightarrow \rr_m$ preserves the scalar Hankel function and reverses $\hat{\rr}_{nm}\to-\hat{\rr}_{nm}$, leaving the dyadic term $\hat{\rr}_{nm}\hat{\rr}_{nm}^{\!\top}$ invariant. Thus,
\begin{equation}
  G^{\mathrm{TM}} \;=\; \big(G^{\mathrm{TM}}\big)^{\!\top},
  \qquad
  G^{\mathrm{TE}} \;=\; \big(G^{\mathrm{TE}}\big)^{\!\top}.
  \label{eq:reciprocity}
\end{equation}
This complex-symmetric reciprocity ($G=G^{\!\top}$, $G\ne G^\dg$) is verified numerically to machine precision. It constrains the Hermitization blocks described in Section~\ref{sec:three}, eliminating several candidate landscapes as informationally redundant (see Appendix~\ref{app:survey}).

Three consequences used throughout the paper follow immediately from $G=G^{\!\top}$. First, the singular value decomposition of $W(z)=G-zI$ takes the Takagi form $W=\bar V\,\Sigma\,V^\dagger$: the left and right singular vectors are related by complex conjugation, $u_k=\bar v_k$, and the right singular vectors $v_k$ carry all the spatial information. Second, the two diagonal blocks of $B^{-1}$ are related by $X=\bar Y$ (Eq.~\eqref{eq:XYdeg}); in particular $X_{jj}=Y_{jj}$ at every site, so the left-eigenvector LDOS of $G$ coincides with the right-eigenvector LDOS and no independent $X$ landscape need be considered. Third, the biorthogonal left and right eigenvectors satisfy $\psi_n=\bar\phi_n$, hence $|\psi_n(j)|=|\phi_n(j)|$ pointwise; the eigenvalue condition number simplifies to $\kappa_n=\|\phi_n\|^2/|\phi_n^{\!\top}\phi_n|$, with the Petermann factor $K_n=\kappa_n^2$ measuring how far the unconjugated product $\phi_n^{\!\top}\phi_n$ falls below $\phi_n^\dagger\phi_n$.

\paragraph{Mode-vs.-pseudomode distinction.}
The right eigenvectors $\{v_n^R\}$ and left eigenvectors $\{v_n^L\}$ of $G$ form a biorthogonal system, $\langle v_n^L,v_m^R\rangle = \delta_{nm}$. Physically relevant modes are the long-lived eigenmodes. Following~\cite{skipetrov2014absence,skipetrov2018finite}, using complex eigenvalue coordinates $(x_s,y_s)=(-\Realp(z)/2,\,\Imag(z))$, these correspond to eigenvalues with minimal $y_s$. For any complex shift $z\in\C$, the singular value decomposition
\begin{equation}
  G-zI \;=\; U(z)\,\Sigma(z)\,V(z)^\dg,
  \qquad
  \Sigma(z) = \diag\big(\sigma_1(z),\ldots,\sigma_n(z)\big),
  \label{eq:svd}
\end{equation}
identifies the right singular vectors $\{V_{:,k}(z)\}$ associated with minimal $\sigma_k(z)$ as the \emph{pseudomodes} of $G$ at frequency $z$. Pseudomodes coincide with true eigenvectors only if $z$ is an exact eigenvalue and $G$ is normal; generally, they exhibit richer spatial structure and serve as the correct target for generalized landscape analysis~\cite{trefethen2005spectra}. All Schur-block landscapes in the main text act on pseudomodes by construction. The wavenumber is fixed via $k_0=2\pi/\sqrt{\mathrm{OD}/\rho}$, where $\rho$ is the site density within the enclosing disk.

\section{Survey and elimination of candidate landscapes}
\label{app:survey}
Throughout, $W(z) = G - zI$ and the resolvent is
$R(z) = (zI-G)^{-1} = -W(z)^{-1}$ when $z\notin\sigma(G)$, matching
the convention of Appendix~\ref{app:rigorous}. Write the singular value decomposition
$W(z) = U\Sigma V^\dagger$ with singular values $\sigma_1\ge
\sigma_2\ge\cdots\ge\sigma_{\min} \ge 0$ and right singular vectors $v_m =
V e_m$.

We enumerate candidate landscape constructions prior to selecting the final operators. Here $W=G-zI$, $\eta>0$ denotes the regulator, and a \emph{landscape} refers to a per-site real non-negative field $u(j)$ derived from Schur algebra vectors or matrices. For TE cases, per-site reduction involves summing $|u^{(x)}(j)|^2+|u^{(y)}(j)|^2$.

\paragraph{Direct FM transcription.}
Solving $G u=\one$ yields a complex vector $u$. While $|u|^2$ is positive, it lacks sign-stable correlation with mode support across patterns due to destructive phase interference in the resonance directions. This construction is unsuitable.

\paragraph{Biorthogonal landscape $|u_L|\!\cdot\!|u_R|$.}
With $u_R=G^{-1}\one$ and $u_L=(G^\dg)^{-1}\one$, the geometric mean $|u_L|\!\cdot\!|u_R|$ is real positive~\cite{chalker1998eigenvector}. However, absolute values discard phase information, and the construction lacks pseudospectral shift dependence.

\paragraph{Gramian inverse $\Bop=|(G^\dg G+\eta^2 I)^{-1}\one|^2$ }
This represents a Tikhonov-regularized Gramian inverse. Using SVD \eqref{eq:svd} with $z=0$,
\begin{equation}
  (G^\dg G+\eta^2 I)^{-1}
  \;=\; V_0\,\diag\!\Big(\tfrac{1}{\sigma_k(0)^2+\eta^2}\Big)\,V_0^\dg,
  \label{eq:gramSVD}
\end{equation}
$\Bop$ amplifies sites overlapping small-$\sigma$ singular vectors. As the simplest $z$-independent family member, $\Bop$ provides a spectrum-blind reference landscape.

\paragraph{Diagonal blocks of $B^{-1}$: $\SY=|Y\one|^2$, $|X\one|^2$ }
The diagonal blocks of $B^{-1}$ in \eqref{eq:Binvblocks} are, up to the factor $-\mathrm{i}\eta$, $Y=(W^\dg W+\eta^2 I)^{-1}$ and $X=(WW^\dg+\eta^2 I)^{-1}$. Reciprocity \eqref{eq:reciprocity} implies $WW^\dg=(W^\dg W)^{\!\top}$, hence $X=\overline{Y}$ and
\begin{equation}
  |X\one| \;=\; |\overline{Y}\one| \;=\; |Y\one|.
  \label{eq:XYdeg}
\end{equation}
For reciprocal media, $\{X,Y\}$ do not yield independent landscapes. We retain $\SY$ for definiteness.

\paragraph{Off-diagonal blocks of $B^{-1}$: $|(WY)\one|^2$, $|(YW^\dg)\one|^2$ }
The identity $WY=XW$ combined with $X=\overline{Y}$ implies magnitude degeneracy: $|(WY)\one|=|(YW^\dg)\one|$. Empirical results indicate these do not outperform $\SY$ or the full hermitized inverse $\SBf$. Thus, they are incorporated into $\SBf$ rather than treated separately.

\paragraph{Full hermitized inverse $\SBf=|B^{-1}[\one;\one]|^2$ }
This applies the full block inverse of \eqref{eq:B} to the doubled source $[\one;\one]$, summing intensities from both $N$-half-blocks. It incorporates contributions from all four blocks of $B^{-1}$, avoiding projection bias.

\paragraph{Source-free diagonal of $Y$ ($\Ydiag$, retained as TE cross-check).}
The polarization-traced diagonal
\begin{equation}
  [\Ydiag]_{\,j}
  \;=\; \sum_{\alpha\in\{x,y\}} Y_{(j,\alpha),(j,\alpha)}(z,\eta)
  \;=\; \sum_{\alpha\in\{x,y\}}
        \sum_n \frac{|\langle e_{j,\alpha},v_n\rangle|^2}
                    {\sigma_n^2+\eta^2},
  \label{eq:ydiagsvd}
\end{equation}
represents a partial local density of states for $W$ pseudomodes. It serves as a source-free counterpart to $\SY$ and provides a polarization-even cross-check for TE (Appendix~\ref{app:tepol}).

Consequently, only three independent operators remain --- $\Bop$, $\SY$, and $\SBf$ --- forming the working set of the main text.

\section{Rigorous diamagnetic confinement bound for \texorpdfstring{$\Bop$}{Gram}}
\label{app:diamag}
In the limit $\eta\to0$, $\Bop$ reduces to the landscape $v_H=H^{-1}\one$ of $H=G^\dg G$. Since $G$ contains complex scattering phases, $H$ is a complex Hermitian matrix with real positive eigenvalues $\lambda=\sigma^2$ but complex eigenmodes $\varphi$. Thus $v_H$ cannot be reciprocated directly into a potential. We establish a rigorous bound using the discrete diamagnetic inequality~\cite{GueryOdelin2026,Poggi2024,Mugnolo2024,lyra2015dual}. 

Introduce a real-valued, positivity-preserving M-matrix $H_0$ dominating $H$, constructed from absolute scattering magnitudes. Applying the diamagnetic inequality to the resolvent,
\begin{equation}
  |\varphi| \;=\; \sigma^2\,|H^{-1}\varphi|
             \;\le\; \sigma^2\,H_0^{-1}\,|\varphi|.
\end{equation}
Bounding $|\varphi|$ by its supremum norm and defining the real positive bounding landscape $v_0=H_0^{-1}\one$,
\begin{equation}
  |\varphi| \;\le\; \sigma^2\,\|\varphi\|_\infty\,H_0^{-1}\one
            \;=\; \sigma^2\,\|\varphi\|_\infty\,v_0.
\end{equation}
Defining the effective potential $W_0=1/v_0$, we derive the pointwise bound
\begin{equation}
  \frac{|\varphi_i|}{\|\varphi\|_\infty} \;\le\;
    \frac{\sigma^2}{W_{0,i}},
  \label{eq:diamagbound}
\end{equation}
indicating that complex singular mode decay is controlled by the valleys of the real potential $W_0$~\cite{Poggi2024}. This provides $\Bop$ with a rigorous counterpart: the Tikhonov regulator $\eta$ ensures conditioning, while Eq.~\eqref{eq:diamagbound} certifies confinement in the $\eta\to0$ limit, offering an FM-style justification absent for other landscapes.

\paragraph{Scope of the hypothesis.}
The bound \eqref{eq:diamagbound} rests on the existence of a nonsingular, real, entrywise-positive $M$-matrix $H_0$ that dominates $H=G^\dg G$ in the diamagnetic sense; the Ostrowski comparison inequality $|H^{-1}\varphi|\le H_0^{-1}|\varphi|$ requires this hypothesis~\cite{GueryOdelin2026,Poggi2024,Mugnolo2024}. The standard candidate for $H_0$ is the comparison matrix $\langle H\rangle$ with diagonal $|H_{ii}|$ and off-diagonal entries $-|H_{ij}|$. For the dense, slowly decaying coupled-dipole Gramian $G^\dg G$ considered here we have not been able to exhibit such an $H_0$: the mean off-diagonal absolute row sum of $\langle H\rangle$ at $N=1500$ far exceeds its mean diagonal, so $\langle H\rangle$ is not diagonally dominant and in particular not an $M$-matrix. Equation~\eqref{eq:diamagbound} should therefore be read as an FM-style envelope valid wherever a dominating $M$-matrix exists; it is not the primary rigorous guarantee for $\Bop$ on the arrays studied in this paper. The algebraic guarantees that \emph{do} apply uniformly --- the resolvent-norm identity \eqref{eq:norminv}, the Cauchy--Schwarz envelopes \eqref{eq:SYenvelope}--\eqref{eq:Bfenvelope}, and the pointwise bound \eqref{eq:fm-bound} on true eigenvectors --- are proved in Appendix~\ref{app:rigorous} without any $M$-matrix hypothesis.


\section{Foundations of the \texorpdfstring{$\SY$}{Spec} and \texorpdfstring{$\SBf$}{Herm} landscapes}
\label{app:rigorous}

The diamagnetic bound of Appendix~\ref{app:diamag} certifies $\Bop$ as
a confinement envelope in the $\eta\to 0$ limit, but it is
spectrum-blind: it does not explain why $\SY$ and $\SBf$ sharpen the
mode localization once the spectral shift $z$ is tuned to a long-lived
eigenvalue of $G$. The purpose of this appendix is to assemble that
justification from three logically independent ingredients:
\emph{(i)} the exact operator-norm identity that links the resolvent
norm, the smallest singular value of $G-zI$, and the eigenvalue
condition number of a diagonalisable
$G$~\cite{wilkinson1965algebraic,bauer1960norms,trefethen2005spectra};
\emph{(ii)} the Metz--Neri--Rogers (MNR) hermitization
identity~\cite{metz2019spectral}, which recovers the biorthogonal
projector of $G$ from the diagonal blocks of $B^{-1}(\lambda_n,\eta)$
without ever diagonalising $G$ directly; and \emph{(iii)} a physical
argument that converts small $\sigma_{\min}(W)$ into compact
spatial support for the aperiodic arrays studied here. We treat the
three in turn.

\subsection{Resolvent norm, smallest singular value, and non-normality}
\label{app:rigorous:norm}

\paragraph{Hermitian reference case.}
For a Hermitian $H = H^{\dg}$ with real eigenvalues $\{E_n\}$ and
orthonormal eigenvectors $\{\lvert\psi_n\rangle\}$, the resolvent
 $R(z) = (zI-H)^{-1} = \sum_n \lvert\psi_n\rangle\langle\psi_n\rvert/(z-E_n)$;
each term is an orthogonal projector. Its operator norm is
controlled by the distance to the nearest eigenvalue,
$\|R(z)\|_2 = 1/\mathrm{dist}(z,\sigma(H))$.
For a non-Hermitian $G$ neither
step survives in this form: eigenvectors are not orthogonal, the
projectors are rank-one but not self-adjoint, and $\|R(z)\|$ can be
orders of magnitude above $1/\mathrm{dist}(z,\sigma(G))$. The
pseudospectrum formalism of Trefethen and
Embree~\cite{trefethen2005spectra} is the standard framework for this
non-normal regime; we review three properties that are needed
below.

\paragraph{Resolvent norm as inverse smallest singular value.}
Let $W(z) = G - zI$ and $R(z)=(zI-G)^{-1}=-W(z)^{-1}$. For any invertible
$M\in\C^{N\times N}$ the substitution $u = M^{-1}v$ gives
\begin{equation}
\|M^{-1}\|_2
= \max_{v\neq 0}\frac{\|M^{-1}v\|}{\|v\|}
= \max_{u\neq 0}\frac{\|u\|}{\|Mu\|}
= \left(\min_{u\neq 0}\frac{\|Mu\|}{\|u\|}\right)^{-1}
= \frac{1}{\sigma_{\min}(M)},
\label{eq:norminv}
\end{equation}
so that $\|R(z)\|_2 = 1/\sigma_{\min}(G - zI)$. This is a
pure linear-algebra identity; it holds for every matrix and makes no
reference to spectrum. Two consequences follow immediately. First, the
$\varepsilon$-pseudospectrum
$\sigma_\varepsilon(G) = \{z\,:\,\|R(z)\|_2 > \varepsilon^{-1}\}$ is
the sub-level set $\{z\,:\,\sigma_{\min}(G-zI) < \varepsilon\}$: deep
pseudospectral poles are literally sites where $\sigma_{\min}$ is
small. 
Second, the matrix $B(z,\eta)$ of Eq.~\eqref{eq:B} has singular
values $\sqrt{\sigma_m(W)^2+\eta^2}$
(Sec.~\ref{app:rigorous:MNR}), the smallest of which is
$\sqrt{\sigma_{\min}(W)^2+\eta^2}\ge\eta$. The Tikhonov regulator
$\eta>0$ therefore keeps $B$ invertible, with
$\|B^{-1}\|_2\le 1/\eta$, even when $z\in\sigma(G)$.  

\paragraph{Biorthogonal spectral representation.}
Suppose $G$ is diagonalisable with right/left eigenvectors
$G\lvert v_n\rangle = \lambda_n\lvert v_n\rangle$,
$\langle u_n\rvert G = \lambda_n\langle u_n\rvert$, biorthogonal to
$\langle u_m\lvert v_n\rangle = \delta_{mn}$. The resolvent then
decomposes as
\begin{equation}
R(z) = \sum_{m=1}^{N}\frac{\lvert v_m\rangle\langle u_m\rvert}{z-\lambda_m}.
\label{eq:specres}
\end{equation}
Isolating the $m=n$ term near a simple eigenvalue, with $z = \lambda_n + \delta$,
yields the Laurent form
\begin{equation}
R(z) = \frac{P_n}{\delta} + R_{\mathrm{reg}}(z),
\qquad P_n = \lvert v_n\rangle\langle u_n\rvert,
\label{eq:laurent}
\end{equation}
where $R_{\mathrm{reg}}$ is analytic at $\lambda_n$ and $P_n$ is the
rank-one projector onto the one-dimensional eigenspace of $\lambda_n$.
Unlike the Hermitian case, $P_n$ is \emph{not} self-adjoint
whenever $\langle u_n\rvert\neq\lvert v_n\rangle^{\dg}$, and its
operator norm is at least one.

\paragraph{Eigenvalue condition number and Petermann factor.}
The operator norm of the rank-one projector $P_n$ is
\begin{equation}
\kappa_n \;\equiv\; \|P_n\|_2 \;=\;
\frac{\|v_n\|\,\|u_n\|}{\lvert\langle u_n\lvert v_n\rangle\rvert}\;\ge 1,
\label{eq:kappadef}
\end{equation}
written here in a form that does not depend on how the eigenvectors
are normalized. It is
the \emph{eigenvalue condition number} of $\lambda_n$ in the sense of
Wilkinson~\cite{wilkinson1965algebraic} and
Bauer--Fike~\cite{bauer1960norms}. Combining
Eqs.~\eqref{eq:norminv} and \eqref{eq:laurent},
{ \small
\begin{equation}
 \|R(z)\|_2=\frac{1}{\sigma_{\min}(W(z))}
  =\frac{\kappa_n}{|z-\lambda_n|}+O(1),
  \qquad
  \sigma_{\min}(W(z))=\frac{|z-\lambda_n|}{\kappa_n}
  +O\bigl(|z-\lambda_n|^2\bigr)
  \quad (z\to\lambda_n).
\label{eq:kappa_scaling}
\end{equation}
}
In addition, $\sigma_{\min}(W(z))\le|z-\lambda_n|$ holds exactly for
every $z$ and every $n$, because
$\|W(z)v_n\|=|\lambda_n-z|\,\|v_n\|$.
Three limits of \eqref{eq:kappadef}--\eqref{eq:kappa_scaling} fix the
physical picture. (a) When $G$ is normal ($G G^{\dg} = G^{\dg} G$),
$\langle u_n\rvert = \lvert v_n\rangle^{\dg}$ and $\kappa_n=1$: the
resolvent norm equals the inverse distance to the spectrum and the
$\varepsilon$-pseudospectrum is a tube of radius $\varepsilon$ about
$\sigma(G)$~\cite{trefethen2005spectra}. (b) When $G$ is highly
non-normal, left and right eigenvectors can be nearly orthogonal, so
$\kappa_n\gg 1$ and the $\varepsilon$-pseudospectrum inflates to
radius $\kappa_n\varepsilon$ around each
$\lambda_n$~\cite{horn2013matrix}. (c) At an exceptional point two
eigenvectors coalesce, $\langle u_n\lvert v_n\rangle\to 0$, and
$\kappa_n\to\infty$: the Laurent expansion \eqref{eq:laurent} then
inherits higher-order poles and the identity
$\|R\|\sim \kappa_n/|\delta|$ in \eqref{eq:kappa_scaling} is
superseded by the Jordan-chain scaling $\|R\|\sim 1/|\delta|^k$ for a
Jordan block of size $k$~\cite{kato1995perturbation}; defective
spectra are not generic in the dipole arrays we study and we work
throughout in the simple-eigenvalue regime.

The square $\kappa_n^2$ is the
Petermann excess-noise factor
$K_n = \|u_n\|^2\|v_n\|^2 / |\langle u_n\lvert v_n\rangle|^2$ familiar
from open-cavity and random-laser
analyses~\cite{petermann1979calculated,patra1999excess}.
The scaling $\|R(z)\|\sim\sqrt{K_n}/|\delta|$ is therefore the
non-Hermitian counterpart of Petermann amplification. For the
reciprocal Green's matrices considered here, $G=G^{\top}$
\eqref{eq:reciprocity}, the left eigenvector is the complex conjugate
of the right one, $|u_n\rangle\propto\overline{|v_n\rangle}$, so that
\begin{equation}
  \kappa_n=\frac{\|v_n\|^2}{|v_n^{\top}v_n|},
  \qquad |u_n(j)|\propto|v_n(j)|\ \text{ at every site } j:
  \label{eq:kappa-recip}
\end{equation}
non-normality is measured by how far the unconjugated product
$v_n^{\top}v_n$ falls below $v_n^\dagger v_n$, and left and right
eigenvectors have the same spatial profile.
We stress that $K_n$
quantifies biorthogonality collapse, that is, eigenvalue sensitivity
and proximity to exceptional points, and not the spatial support of
$v_n$. 
By \eqref{eq:kappa_scaling}, the depth
of a pseudospectral pole probed by Spec and Herm combines a small
spectral distance $|\delta|$ with the condition-number amplification
$\kappa_n$; the distribution of $K_n$ over the modes of each pattern
is reported in Table~\ref{tab:indicators}.
%

\subsection{Eigenvector reconstruction via the Hermitisation}
\label{app:rigorous:MNR}

\paragraph{Why hermitise.}
Evaluating \eqref{eq:laurent} directly requires knowledge of the
biorthogonal pair $(\lvert u_n\rangle,\lvert v_n\rangle)$ of $G$,
which is numerically costly for large $N$ and conceptually fragile in
the non-normal regime. 
The Feinberg--Zee construction~\cite{feinberg1997nonhermitian}
replaces this problem by a Hermitian one on a doubled space. 
%
The Hermitisation of $W=G-zI$ is the $2N\times 2N$ matrix
\begin{equation}
  H_0(z)=\begin{pmatrix}0&W\\W^\dagger&0\end{pmatrix}=H_0(z)^\dagger,
  \label{eq:app-H0}
\end{equation}
whose eigenvalues are $\pm\sigma_m(W)$: the singular values of the
non-normal matrix $W$ become the eigenvalues of a Hermitian one. The
matrix $B(z,\eta)$ of \eqref{eq:B} is this Hermitisation shifted
by the regulator, $B=H_0+i\eta I$. It is normal but not Hermitian,
with eigenvalues $\pm\sigma_m(W)+i\eta$ and singular values
$\sqrt{\sigma_m(W)^2+\eta^2}$. 
Its inverse is therefore the resolvent of the Hermitian matrix $H_0$
evaluated at the point $-i\eta$ off the real axis,
\begin{equation}
  B^{-1}(z,\eta)=\bigl(H_0(z)+i\eta I\bigr)^{-1},
  \qquad
  \|B^{-1}\|_2=\frac{1}{\sqrt{\sigma_{\min}(W)^2+\eta^2}}\le\frac1\eta .
  \label{eq:Binv-norm}
\end{equation}
The smallest singular value of $W$ thus controls the norm of $B^{-1}$,
and $B$ is invertible for every $\eta>0$, including at
$z\in\sigma(G)$ where $W$ itself is singular. The pseudospectral
information carried by the non-normal resolvent of $G$, namely
$\sigma_{\min}(G-zI)$ and its singular vectors, is in this way encoded
in the resolvent of a Hermitian matrix, to which the standard spectral
tools apply.

\paragraph{Block structure of $B^{-1}$.}
Writing the inverse of \eqref{eq:B} in block form,
\begin{equation}
B^{-1}(z,\eta)
=
\begin{pmatrix}
-\mathrm{i}\eta\,X & W\,Y \\[2pt]
Y\,W^{\dg} & -\mathrm{i}\eta\,Y
\end{pmatrix},
\qquad
X = (WW^{\dg}+\eta^2 I)^{-1},
\quad
Y = (W^{\dg}W+\eta^2 I)^{-1},
\label{eq:Binvblocks}
\end{equation}
exhibits the two Hermitian positive-definite matrices $X$ and $Y$, the regularized inverses of the Gram matrices $WW^\dagger$ and $W^\dagger W$. 
For $z\notin\sigma(G)$ the lower off-diagonal block of $B^{-1}$ is the
regularized resolvent of $G$:
\begin{equation}
  Y(z,\eta)\,W(z)^\dagger = W(z)^{-1}+O(\eta^2) = -R(z)+O(\eta^2),
  \label{eq:offdiag}                                   
\end{equation}
while the trace of the same block delivers the empirical spectral density
\begin{equation}
  \rho_G(z)=-\lim_{\eta\to 0^+}\frac{1}{N\pi}\,
  \partial_{\bar z}\sum_{j=1}^{N}\bigl[B^{-1}(z,\eta)\bigr]_{j+N,\,j}
  \label{eq:MNRdos}
\end{equation}
through the standard Feinberg--Zee electrostatic analogy, $\rho_G=\lim_{\eta\to0}(\pi N)^{-1}\partial_{\bar z}\partial_z \ln\det(W^\dagger W+\eta^2 I)$~\cite{metz2019spectral}. The resolvent and the
spectral density of $G$ are therefore \emph{both} encoded in the
blocks of $B^{-1}$: the content of the landscape construction is that
these same blocks also encode the \emph{eigenvectors}.

\paragraph{MNR projector reconstruction.}
At a non-degenerate eigenvalue $z=\lambda_n\in\sigma(G)$ the matrix
$W$ is singular, with $W|v_n\rangle=0$ and $W^\dagger|u_n\rangle=0$.
The Hermitisation $H_0(\lambda_n)$ therefore has a two-fold zero
eigenvalue, and $B(\lambda_n,\eta)=H_0+i\eta I$ has the two-fold
eigenvalue $i\eta$~\cite{metz2019spectral}, with eigenvectors
\begin{equation}
 \begin{pmatrix}|\hat u_n\rangle\\ 0\end{pmatrix},
  \qquad
  \begin{pmatrix}0\\ |\hat v_n\rangle\end{pmatrix},
  \qquad
  |\hat u_n\rangle=\frac{|u_n\rangle}{\|u_n\|},\quad
  |\hat v_n\rangle=\frac{|v_n\rangle}{\|v_n\|},
\label{eq:MNRbnpm}
\end{equation}
built from the biorthogonal pair of $G$, each normalized to unit
length. (The combinations
$|b_n^\pm\rangle=\tfrac{1}{\sqrt2}(\pm|\hat u_n\rangle,|\hat v_n\rangle)^{\top}$
span the same eigenspace.) 
The reason the construction succeeds is the
chiral (sublattice) symmetry of $H_0$, generated by $\sigma_z\otimes I$:
it forces the eigenvalues of $H_0$ to come in $\pm$ pairs and lets the
zero modes be chosen on a single sublattice each, which separates the
left eigenvector (upper half) from the right eigenvector (lower half).
Since $B^{-1}$ acts as $1/(i\eta)$ on this eigenspace and as
$1/(\pm\sigma_m+i\eta)$ on the others, the zero modes saturate the
inverse for small $\eta$,
\begin{equation}
  i\eta\,B^{-1}(\lambda_n,\eta)=
  \begin{pmatrix}
    |\hat u_n\rangle\langle\hat u_n| & 0\\
    0 & |\hat v_n\rangle\langle\hat v_n|
  \end{pmatrix}
  +O\!\left(\frac{\eta}{\sigma_{N-1}}\right),
  \label{eq:MNRproj}                                 
\end{equation}
where $\sigma_{N-1}>0$ is the second-smallest singular value of
$W(\lambda_n)$. The right-hand side is the orthogonal projector onto
the left and right eigenvectors of $G$, which~\cite{metz2019spectral}
identify with the spectral information of the biorthogonal pair.

\paragraph{LDOS identity for the diagonal blocks.}
Reading \eqref{eq:MNRproj} block by block in the inverse
\eqref{eq:Binvblocks}, whose diagonal blocks are
$-\mathrm{i}\eta\,X$ and $-\mathrm{i}\eta\,Y$, yields the pointwise
identities
\begin{equation}
\eta^{2}\,Y_{jj}(\lambda_n,\eta) \xrightarrow[\eta\to 0^{+}]{}
|\hat v_n(j)|^2,
\qquad
\eta^{2}\,X_{jj}(\lambda_n,\eta) \xrightarrow[\eta\to 0^{+}]{}
|\hat u_n(j)|^2,
\label{eq:LDOSidentity_app}
\end{equation}
valid for $\eta\ll\sigma_{N-1}$, where $\sigma_{N-1}>0$ is the
second-smallest singular value of $W(\lambda_n)$ and
$|\hat v_n\rangle=|v_n\rangle/\|v_n\|$,
$|\hat u_n\rangle=|u_n\rangle/\|u_n\|$ are the right and left
eigenvectors of $G$ each normalized to unit length. Up to the
regulator, and in the limit $z\to\lambda_n$, the diagonal of $Y$ is
therefore the local density of states (LDOS) of the right eigenvector
of $G$ at frequency $z$, and the diagonal of $X$ that of the left
eigenvector; by \eqref{eq:kappa-recip} the two coincide for
reciprocal $G$, $X_{jj}=Y_{jj}$. Away from this limit the same
quantity is a windowed sum over the right singular vectors of $W$
\eqref{eq:svd},
\begin{equation}
\eta^{2}\,Y_{jj}(z,\eta) \;=\; \sum_{k=1}^{N}
\frac{\eta^{2}}{\sigma_k(z)^2+\eta^{2}}\,|V_{jk}(z)|^2 ,
\label{eq:window}
\end{equation}
in which every pseudomode with $\sigma_k\lesssim\eta$ contributes with
weight close to one. This is the non-Hermitian counterpart of the
standard Hermitian identity
$-\pi^{-1}\,\mathrm{Im}\,R_{jj}(E+\mathrm{i}0^+) = \sum_n|\psi_n(j)|^2
\delta(E-E_n)$, with the regulator $\eta$ playing the role of the
imaginary infinitesimal, that is, of the spectral resolution, and with
the two LDOS branches now separated by the Hermitisation.

\paragraph{Landscapes as Cauchy--Schwarz envelopes of the LDOS.}
The source-driven landscapes of \eqref{eq:specdef}--\eqref{eq:hermdef}
are bounded by the LDOS through a Cauchy--Schwarz step. For $\SY$,
since $Y$ is Hermitian with $0\prec Y\preceq\eta^{-2}I$, we have
$\sum_k|Y_{jk}|^2 = (Y^2)_{jj}\le \eta^{-2}Y_{jj}$ and therefore
\begin{equation}
u_{\SY}(j)
= \lvert[Y\one]_j\rvert^2
= \Bigl|\sum_k Y_{jk}\Bigr|^2
\;\le\; N\sum_k |Y_{jk}|^2
\;\le\; \frac{N}{\eta^{2}}\,Y_{jj}(z,\eta).
\label{eq:SYenvelope}
\end{equation}
For $\SBf$, the identity
$B^{-1}B^{-\dg}=(B^\dg B)^{-1}=\operatorname{diag}(X,Y)$ shows that the
squared row norms of $B^{-1}$ are exactly $X_{jj}$ in the upper half
and $Y_{jj}$ in the lower half, and Cauchy--Schwarz against the
$2N$-vector $(\one;\one)$ gives, with no remainder term,
\begin{equation}
u_{\SBf}(j) \;\le\; 2N\,\bigl[X_{jj}(z,\eta)+Y_{jj}(z,\eta)\bigr]
\;=\; 4N\,Y_{jj}(z,\eta),
\label{eq:Bfenvelope}
\end{equation}
where the last equality uses $X_{jj}=Y_{jj}$ from
\eqref{eq:kappa-recip}. At $z=\lambda_n$,
\eqref{eq:LDOSidentity_app} converts
\eqref{eq:SYenvelope}--\eqref{eq:Bfenvelope} into
$\eta^{4}\,u_{\SY}(j)\le N\,|\hat v_n(j)|^2+O(\eta^2)$ and
$\eta^{2}\,u_{\SBf}(j)\le 2N\bigl(|\hat u_n(j)|^2+|\hat v_n(j)|^2\bigr)+O(\eta^2)$:
a site at which the eigenvector amplitude vanishes is also a zero of
$\SY$ and $\SBf$ to leading order. Note the direction of
\eqref{eq:SYenvelope}--\eqref{eq:Bfenvelope}: they bound the
landscape by the LDOS, whereas the Filoche--Mayboroda
inequality~\cite{filoche2012universal,arnold2016effective} bounds the
eigenfunction by the landscape. The bound in the Filoche--Mayboroda
direction, for the LDOS landscape $Y_{jj}$ itself, is the subject of
the next paragraph.

\paragraph{A pointwise bound on the eigenvectors.}
The bound in the Filoche--Mayboroda direction holds for the diagonal
of $Y$ itself. For any $z\in\C$, any $\eta>0$ and any unit-norm right
eigenvector $|\hat v_n\rangle$ of $G$,
\begin{equation}
|\hat v_n(j)|^2 \;\le\; \bigl(|\lambda_n-z|^2+\eta^2\bigr)\,
Y_{jj}(z,\eta).
\label{eq:fm-bound}
\end{equation}
Indeed, $W\hat v_n=(\lambda_n-z)\hat v_n$ and $Y(W^\dg W+\eta^2 I)=I$
give $\hat v_n=(\lambda_n-z)\,YW^\dg\hat v_n+\eta^2\,Y\hat v_n$, hence
$|\hat v_n(j)|\le|\lambda_n-z|\,\|WYe_j\|+\eta^2\|Ye_j\|$ with
$\|WYe_j\|^2=Y_{jj}-\eta^2(Y^2)_{jj}$ and $\|Ye_j\|^2=(Y^2)_{jj}$; a
second Cauchy--Schwarz inequality on the pairs
$(|\lambda_n-z|,\eta)$ and
$\bigl(\sqrt{Y_{jj}-\eta^2(Y^2)_{jj}},\,\eta\sqrt{(Y^2)_{jj}}\bigr)$
gives \eqref{eq:fm-bound}. The bound involves no condition number:
every eigenmode with $|\lambda_n-z|\lesssim\eta$ can be large only
where $Y_{jj}(z,\eta)$ is large, however non-normal $G$ is.

\subsection{A physical argument for spatial localization}
\label{app:rigorous:chain}

Eqs.~\eqref{eq:kappa_scaling} and \eqref{eq:LDOSidentity_app} tie
together the resolvent norm, the smallest singular value of $W(z)$,
and the right-eigenvector LDOS of $G$. Neither of them, however,
forces the modes picked out by $\SY$ and $\SBf$ to be \emph{spatially
localized}: in a translation-invariant array those modes would be
delocalized Bloch waves whose Y-diagonal is uniform. The spatial
compactness we observe in aperiodic arrays comes from the physics of
the operator $\Imag G$, which represents the coherent radiative loss
of the Green's matrix. The argument proceeds in four short steps.

\paragraph{Step (i)  Depth of a pseudospectral pole requires a long-lived mode.}
From Eq.~\eqref{eq:kappa_scaling}, the smallest singular value along a
real-$z$ slice satisfies
$\sigma_{\min}(W(z)) \gtrsim |\Imag\lambda_n|/\kappa_n$ for the
nearest eigenvalue $\lambda_n$. For $\sigma_{\min}(W(z))$ to drop far below the typical eigenvalue
spacing therefore requires either a mode with small $\Imag\lambda_n$
(i.e.\ a long radiative lifetime) or, through
Eq.~\eqref{eq:kappadef}, a large condition number $\kappa_n$. For the coupled-dipole Green's matrix on an
aperiodic array both conditions coincide: the subradiant branch of
$G$ is simultaneously long-lived and strongly
non-normal~\cite{Rusek2000,skipetrov2014absence,skipetrov2018finite}.

\paragraph{Step (ii) A long lifetime forces alignment with the kernel of $\Imag G$.}
The Green's matrix admits the Hermitian decomposition
$G = \Realp G + \mathrm{i}\,\Imag G$, in which $\Imag G$ is the
coherent radiation operator of the array. By reciprocity
\eqref{eq:reciprocity} $G=G^{\top}$, so $\Realp G$ and $\Imag G$ are
both real symmetric and in particular Hermitian. The Rayleigh
quotient of a right eigenvector therefore separates cleanly into
real and imaginary parts and depends only on $v_n$,
\begin{equation}
\Imag\lambda_n
\;=\; \frac{\langle v_n\rvert\,\Imag G\,\lvert v_n\rangle}
          {\langle v_n\lvert v_n\rangle}
\;\ge\; 0,
\label{eq:ImLambda_app}
\end{equation}
without reference to the left eigenvector. The condition
$\Imag\lambda_n\to 0$ therefore forces $|v_n\rangle$ into the
near-kernel of $\Imag G\succeq 0$: the mode itself must lie almost
entirely in the subspace of configurations that the radiation
operator nearly annihilates.

\paragraph{Step (iii) The near-kernel of $\Imag G$ consists of weakly-radiating configurations.}
For the TM Green's matrix the entries of $\Imag G$ are
$(\Imag G)_{nm}=J_0(k_0 r_{nm})$, and the plane-wave representation
$J_0(k_0|\mathbf r|)=(2\pi)^{-1}\!\int_0^{2\pi}
e^{\mathrm{i}k_0\hat{\mathbf k}(\theta)\cdot\mathbf r}\,\mathrm{d}\theta$
turns \eqref{eq:ImLambda_app} into the exact angular integral of the
far-field intensity along the Ewald circle,
\begin{equation}
\Imag\lambda_n
\;=\;
\frac{1}{2\pi\,\|v_n\|^2}
\int_0^{2\pi}\lvert A_{v_n}(\theta)\rvert^2\,\mathrm{d}\theta,
\qquad
A_v(\theta) \;=\; \sum_{j=1}^{N} v_j\,
e^{-\mathrm{i}k_0 \hat{\mathbf k}(\theta)\cdot \mathbf r_j},
\label{eq:farfield_app}
\end{equation}
with $|\mathbf k| = k_0$~\cite{facchinetti2016storing,guerin2017light}.
The decay rate of a mode is therefore its angle-averaged far-field
intensity, and $\Imag\lambda_n\to 0$ is equivalent to approximate
vanishing of $A_{v_n}$ on the entire Ewald circle: a configuration
in the near-kernel of $\Imag G$ is one whose dipoles interfere
destructively into every propagating direction and so couples only
weakly to the radiation continuum.

\paragraph{Step (iv) On aperiodic arrays these configurations are compactly supported.}
The vanishing of $A_v$ along the Ewald circle constrains the discrete
Fourier transform of $v$, sampled at the array positions, to vanish on
the shell $|\mathbf k| = k_0$. Two mechanisms can enforce this
constraint. In a periodic array with lattice constant below
$\lambda_0/2$, the Bragg peaks of the structure factor lie outside the
Ewald circle and extended Bloch modes satisfy it globally. In an
aperiodic array (pinwheel, coprime, Thue--Morse, Fibonacci,
Rudin--Shapiro) the absence of translational symmetry removes this
global Bragg mechanism, so the cancellation can only be achieved
locally, through tightly packed sub-wavelength clusters whose
intra-cluster phases interfere destructively on the Ewald
circle~\cite{jen2016cooperative,facchinetti2016storing,guerin2017light}.
Extended, slowly varying mode profiles produce sharp diffractive peaks
in $A_v$ and cannot satisfy the vanishing condition. In the aperiodic
arrays considered here the modes with the smallest $|\Imag\lambda_n|$
are therefore necessarily supported on compact cancellation-tuned
clusters.

\paragraph{Summary.}
Steps (i) and (ii)$\Leftrightarrow$(iii) are exact identities for
reciprocal $G$; step (iv) is a physical claim, not a theorem, and
fails in sub-wavelength periodic lattices where extended Bloch modes
satisfy the Ewald-circle vanishing globally. On the aperiodic arrays
considered here, evaluating $\SY$ and $\SBf$ at the longest-lived
eigenvalue of $G$ selects the smallest singular value of $W(z)$,
which through Eqs.~\eqref{eq:kappa_scaling}--\eqref{eq:LDOSidentity_app}
is the regularized right-eigenvector LDOS and which, by the argument
above, is forced to be a spatially compact, cancellation-tuned
cluster mode. Appendix~\ref{app:diamag} supplies the complementary
spectrum-blind guarantee for $\Bop$, and together they justify the
progression Gram $\to$ Spec $\to$ Herm introduced in
Section~\ref{sec:three}.

The algebraic skeleton of Secs.~\ref{app:rigorous:norm}--\ref{app:rigorous:MNR}
--- the operator-norm identity, the Laurent expansion, the Petermann
factor, and the MNR projector reconstruction --- holds for any
diagonalisable complex square matrix with simple eigenvalues and
transposes verbatim to other non-Hermitian operators. The localization
chain of Sec.~\ref{app:rigorous:chain} additionally requires the
physical input that $\Imag G\succeq 0$ is the radiation operator of
an open scattering system and that its kernel admits only
compactly-supported realisations on the aperiodic point sets of
interest here, Note that the first condition fails for lattice models with
gain--loss or non-reciprocal hopping rates, where other mechanisms take over the role of
geometric cancellation.


\section{Polarization-isotropic adjustment for TE}
\label{app:tepol}
The operators defined in Section~\ref{sec:three} are polarization-agnostic at the algebraic level ($W$, $W^\dg W$, $Y$, $B^{-1}$). However, the source vector $\one_{2N}$ is not. We address this sensitivity and provide a polarization-isotropic adjustment.

\subsection{Source vector dependence in TE}

In the TE block layout, the $2N$-vector at site $j$ comprises two components, $[u^{(x)}_j;u^{(y)}_j]$, ordered as $[\mathbf{u}^{(x)};\mathbf{u}^{(y)}]$. Thus,
\begin{equation}
  \one_{2N}
  \;=\; \begin{pmatrix} \one_N \\ \one_N \end{pmatrix}
  \;=\; \sum_{j=1}^N \big(\,\mathbf{e}_j^{(x)}+\mathbf{e}_j^{(y)}\,\big),
  \label{eq:onesdecomp}
\end{equation}
where $\mathbf{e}_j^{(\alpha)}$ is the unit vector for component $\alpha$. Driving with Eq.~\eqref{eq:onesdecomp} corresponds to a uniform $45^\circ$ linearly polarized dipole at every site. TE eigenmodes exhibit spatially varying in-plane polarization $\hat{p}_n(\rr)$. A fixed-polarization probe couples proportionally to the local overlap with $\hat{p}_n$, systematically suppressing orthogonal components. In our pattern sets, this induces a pattern-dependent sign flip in the Spearman correlation between source-driven landscapes and long-lived mode support. This discrepancy arises from the source term, not the operators.

\subsection{Polarization-isotropic source pair}

We replace $\one_{2N}$ with an orthogonal probe pair:
\begin{equation}
  \mathbf{s}^{(x)}
  \;=\; \begin{pmatrix} \one_N \\ \zerov_N \end{pmatrix},
  \qquad
  \mathbf{s}^{(y)}
  \;=\; \begin{pmatrix} \zerov_N \\ \one_N \end{pmatrix},
  \label{eq:isoprobes}
\end{equation}
summing per-site intensities. This polarization-isotropic source maintains the original $L^1$ norm while distributing energy equally across $x$ and $y$ without prescribing phase relations. For any operator $\mathcal{O}$ acting on $\C^{2N}$,
\begin{equation}
  u^{\mathrm{iso}}(j)
  \;=\; \sum_{\alpha\in\{x,y\}} \sum_{\beta\in\{x,y\}}
        \big|\,[\mathcal{O}^{-1}\mathbf{s}^{(\alpha)}]_{(j,\beta)}\,\big|^2.
  \label{eq:iso}
\end{equation}
This sums responses to $x$ and $y$ probes equally, ensuring invariance under in-plane rotations. In TM, Eq.~\eqref{eq:isoprobes} collapses to $\one_N$, yielding $\BBop=\Bop$ and $\SBBf=\SBf$. The iso variants are operator-identical to the originals; they differ only in the right-hand side excitation.

Applied to the source-driven landscapes, the iso versions are
\begin{align}
  u_{\BBop}(j)
    &\;=\; \sum_{\alpha\in\{x,y\}}
           \big\|[(G^\dg G+\eta^2 I)^{-1}\mathbf{s}^{(\alpha)}]_{(j,\cdot)}\big\|^2,
  \label{eq:gramiso} \\
  u_{\SBBf}(j)
    &\;=\; \sum_{\alpha\in\{x,y\}}
           \Big(\big\|\mathbf{a}^{(\alpha)}_j\big\|^2
                + \big\|\mathbf{b}^{(\alpha)}_j\big\|^2\Big),
  \quad
  \begin{pmatrix}\mathbf{a}^{(\alpha)}\\ \mathbf{b}^{(\alpha)}\end{pmatrix}
   = B^{-1}\!\begin{pmatrix}\mathbf{s}^{(\alpha)}\\ \mathbf{s}^{(\alpha)}\end{pmatrix},
  \label{eq:hermiso}
\end{align}
where $\|\cdot\|$ denotes per-site magnitude over polarization components. The spectral landscape $\SY$ responds to the full right singular subspace without source dependence; its natural counterpart is the polarization-traced diagonal of $Y$.

\subsection{Source-free cross-check: \texorpdfstring{$\Ydiag$}{Y\_diag}}

For $\SY$, we utilize the source-free diagonal. Since $Y$ is Hermitian positive semi-definite, the polarization-traced per-site value
\begin{equation}
  u_{\Ydiag}(j)
  \;=\; \sum_{\alpha\in\{x,y\}} Y_{(j,\alpha),(j,\alpha)}
  \label{eq:ydiagdef}
\end{equation}
admits the SVD expansion \eqref{eq:ydiagsvd}, representing a partial local density of states for $W$ pseudomodes. This acts as a non-Hermitian, polarization-traced analogue of the FM ``resolvent-diagonal'' landscape, structurally invariant in polarization. It serves as the natural TE cross-check for $\SY$.

\subsection{Equivalence classes}

Summary of equivalences across polarizations:
\begin{center}
  \renewcommand{\arraystretch}{1.15}
  \begin{tabular}{lll}
    \toprule
    Landscape & TM identity & TE identity \\ \midrule
    $\Bop$ & $\Bop\;=\;\BBop$ & $\Bop\ne\BBop$ \\
    $\SY$  & --- (no iso variant)  & --- (no iso variant) \\
    $\SBf$ & $\SBf\;=\;\SBBf$ & $\SBf\ne\SBBf$ \\
    $X$-block & $|X\one|^2 = |Y\one|^2$ & $|X\one|^2 = |Y\one|^2$ \\
    Off-diagonal & $|WY\one|^2 =|YW^\dg\one|^2$ & $|WY\one|^2 =|YW^\dg\one|^2$ \\
    \bottomrule
  \end{tabular}
\end{center}
The recommended TM set is $\{\Bop,\SY,\SBf\}$. For TE, the set is $\{\BBop,\SY\!+\!\Ydiag,\SBBf\}$, retaining $\SY$ for spectral-density continuity and adding $\Ydiag$ as a polarization-even cross-check.

\section{Dependence on the probe parameter \texorpdfstring{$z$}{z}}

\begin{figure}
\centering
\renewcommand{\arraystretch}{0.4}%
\begin{tabular}{cccccc}
 & \footnotesize $\mathrm{Im}\,z{=}0.00$ & \footnotesize $\mathrm{Im}\,z{=}0.50$ & \footnotesize $\mathrm{Im}\,z{=}1.00$ & \footnotesize $\mathrm{Im}\,z{=}2.00$ & \footnotesize $\mathrm{Im}\,z{=}4.00$ \\
\footnotesize\rotatebox{90}{\textbf{Pinwheel}} & \includegraphics[width=0.18\linewidth]{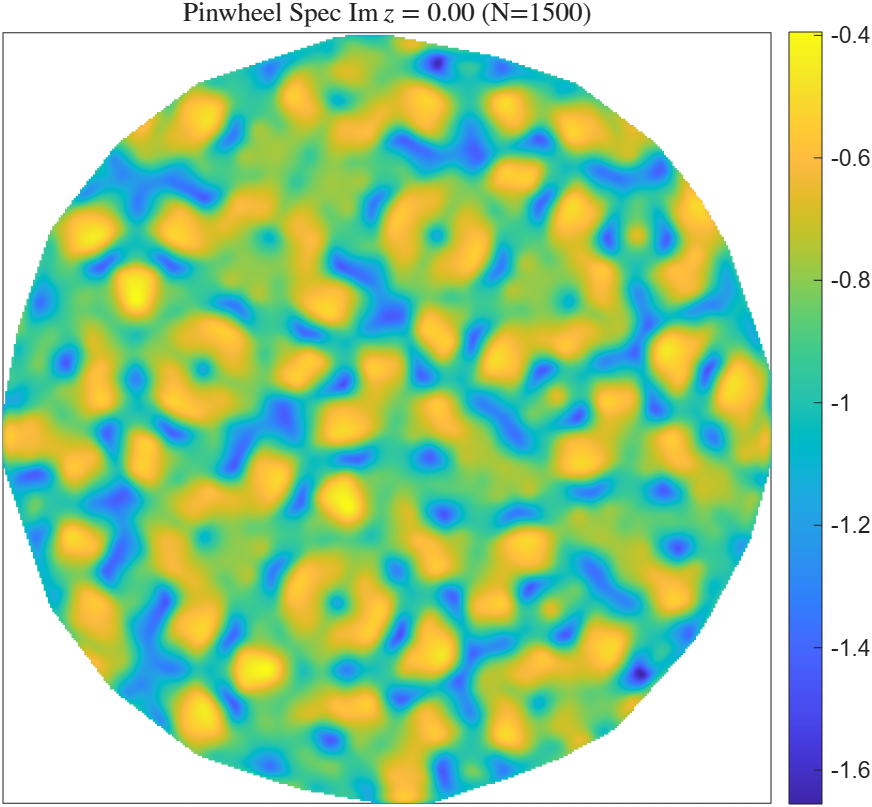} & \includegraphics[width=0.18\linewidth]{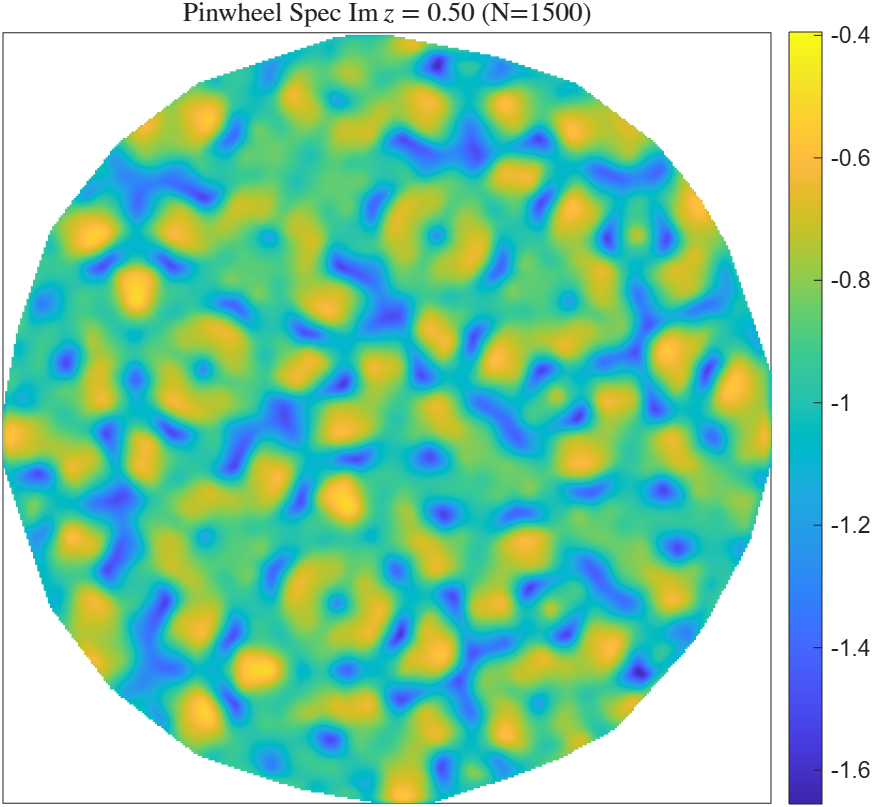} & \includegraphics[width=0.18\linewidth]{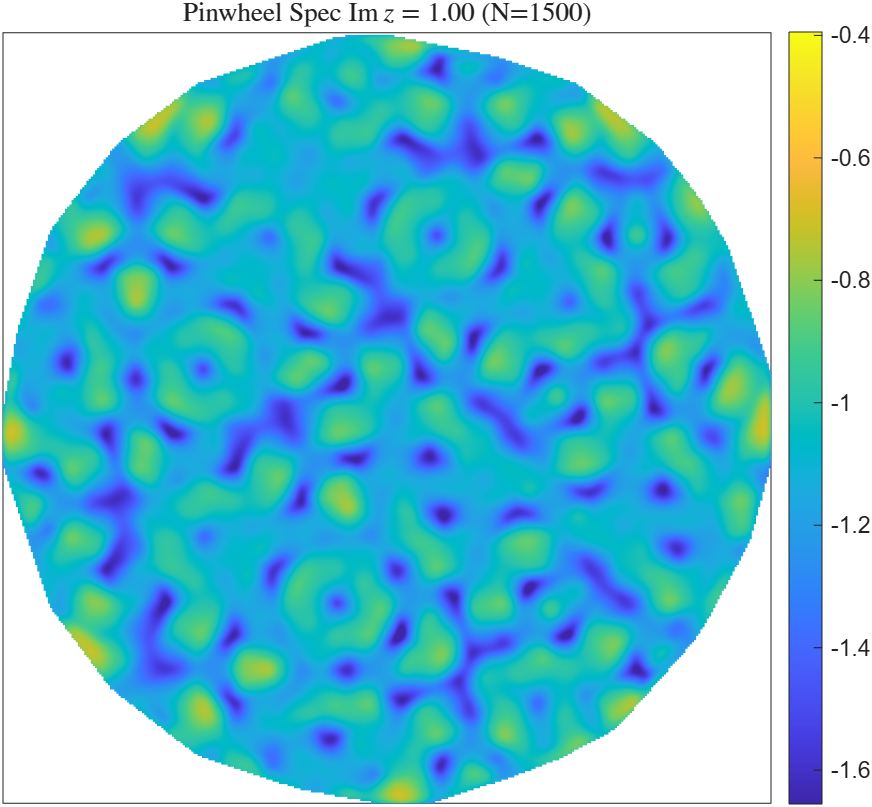} & \includegraphics[width=0.18\linewidth]{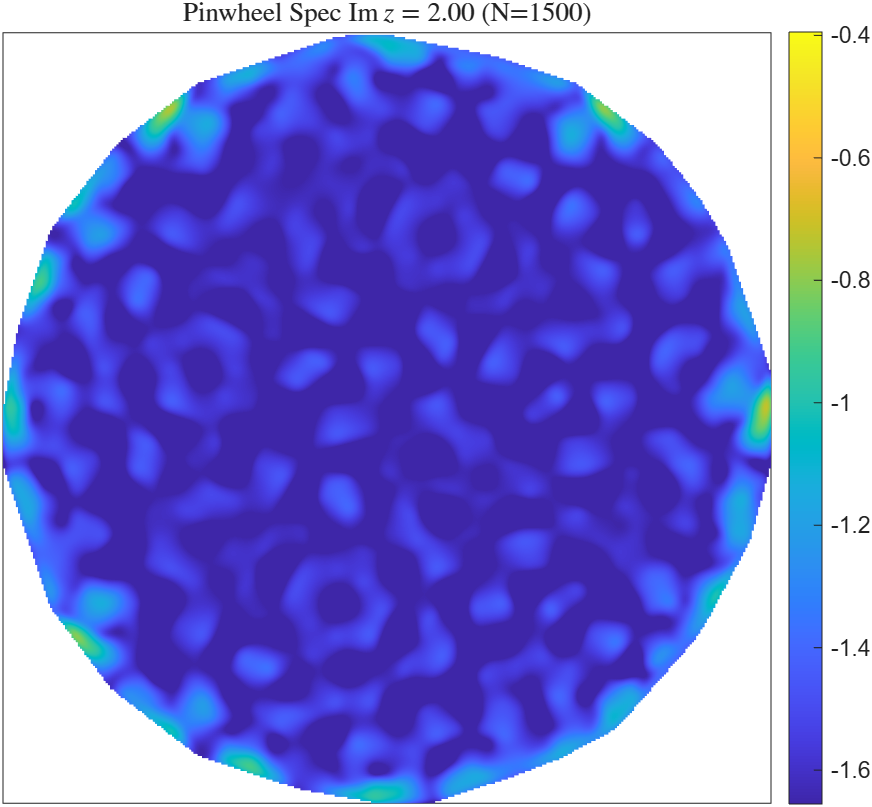} & \includegraphics[width=0.18\linewidth]{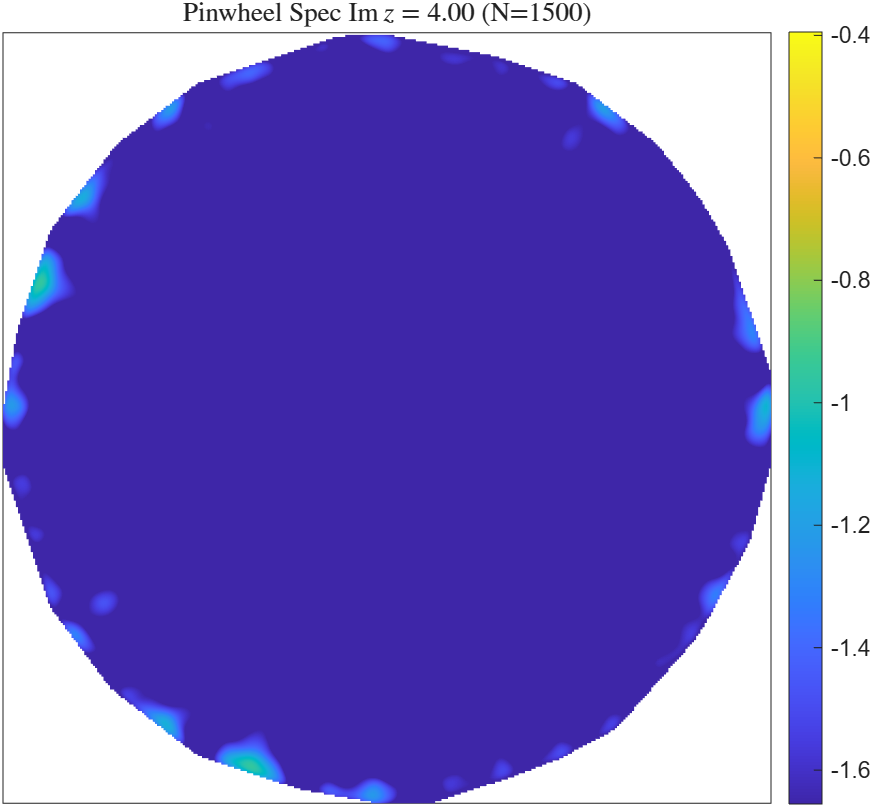} \\
\footnotesize\rotatebox{90}{\textbf{ThueMorse}} & \includegraphics[width=0.18\linewidth]{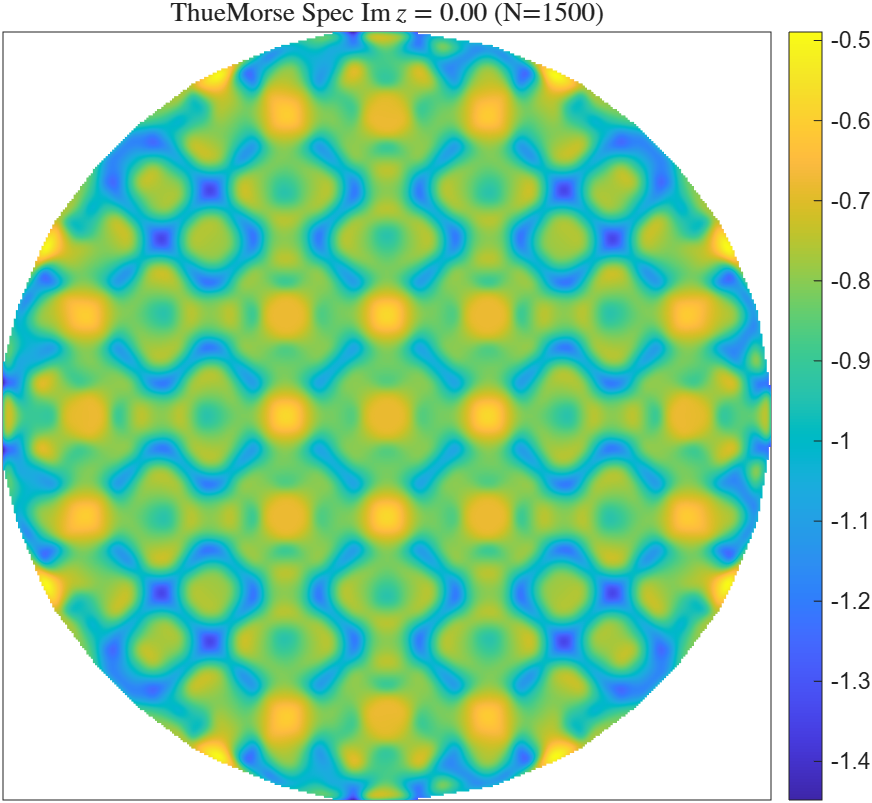} & \includegraphics[width=0.18\linewidth]{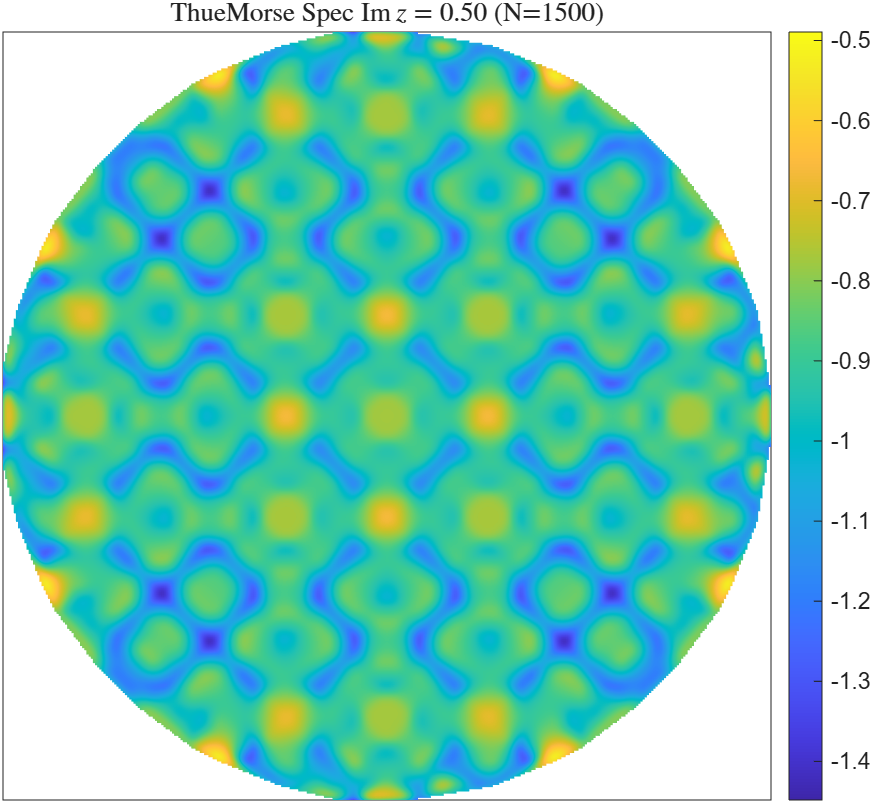} & \includegraphics[width=0.18\linewidth]{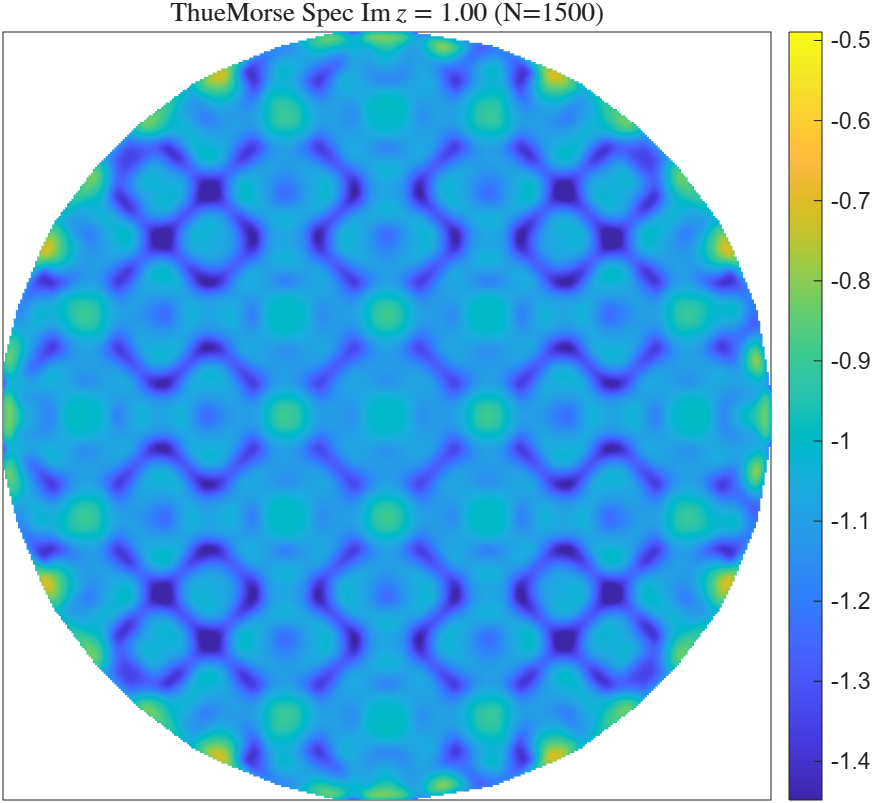} & \includegraphics[width=0.18\linewidth]{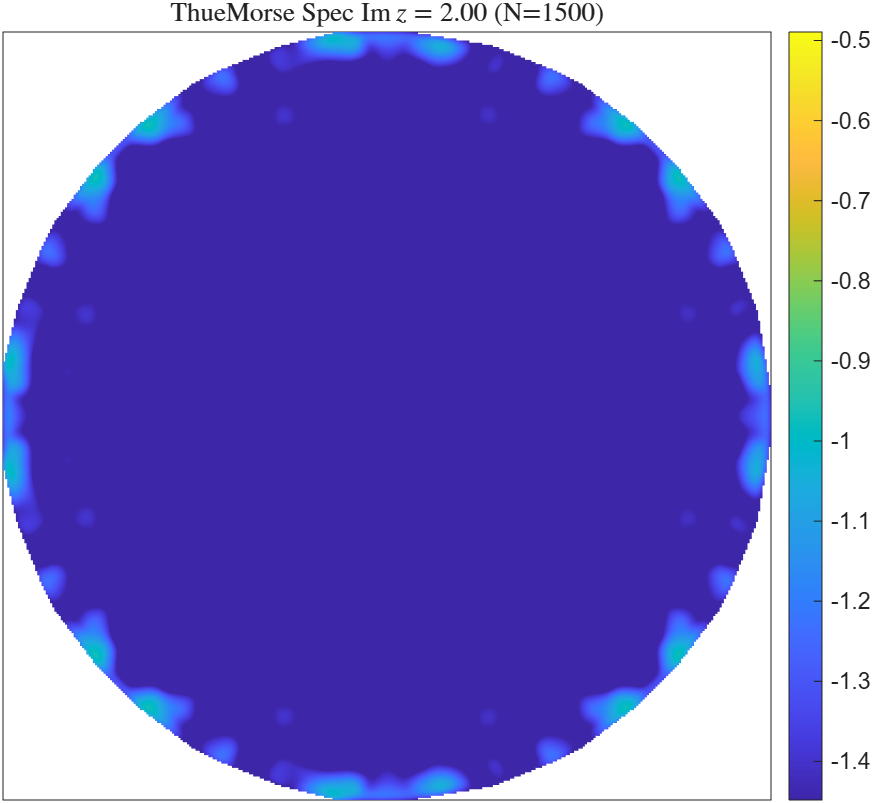} & \includegraphics[width=0.18\linewidth]{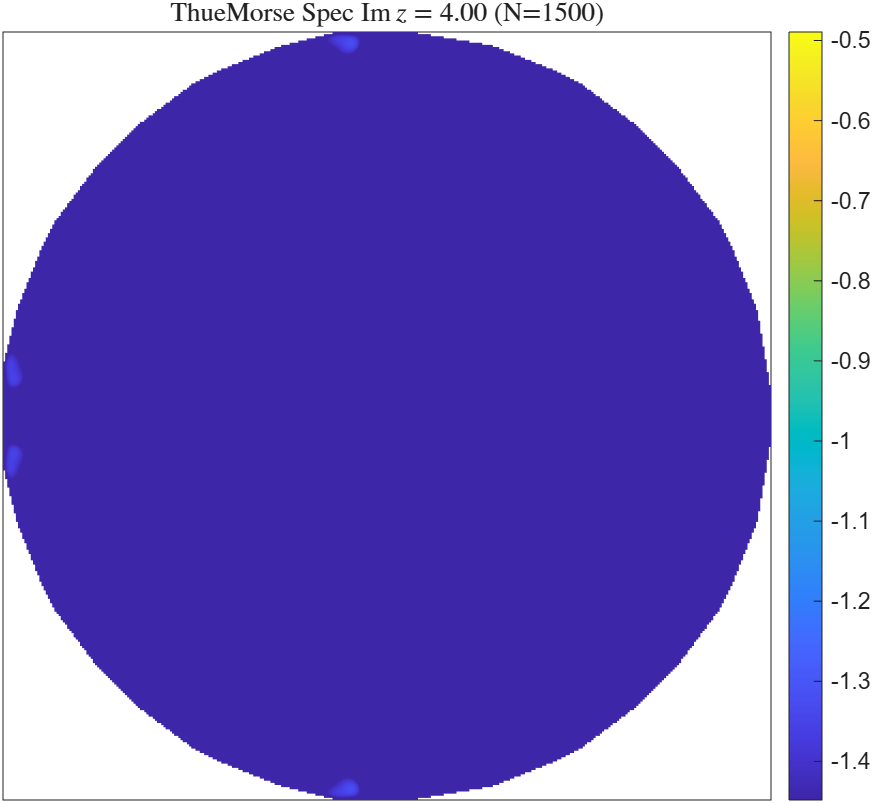}  \\
\end{tabular}
\caption{Spec landscape $\log_{10}u_{\mathrm{Spec}}$ with $Y(z) = (W^\dagger W + \eta^2 I)^{-1}$ and $W = G - zI$, as the probe $z = i\,t$ sweeps the imaginary axis for Pinwheel and Thue-Morse TM patterns ($N\le 1500$ sites, OD\,=\,10, $\eta = 0.005\,\|G\|_F$).  As $|\mathrm{Im}\,z|$ grows the probe moves out of the spectrum, $\sigma_{\min}(G-zI)$ grows like $|\mathrm{Im}\,z|$, the solve becomes well-conditioned, and $u_{\mathrm{Spec}}$ collapses toward the median. }
\label{fig:spec_grid}
\end{figure}

\begin{figure}
\centering
\renewcommand{\arraystretch}{0.4}%
\begin{tabular}{cccccc}
 & \footnotesize $\mathrm{Im}\,z{=}0.00$ & \footnotesize $\mathrm{Im}\,z{=}0.50$ & \footnotesize $\mathrm{Im}\,z{=}1.00$ & \footnotesize $\mathrm{Im}\,z{=}2.00$ & \footnotesize $\mathrm{Im}\,z{=}4.00$ \\
\footnotesize\rotatebox{90}{\textbf{Pinwheel}} & \includegraphics[width=0.18\linewidth]{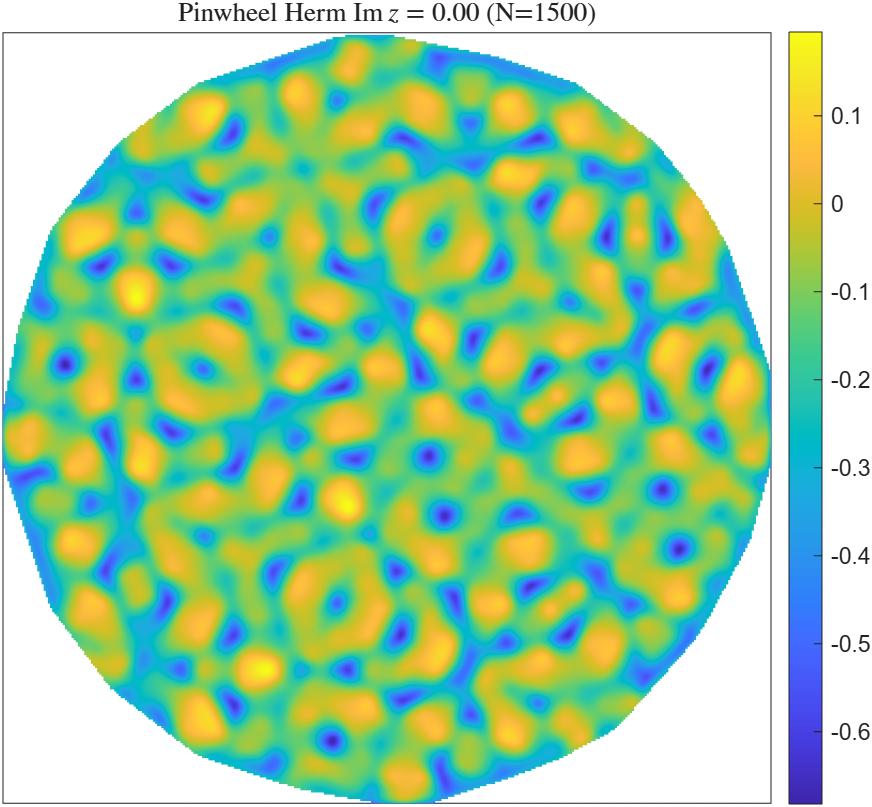} & \includegraphics[width=0.18\linewidth]{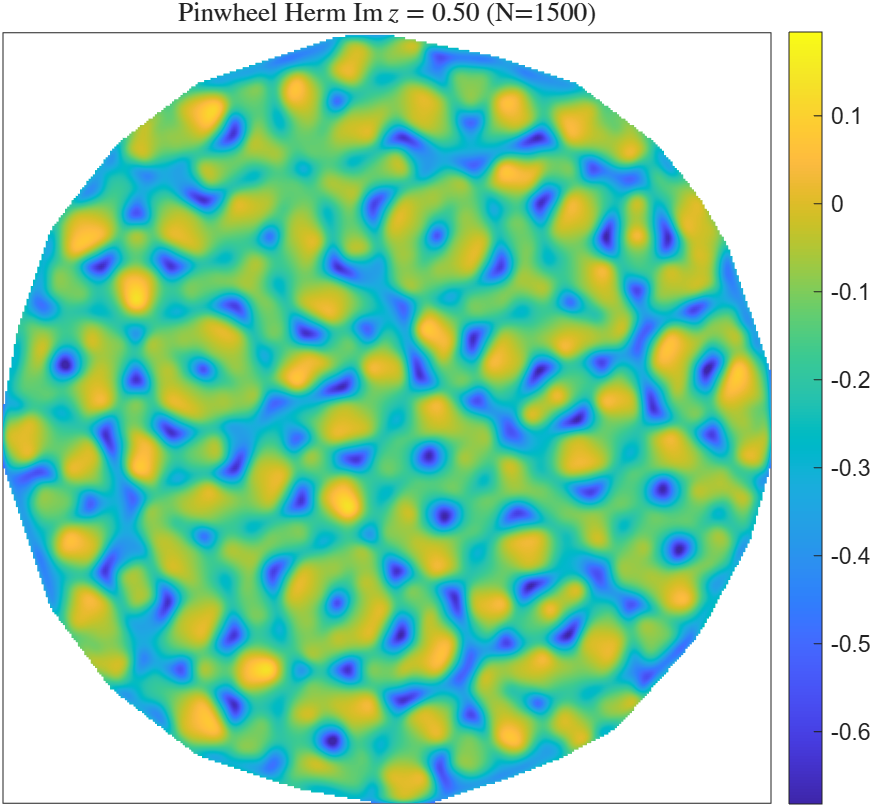} & \includegraphics[width=0.18\linewidth]{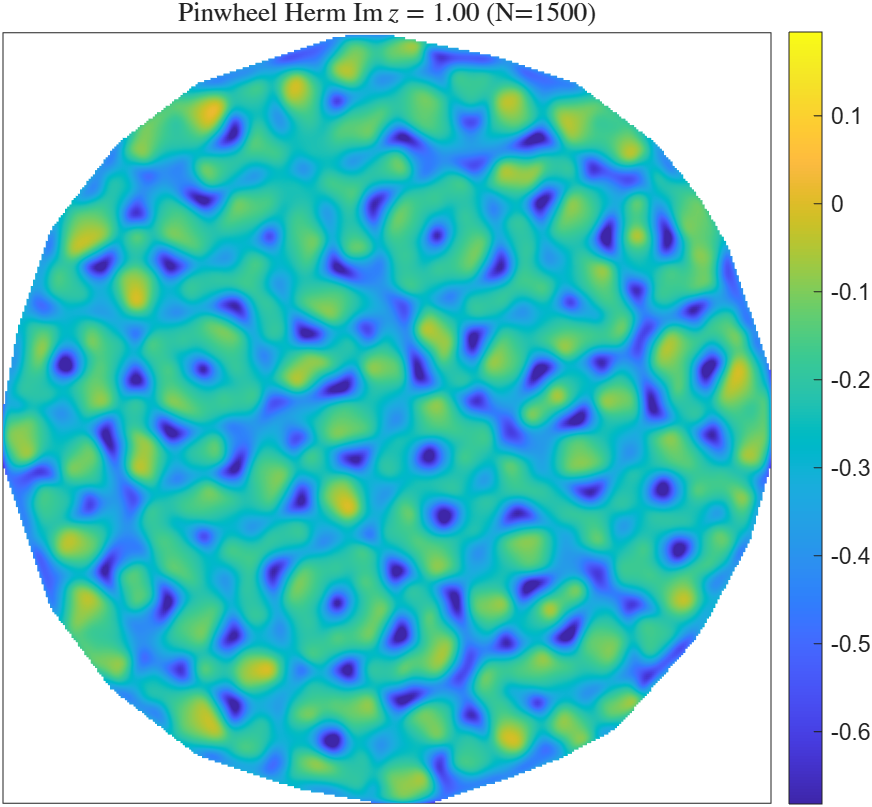} & \includegraphics[width=0.18\linewidth]{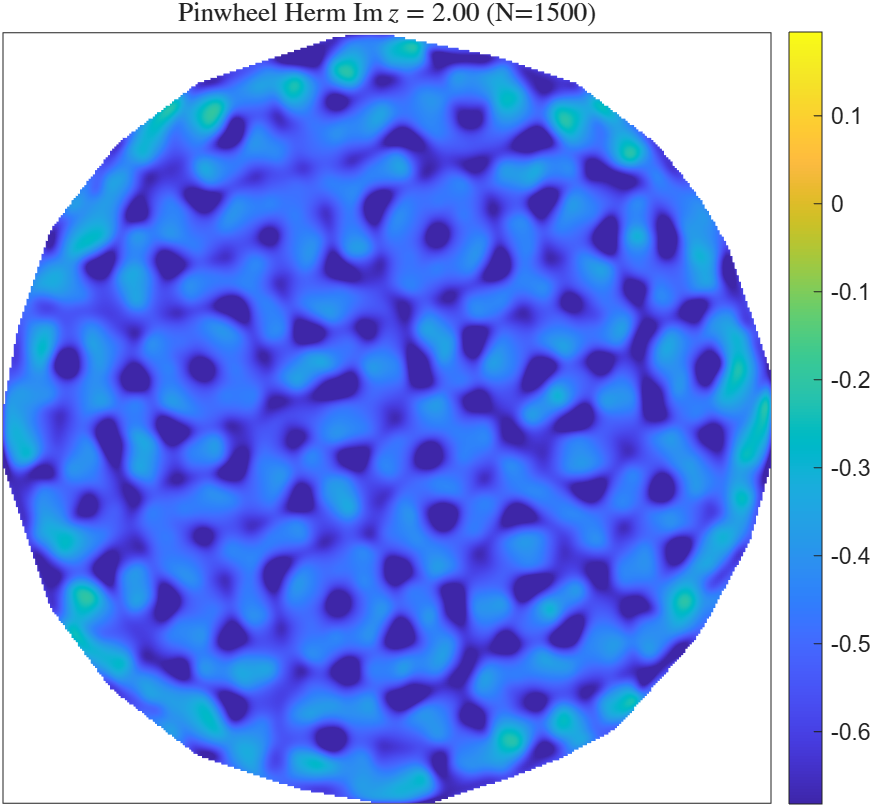} & \includegraphics[width=0.18\linewidth]{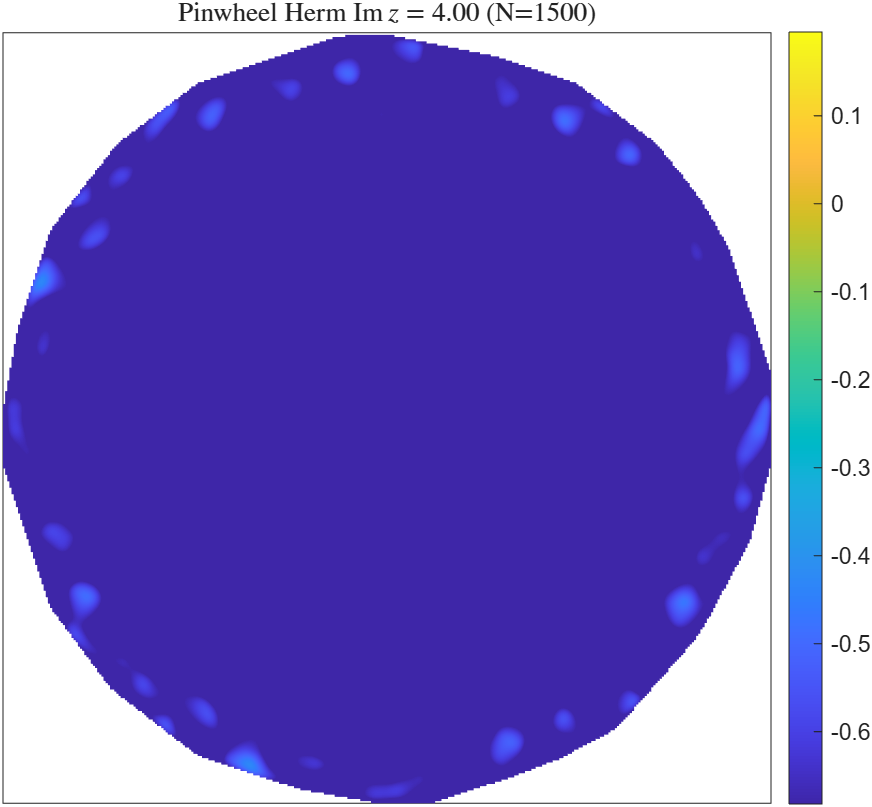} \\
\footnotesize\rotatebox{90}{\textbf{ThueMorse}}  & \includegraphics[width=0.18\linewidth]{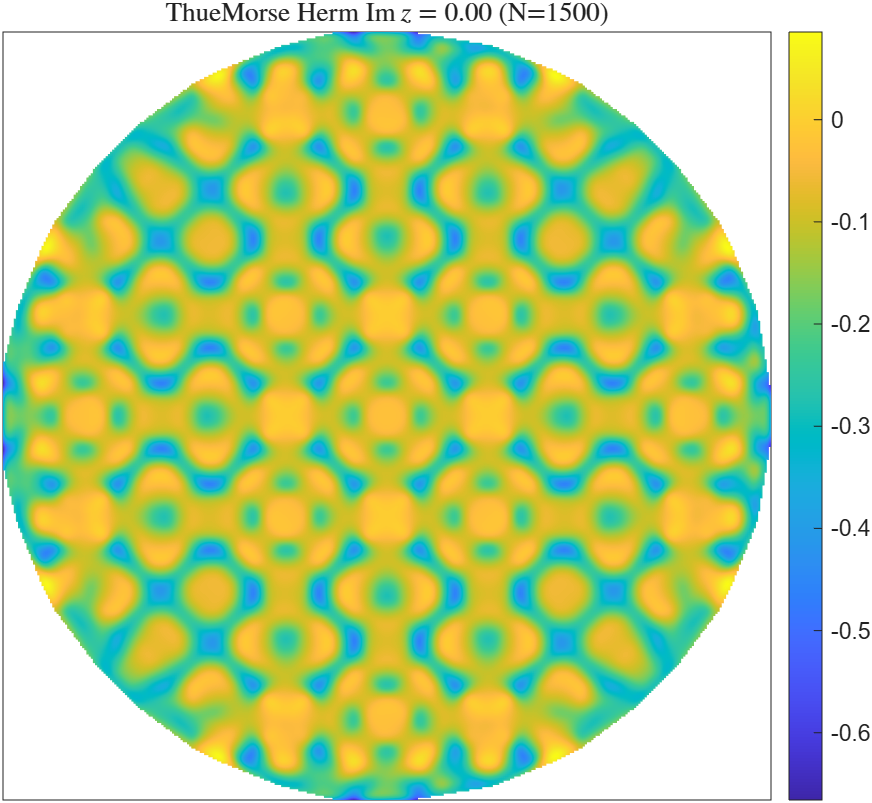} & \includegraphics[width=0.18\linewidth]{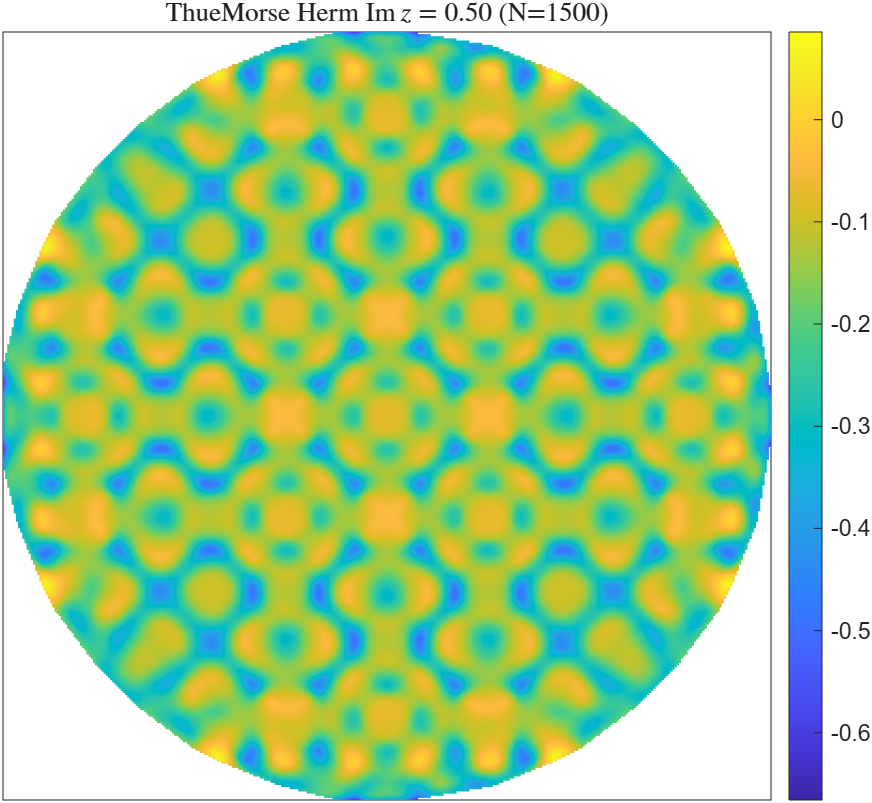} & \includegraphics[width=0.18\linewidth]{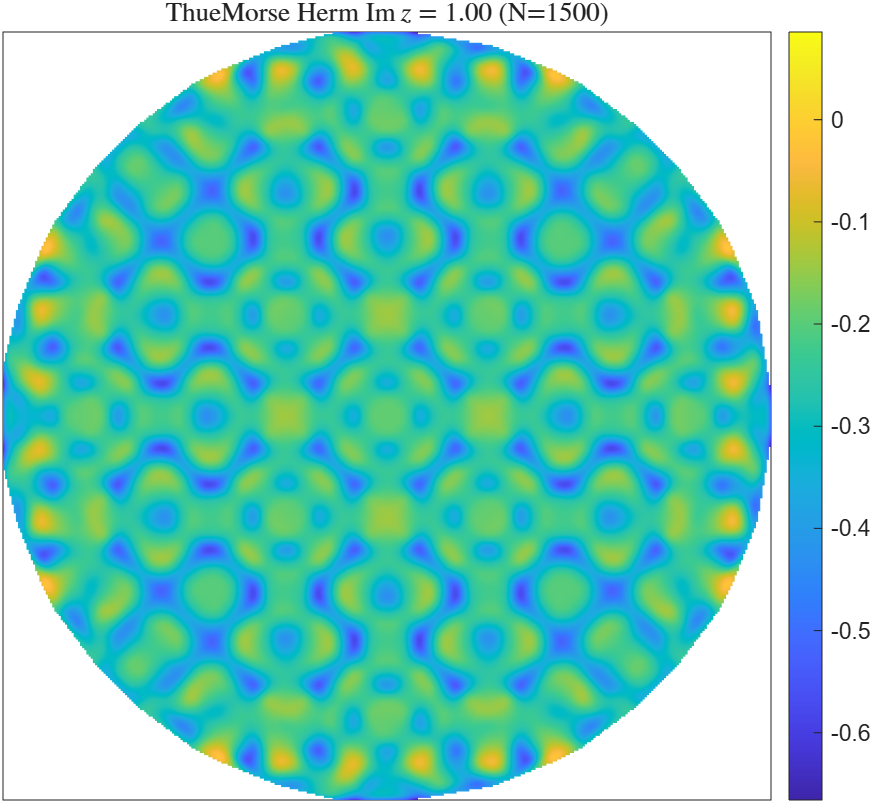} & \includegraphics[width=0.18\linewidth]{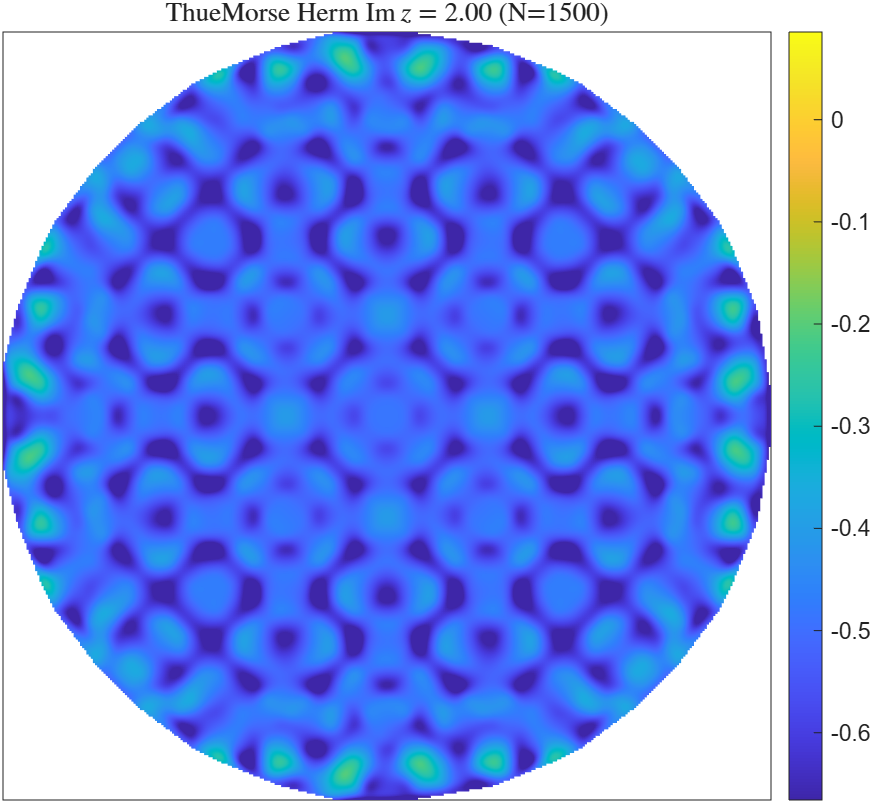} & \includegraphics[width=0.18\linewidth]{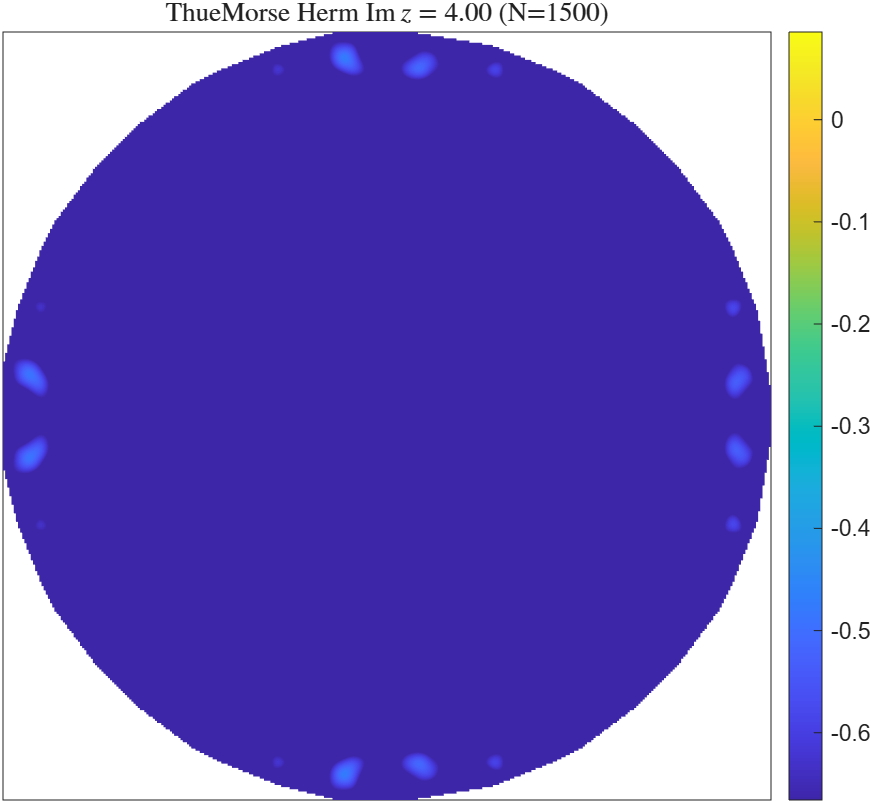}  \\
\end{tabular}
\caption{Herm landscape along the same imaginary-axis sweep. Panels, patterns and colour range match Figure~\ref{fig:spec_grid}; here each landscape is obtained from a single $2N\times 2N$ hermitized solve rather than the explicit $Y$ inverse. The flattening trend with $|\mathrm{Im}\,z|$ is the same, confirming that the loss of localization is a property of the probe location in the complex plane rather than of the specific landscape definition.}
\label{fig:herm_grid}
\end{figure}

\paragraph{Landscape as $z$ is varied.}
The main-text evaluation of $\SY$ and $\SBf$ uses
$z=\lambda_{\mathrm{long}}$, the longest-lived eigenvalue of $G$.
To make the role of this choice explicit we repeat the solve for a
family of probes $z=\mathrm{i}\,t$ moving along the imaginary axis,
holding the pattern, system size and regularizer fixed at the paper's
working values ($N\le 1500$, OD\,$=\,10$,
$\eta=5\!\times\!10^{-3}\|G\|_F$). Figure~\ref{fig:spec_grid} tracks
the resulting $\SY$ landscape on Pinwheel and Thue--Morse arrays, and
Figure~\ref{fig:herm_grid} tracks $\SBf$ on the same arrays over the
same sweep. At $\mathrm{Im}\,z=0$ the probe lies in the physically
relevant subradiant band and both landscapes resolve the compact
$\sim\!\lambda/2$ hot-spots discussed in
Section~\ref{sec:three}. As $|\mathrm{Im}\,z|$ increases, the probe
leaves the spectrum, the smallest singular value grows as
$\sigma_{\min}(W(z))\!\sim\!|\mathrm{Im}\,z|$ via
Eq.~\eqref{eq:kappa_scaling}, the resolvent becomes well conditioned,
and both landscapes flatten toward their site-median. The two
landscape definitions trace the same flattening trend, which is
consistent with their common dependence on $(W^\dagger W+\eta^2 I)^{-1}$
and confirms that the loss of contrast is a property of the probe
location rather than of the specific landscape functional. 

\begin{figure}
\centering
\setlength{\tabcolsep}{2pt}
\renewcommand{\arraystretch}{0.2}
\newcommand{\poisrow}{0.21\linewidth}
\begin{tabular}{@{}c @{\hspace{4pt}} cccc @{}}
 & {\large\textbf{point pattern}} & {\large\textbf{$u_\mathrm{Gram}$ \small($z=0$)}} & {\large\textbf{$u_\mathrm{Spec}$ \small($z=\lambda_\mathrm{long}$)}} & {\large\textbf{$u_\mathrm{Herm}$ \small($z=\lambda_\mathrm{long}$)}} \\[2pt]
{\rotatebox{90}{\textbf{~TM, $N=3394$}}} &
  \includegraphics[width=\poisrow]{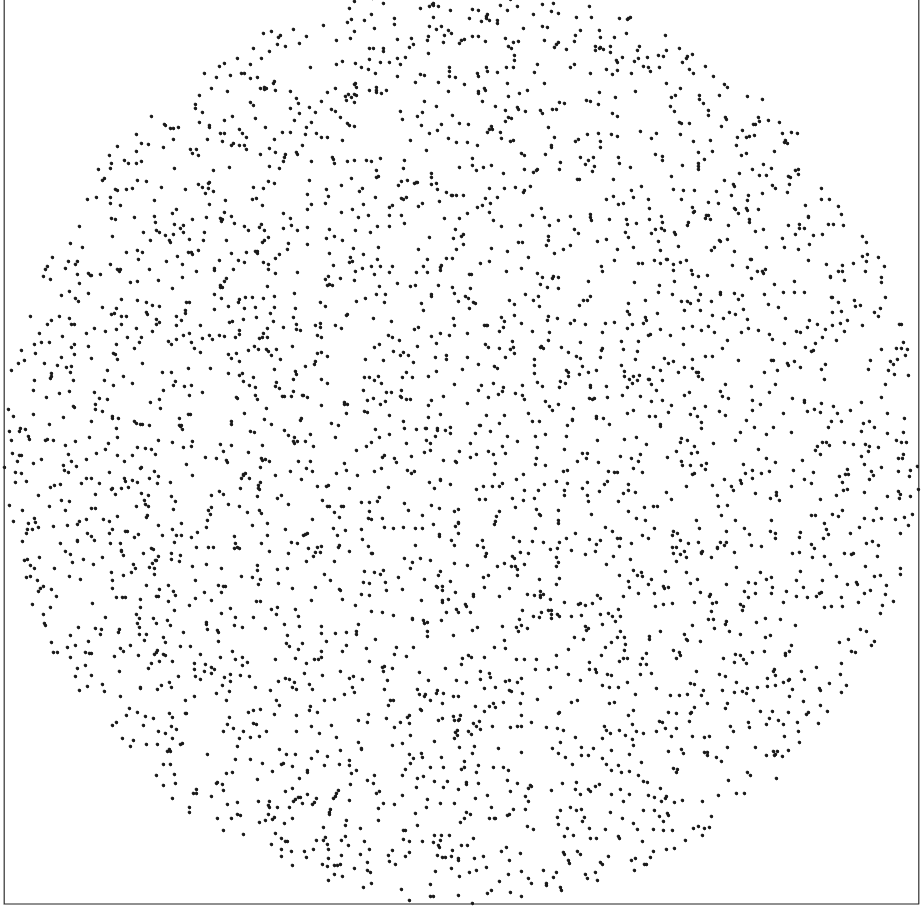} &
  \includegraphics[width=\poisrow]{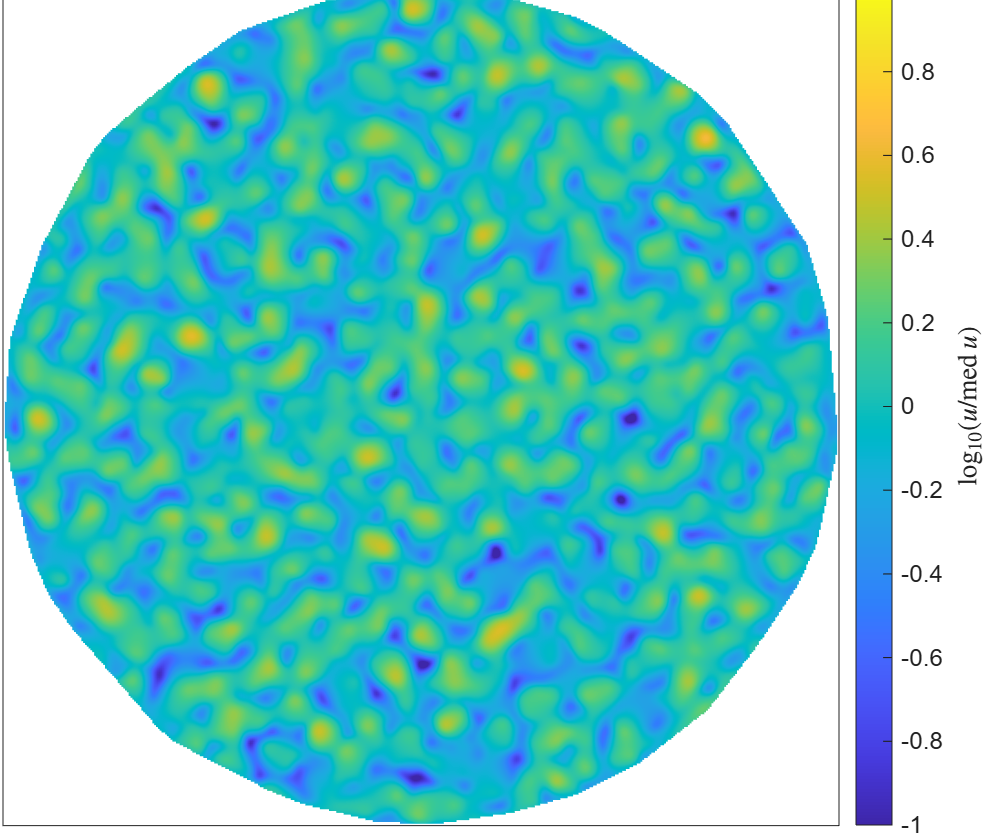} &
  \includegraphics[width=\poisrow]{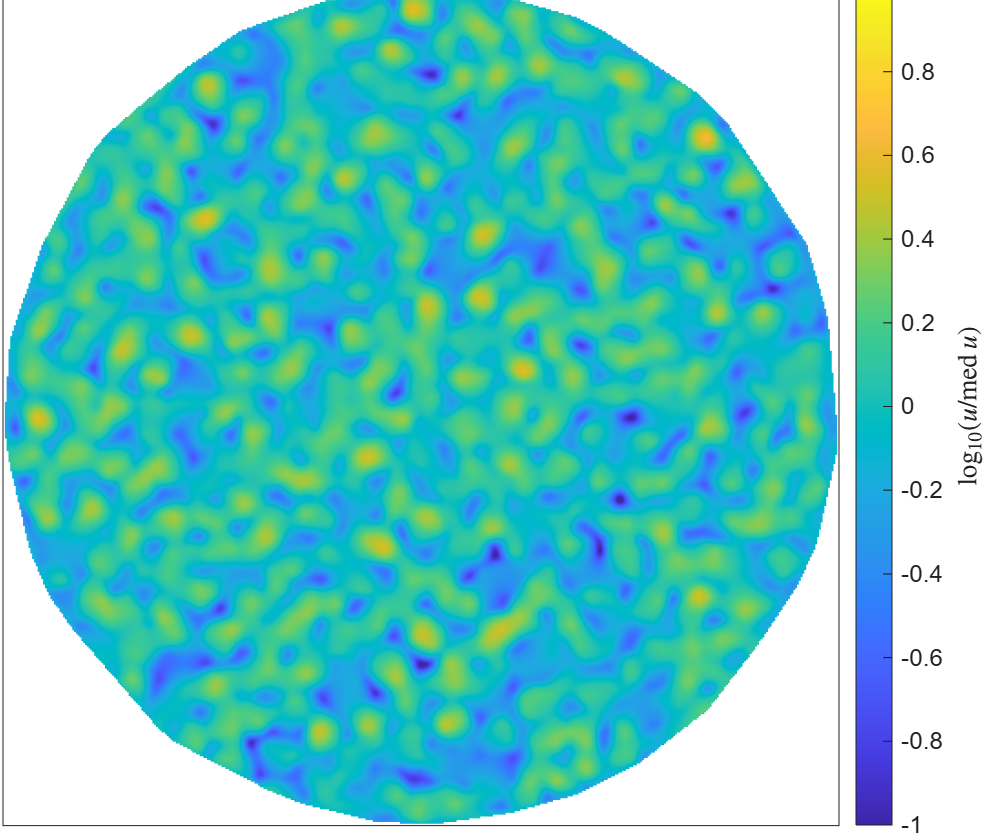} &
  \includegraphics[width=\poisrow]{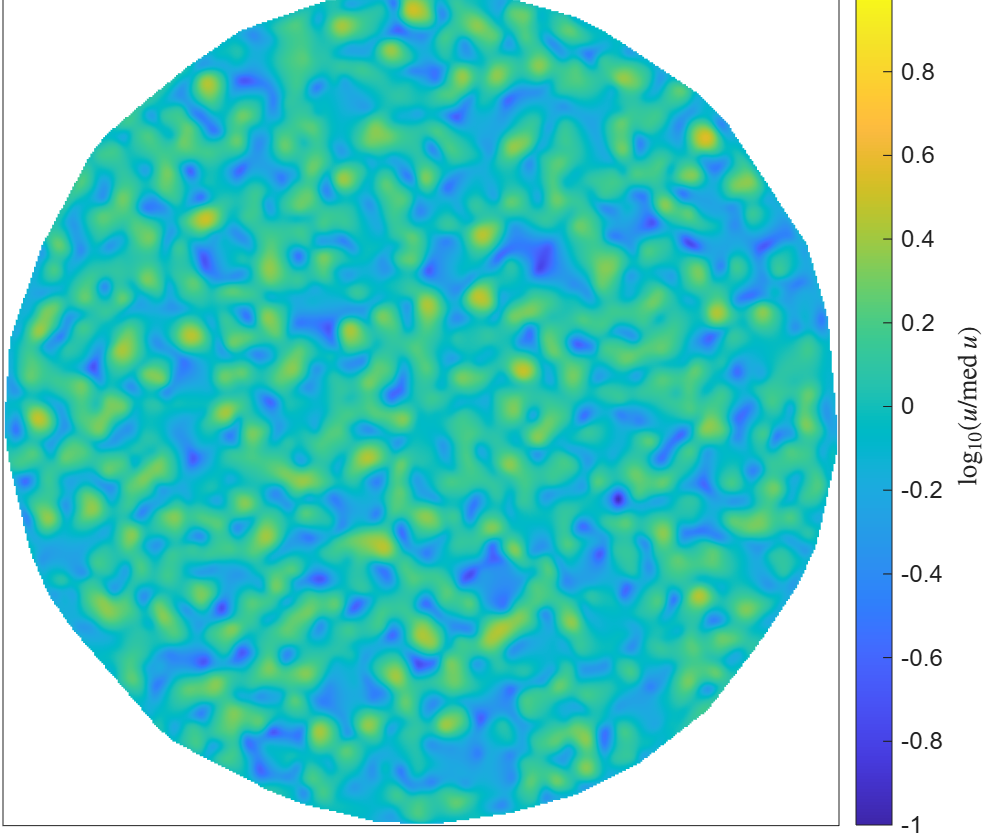} \\
{\rotatebox{90}{\textbf{~TE-iso, $N=2803$}}} &
  \includegraphics[width=\poisrow]{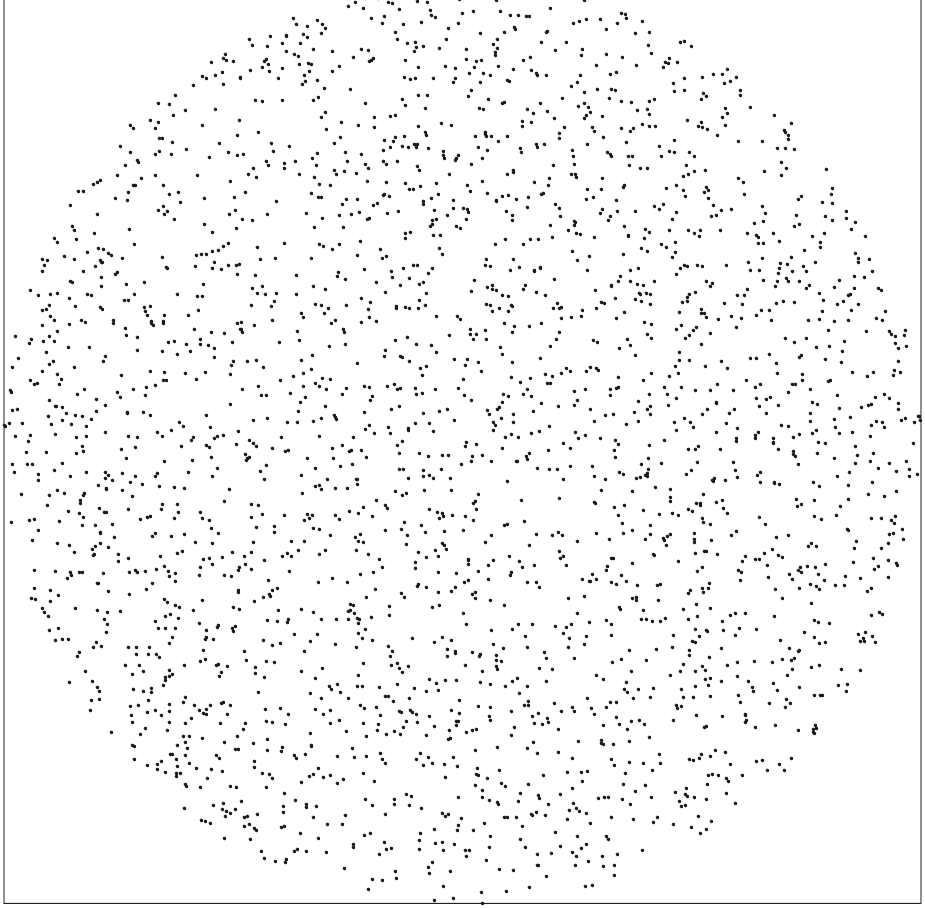} &
  \includegraphics[width=\poisrow]{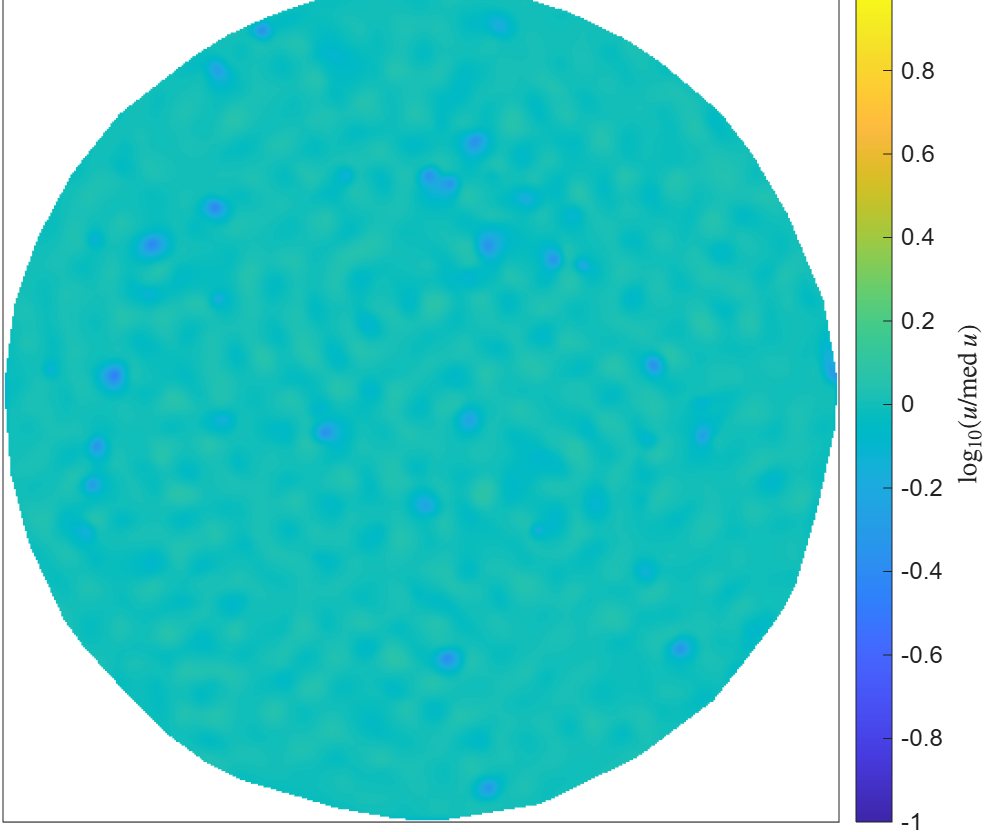} &
  \includegraphics[width=\poisrow]{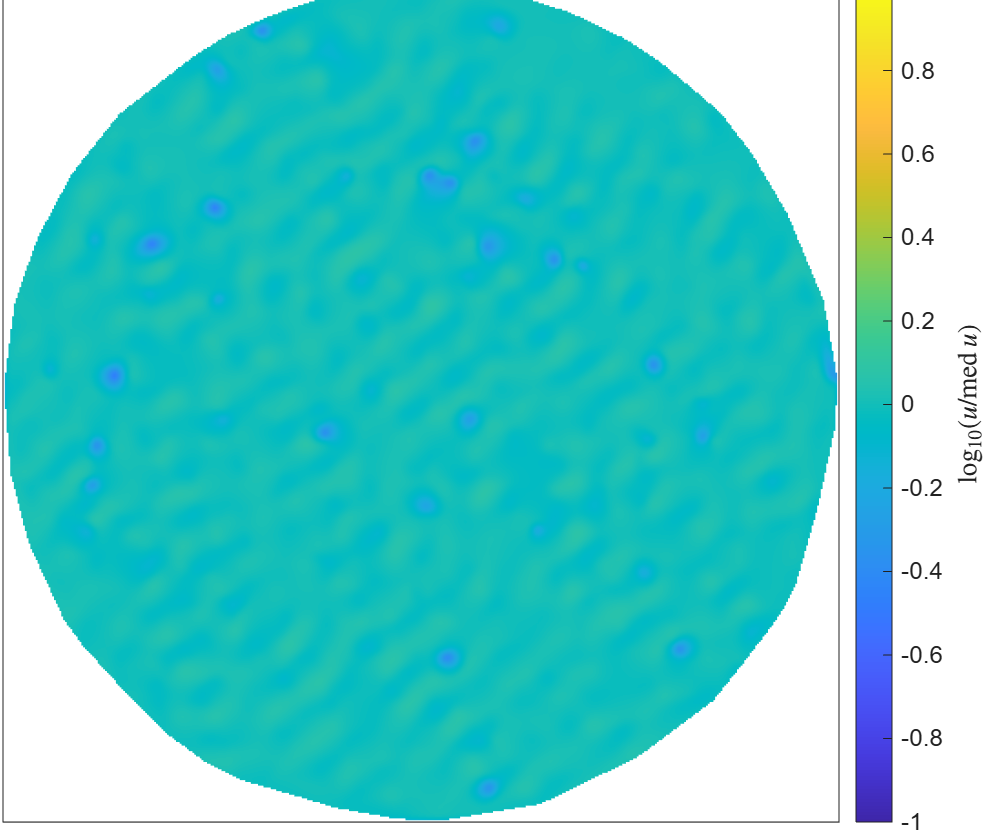} &
  \includegraphics[width=\poisrow]{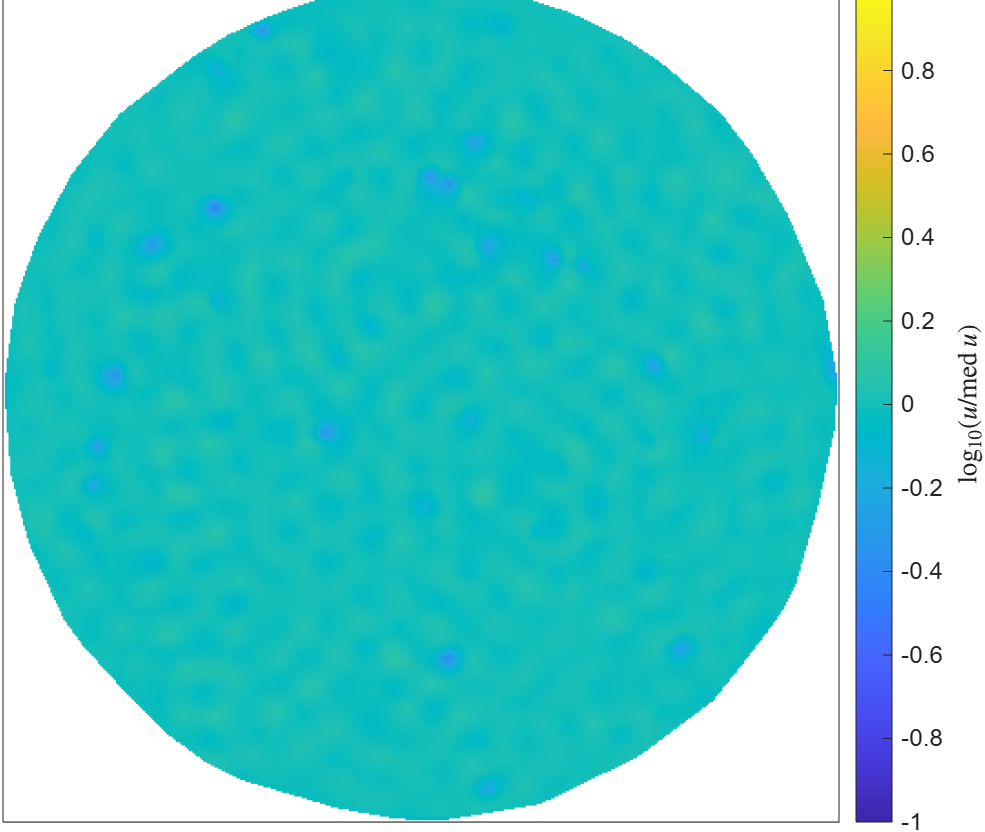} \\
\end{tabular}
\caption{Poisson point patterns and the paper's three landscapes at large $N$. OD\,$=\,10$, $\eta = 0.005\,\|G\|_F$. TE-iso uses the isotropic probe pair $s^{(x)} = (\mathbf{1};\mathbf{0})$, $s^{(y)} = (\mathbf{0};\mathbf{1})$. Column~1: the realized Poisson sample. Columns~2--4: Gram at $z = 0$, Spec and Herm at $z = \lambda_{\mathrm{long}}$ (smallest-$\mathrm{Im}$ eigenvalue of $G$). Landscape colours are $\log_{10}(u/\mathrm{med}\,u)$ clipped at $[-1,+1]$. The TM row shows well-defined $\sim\!\lambda/2$ hot-spots in all three landscapes; the TE-iso row collapses to the median everywhere except at a sparse set of near-field dipole pair-mode sites. This is not a landscape failure: it correctly reports the known absence of TE wave localization in two-dimensional \emph{uncorrelated} random dipole arrays~\cite{skipetrov2014absence,skipetrov2018finite}, with only the near-field pair modes registering above the diffuse background. The correlated and deterministic aperiodic cases of Section~\ref{sec:results} (Bernoulli, Thue--Morse) support TE localization, which the Herm landscape resolves there.}
\label{fig:grid}
\end{figure}

Finally, in Figure \ref{fig:grid} we show the behavior, for both polarizations, of the landscapes computed for a single realization of a Poisson point pattern. We see that the highly fluctuating nature of the landscapes for the TM polarization case demonstrates the onset of spatial wave localization while the spatially uniform landscapes obtained for the TE case is incompatible with wave localization. This behavior, which is clearly predicted by the developed landscapes, is entirely consistent with the well-known absence of localization for TE polarized dipoles in Poisson  arrays due to the detrimental contribution of near-field coupling effects at large optical density \cite{skipetrov2014absence,dalnegro2022waves}.

\section{Implementation and experimental setup}
\label{app:impl}

\paragraph{Landscape solves.}
Given $\eta=\eta_{\mathrm{frac}}\|G\|_F$ (Tikhonov-discrepancy heuristic~\cite{tikhonov1963regularization,hansen2010discrete}) and the long-lived reference spectral parameter $z_{\mathrm{ref}}=\arg\min_{\lambda\in\sigma(G)}y_s(\lambda)$, where $(x_s,y_s)=(-\tfrac12\Realp\lambda,\Imag\lambda)$~\cite{skipetrov2014absence,skipetrov2018finite}, each of $\Bop$, $\SY$, $\SBf$ requires one dense solve at $N=3500$. The $2N{\times}2N$ block for $\SBf$ is assembled and solved directly.

\paragraph{Grid, smoothing, watershed.}
Site values are interpolated to an $N_g{\times}N_g$ grid ($N_g{=}300$) using \texttt{scatteredInterpolant} (natural neighbour). Landscape grids are smoothed with a Gaussian kernel ($\sigma_L{=}4$ pixels); mode intensities use $\sigma_M{=}1$. Watershed segmentation follows the FM construction~\cite{arnold2016effective,arnold2019computing}: the height field $-u(x,y)$ is $h$-minimum--suppressed with $h=0.06\,\mathrm{range}(u)$, processed via \texttt{watershed}, and refined by morphological thinning. The label image $L$ and ridge set $\partial L$ define the basins.

\paragraph{Mode selection.}
Eigenpairs of $G$ are obtained via direct \texttt{eig}. The top-$K$ long-lived modes correspond to minimal $y_s$ within the chosen real-energy window. Their per-site intensities are gridded via \texttt{scatteredInterpolant} and overlaid on watershed basins. Parameter values are listed in Table~\ref{tab:params}.

\begin{table}[h!]
\centering\small
\begin{tabular}{lll}
\toprule
Symbol & Value & Description \\
\midrule
$N$ & 3500 & number of dipole sites per pattern \\
$\mathrm{OD}$ & 10 & target optical density (fixes $k_0$) \\
$\eta_{\mathrm{frac}}$ & $5{\times}10^{-3}$ & Tikhonov regulator $\eta=\eta_{\mathrm{frac}}\|G\|_F$ \\
$K$ & 100 & top long-lived modes overlaid \\
$[E_{\min},E_{\max}]$ & $[-1,1]$ & real-energy window \\
$z_{\mathrm{ref}}$ & long-lived $\lambda$ & spectral parameter for $\SY,\SBf$ \\
$N_g$ & 300 & grid resolution per side \\
$(\sigma_L,\sigma_M)$ & $(4,1)$ & Gaussian smoothing (pixels) \\
$h$-suppress & $0.06\,\mathrm{range}(u)$ & \texttt{imhmin} depth \\
\bottomrule
\end{tabular}
\caption{Parameters used for generating the Figures. Main figures utilize TM polarization; TE-correct variants from Appendix~\ref{app:tepol} require no parameter adjustment.}
\label{tab:params}
\end{table}

\section{Structural properties of the investigated \texorpdfstring{$G$}{G} Matrices}
\label{app:mat-properties}

We present several structural, matrix and mode indicators per pattern in this section. The indicator descriptions are given below followed by the values per pattern in Table~\ref{tab:indicators}.

\begin{table}[!ht]\centering\scriptsize
\caption{Indicators per pattern.}
\label{tab:indicators}
\begin{tabular}{@{}l cccc @{\hspace{3\tabcolsep}} ccccc @{}}\toprule
pattern & CV($d_{nn}$) & CV(dens) & Gini & eff bw & $\Delta_H$ & $K_{\max}$ & $K_{\mathrm{med}}$ & $\overline{\log K}$ & $P(K\!\ge\!10)$ \\ \midrule
Coprime              & 0.06 & 0.27 & 0.14 & 0.284 & 0.761 & 2300    & 4.93 & 0.81 & 0.347 \\
ThueMorse            & 0.13 & 0.26 & 0.11 & 0.283 & 0.782 & 1480    & 3.46 & 0.71 & 0.287 \\
Fibonacci            & 0.16 & 0.23 & 0.10 & 0.285 & 0.789 & 3720    & 5.11 & 0.85 & 0.363 \\
Halton               & 0.27 & 0.25 & 0.12 & 0.276 & 0.779 & 16400 & 2.77 & 0.74 & 0.321 \\
EisensteinPrime      & 0.29 & 0.31 & 0.16 & 0.282 & 0.774 & 1720    & 1.80 & 0.46 & 0.166 \\
GaussianPrime        & 0.23 & 0.33 & 0.17 & 0.282 & 0.774 & 1460    & 2.04 & 0.52 & 0.193 \\
Bernoulli            & 0.12 & 0.29 & 0.15 & 0.277 & 0.781 & 5580    & 4.36 & 0.82 & 0.365 \\
RudinShapiro         & 0.10 & 0.30 & 0.16 & 0.277 & 0.785 & 3320    & 3.85 & 0.75 & 0.313 \\
\bottomrule\end{tabular}\end{table}

\begin{description}\small
\item[$\mathrm{CV}(d_{nn})$] coefficient of variation of the nearest-neighbour distance across all $N=3500$ sites.
\item[$\mathrm{CV}(\mathrm{density})$, Gini(density)] variability of the local point count on a $16\times 16$ grid over the aperture (only occupied cells).
\item[eff bw / N] mean row-participation ratio $\bigl(\sum_i|G_{ji}|^2\bigr)^2/\sum_i|G_{ji}|^4$ divided by $N$: fraction of the row that carries the row's weight (larger $=$ more dense).
\item[Henrici normalized departure] $\Delta_H = \|T-\mathrm{diag}(T)\|_F/\|G\|_F \in [0,1]$, where $T$ is the upper-triangular factor of the complex Schur form of $G$.  $\Delta_H = 0$ iff $G$ is normal.
\item[Petermann factor] $K_n = 1/|\langle L_n | R_n\rangle|^2$ with $L_n, R_n$ the unit-norm left and right eigenvectors of the $n$-th eigenmode.  $K_n=1$ for a normal matrix; large $K_n$ signals biorthogonal near-degeneracy.  Reported: $K_{\max}$, $K_{\mathrm{median}}$, $\overline{\log_{10}K}$, fraction of $K_n\ge 10$.
\end{description}

\section{Pseudospectra of the \texorpdfstring{$G$}{G} Matrices}
\label{app:pseudospectra}


\begin{figure}[t]
  \centering
  \setlength{\tabcolsep}{2pt}
  \renewcommand{\arraystretch}{0.9}
  \begin{tabular}{@{}cc@{}}
    \small Coprime, TM, $N{=}3500$ & \small Thue--Morse, TM, $N{=}3500$ \\
    \includegraphics[width=0.46\linewidth]{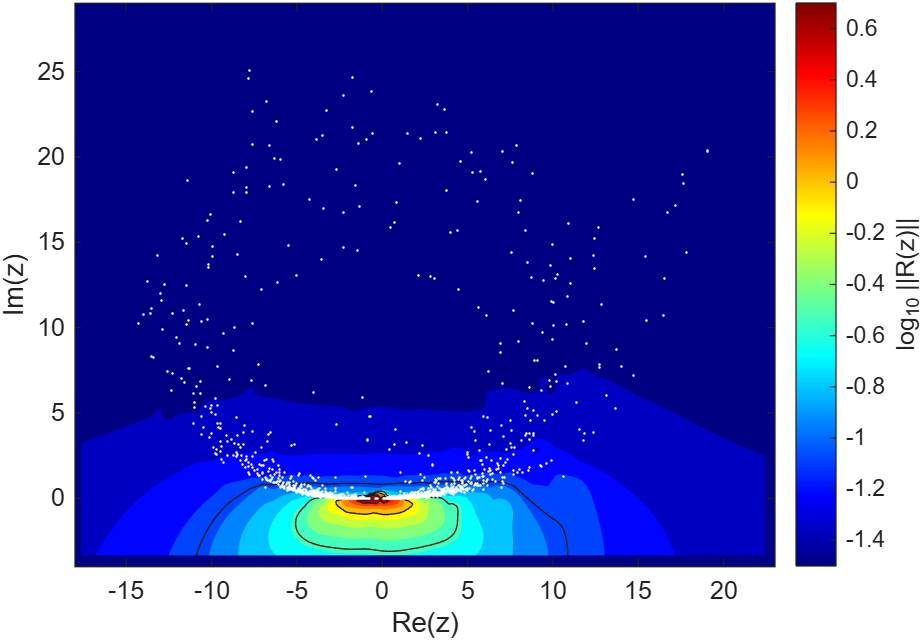} &
    \includegraphics[width=0.46\linewidth]{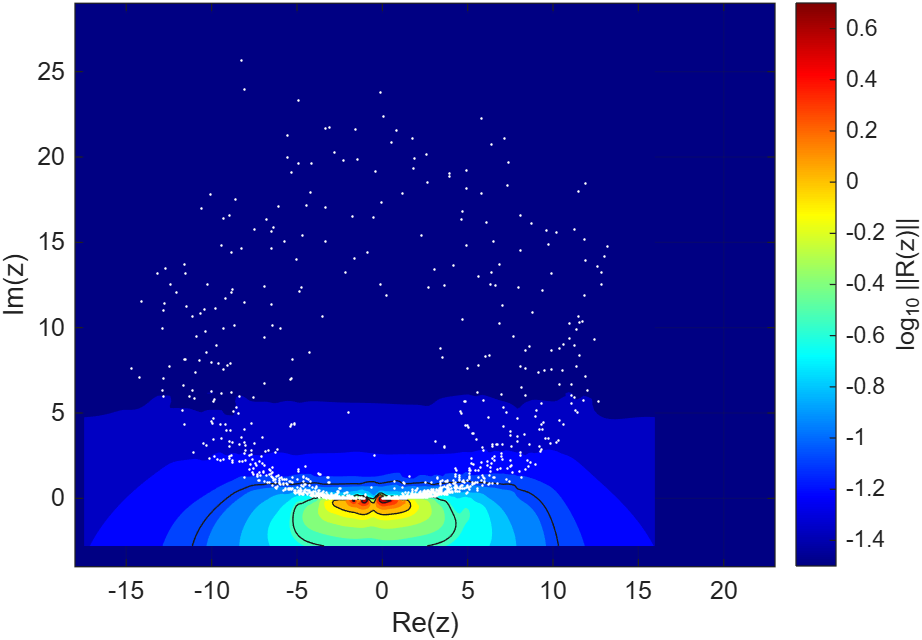} \\[2pt]
    \small zoom on resolvent bulge & \small zoom on resolvent bulge \\
    \includegraphics[width=0.46\linewidth]{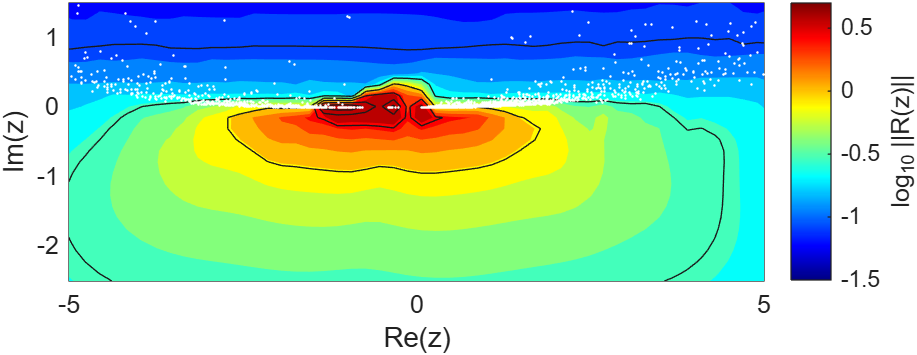} &
    \includegraphics[width=0.46\linewidth]{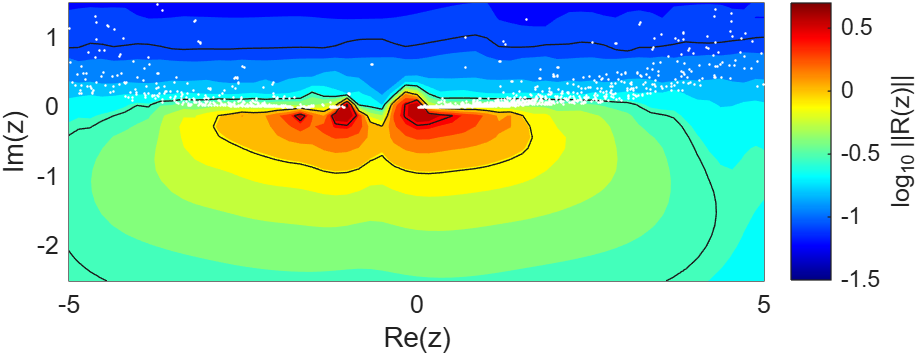} \\
  \end{tabular}
  \caption{Pseudospectrum $\log_{10}\|(zI-G)^{-1}\|_2$ of the TM
    Green's matrix at $N{=}3500$ for two representative patterns, on a
    $200{\times}200$ complex-plane grid with four inverse-iteration
    cycles per point.  Jet fill clamped to $[-1.5, 0.7]$; white dots:
    eigenvalues $\sigma(G)$; black lines: nested
    $\varepsilon$-pseudospectrum contours at
    $\log_{10}\|R\|\in\{-1,-0.5,0,0.5,1,1.5\}$.  \textbf{Top row:} full
    extent.  \textbf{Bottom row:} zoom onto the resolvent bulge
    (Re~$\in[-5,5]$, Im~$\in[-2.5,1.5]$), where the pole structure
    saturates to deep red.  Coprime shows a single tight pole peak;
    Thue--Morse shows the two-lobe peak set by its $C_8$ symmetric
    mode pair.  In both patterns the resolvent is ill-conditioned over
    a region extending well below the eigenvalue arc into the lower
    half-plane.}
  \label{fig:pseudospectra-g100}
\end{figure}

	The coupled-dipole Green's matrices considered here are
	\textbf{non-normal} ($\mathbf{G}\mathbf{G}^\dagger \neq \mathbf{G}^\dagger\mathbf{G}$;
	Henrici departure $\Delta_H\approx 0.78$,
	Table~\ref{tab:indicators}). In such systems, standard eigenvalue
	analysis can misrepresent stability and localization under
	perturbations. Instead, we use the $\epsilon$-\textbf{pseudospectrum}
	$\sigma_\epsilon(\mathbf{G})$, defined as:
	\begin{equation}
		\sigma_\epsilon(\mathbf{G}) = \{ z \in \mathbb{C} \mid \Vert{}(z\mathbb{I} - \mathbf{G})^{-1}\Vert{}_2 > \frac{1}{\epsilon} \}
	\end{equation}
	A perturbation of operator norm $\varepsilon$ can shift
	$\lambda_n$ by about $\kappa_n\varepsilon$, as given by the
	Bauer--Fike theorem~\cite{bauer1960norms,trefethen2005spectra};
	in the strongly non-normal regime this can be much larger than
	the perturbation itself, which is the basic content of the
	pseudospectrum picture.

Figure~\ref{fig:pseudospectra-g100} illustrates this for two
representative TM Green's matrices at $N{=}3500$: an ordered pattern
(Coprime) and a deterministic-aperiodic pattern (Thue--Morse).  Each
panel plots $\log_{10}\|(zI-G)^{-1}\|_2$ on a $200{\times}200$ grid in
the complex plane using the Schur-plus-inverse-iteration method of
Trefethen--Embree~\cite{trefethen2005spectra} (triangular solves make
the per-grid-point cost $O(N^2)$ rather than the $O(N^3)$ of a full
SVD), with eigenvalues $\sigma(G)$ overlaid as white dots and six
auto-spaced nested $\varepsilon$-pseudospectrum contour lines marking
successive levels of resolvent ill-conditioning.  In both cases the
hot spot of the resolvent extends well \emph{below} the eigenvalue
crescent into the lower half-plane.

\section{Extended validation gallery -- TE polarization}
\label{app:validation}

\begin{figure}[!ht]
    \centering
    \noindent
    \begin{tabular}{@{}c@{\hspace{2pt}}ccc@{}}
& \small point cloud & \small Herm landscape + watershed & \small top-100 modes on watershed \\
\rlbl{Bernoulli}          & \pnl{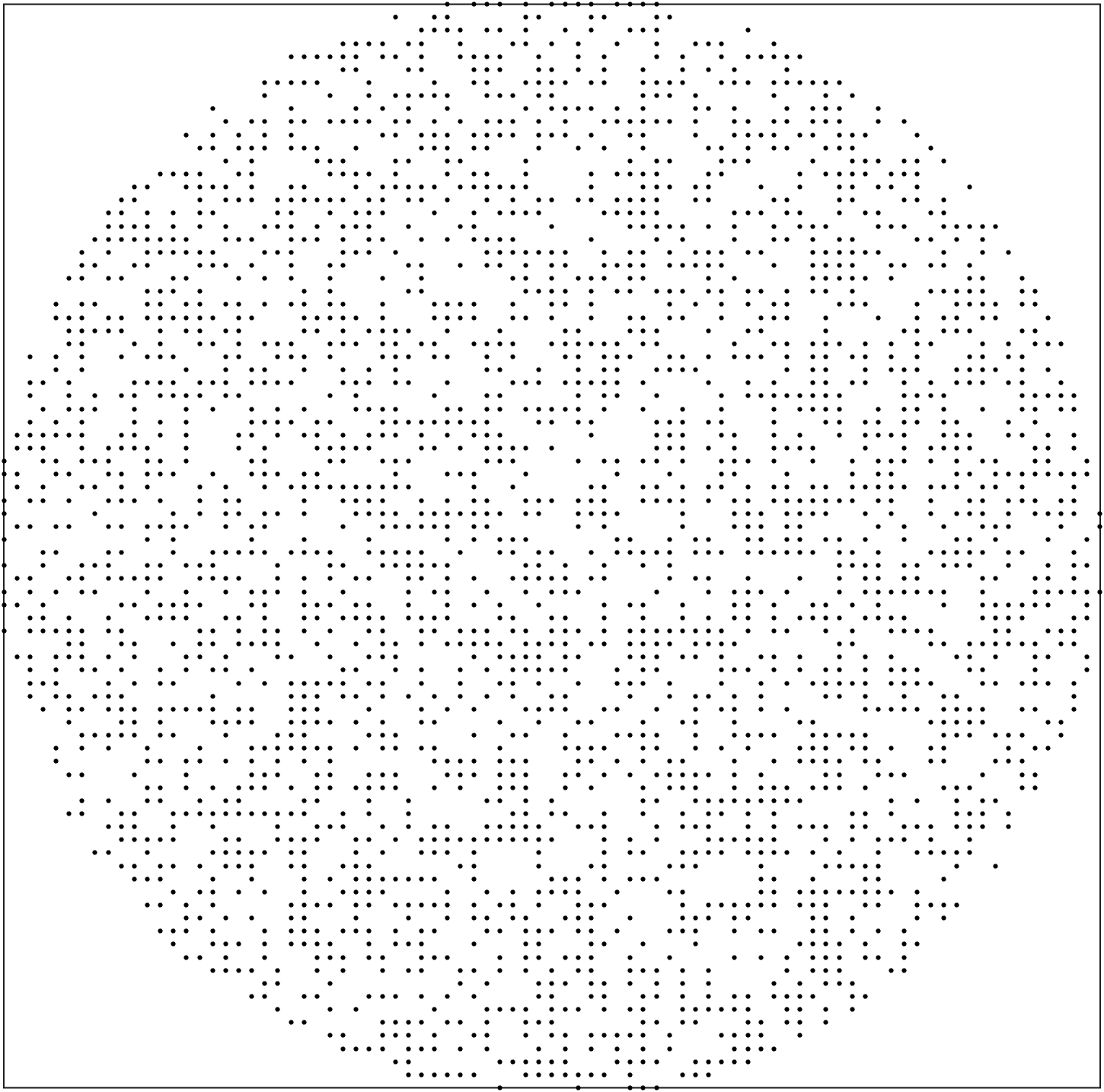}       & \pnl{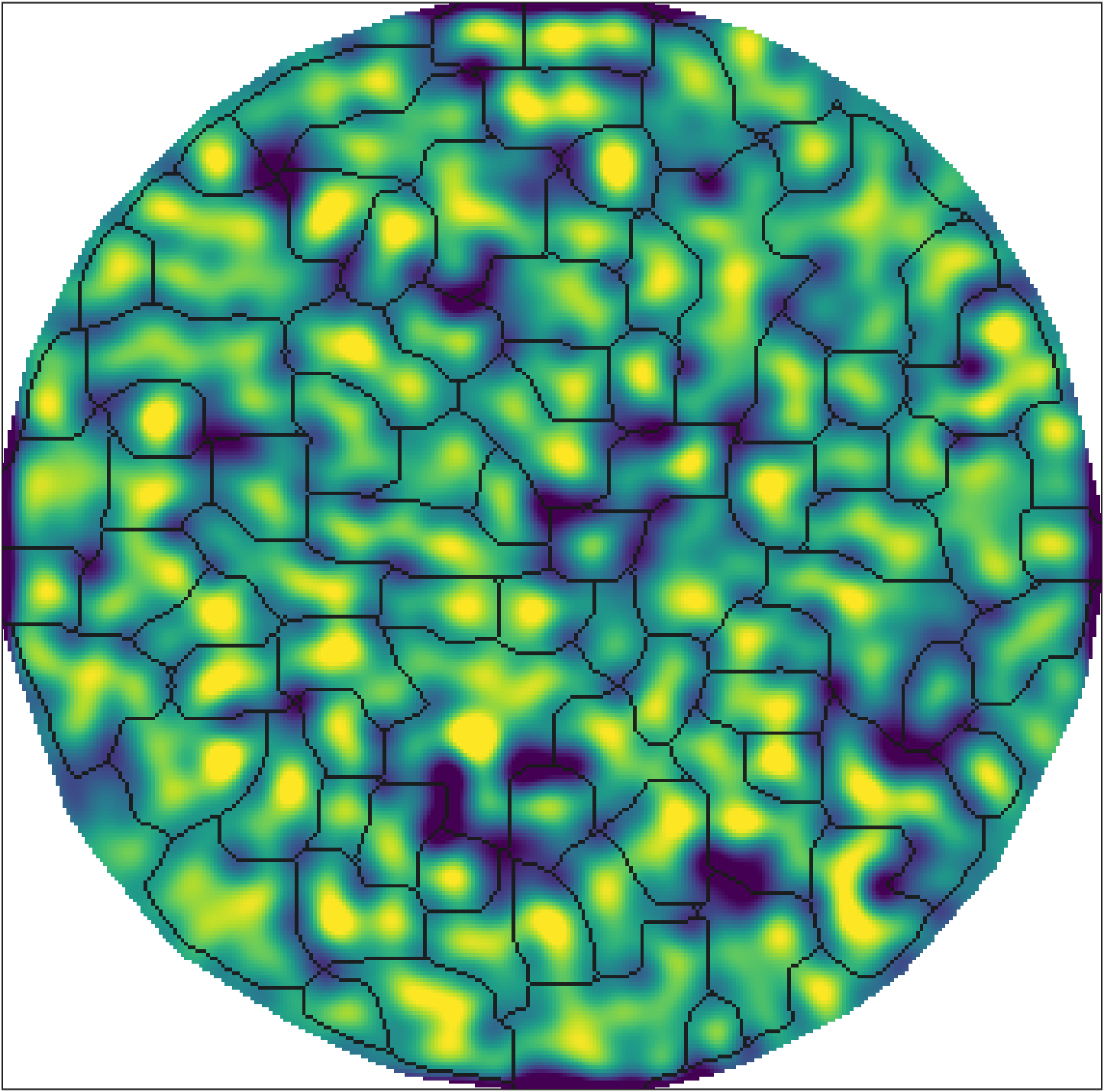}       & \pnl{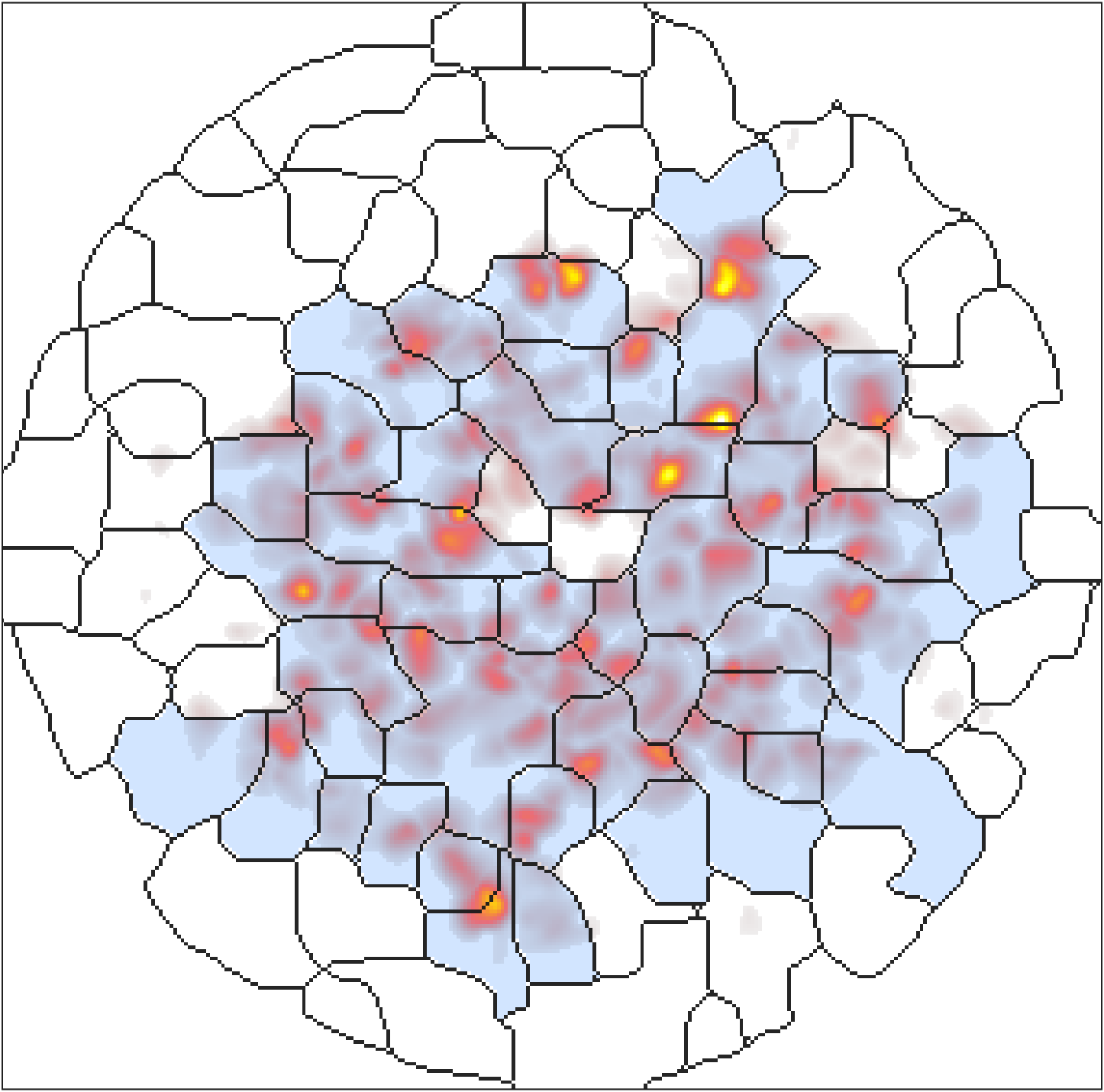}       \\
\rlbl{Coprime}             & \pnl{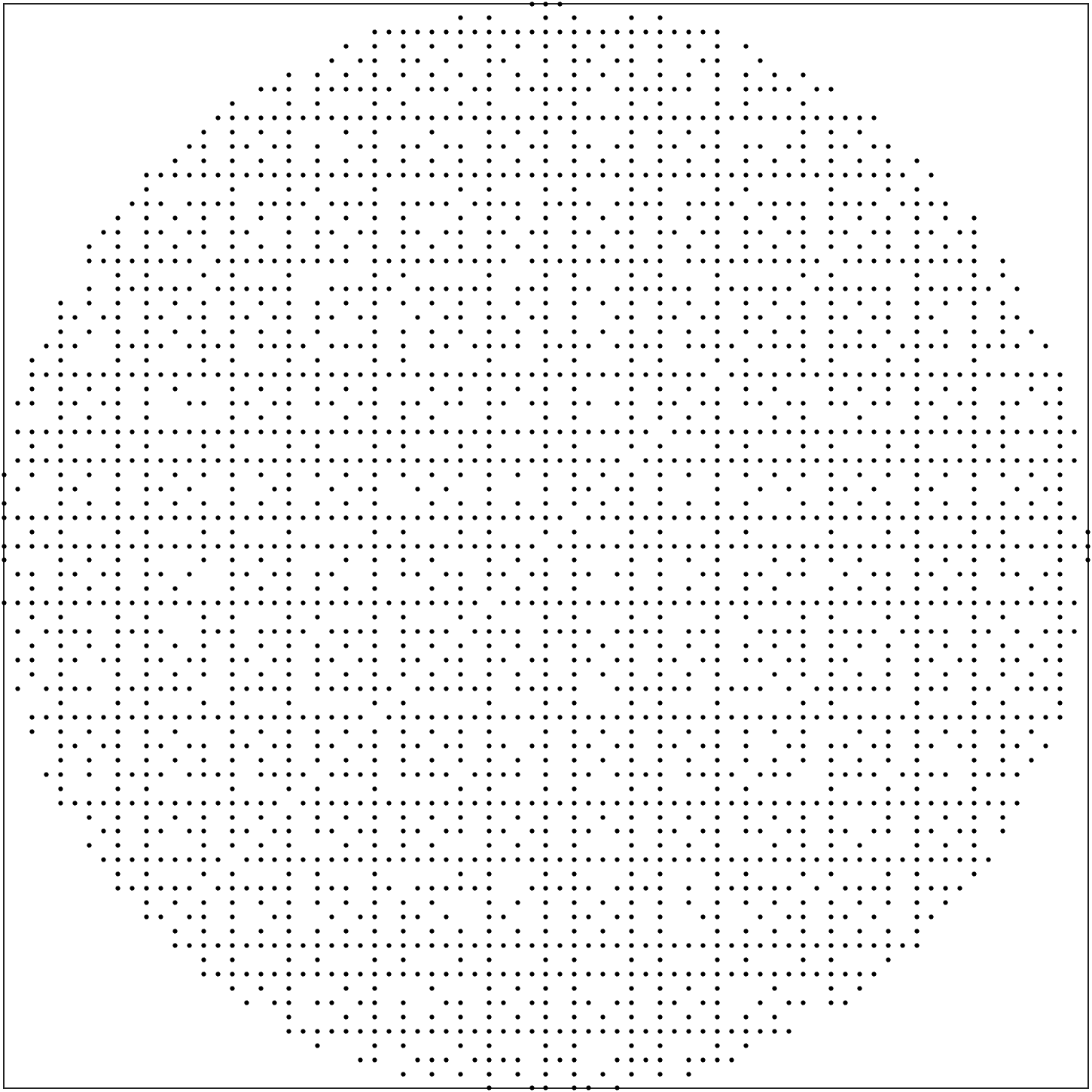}          & \pnl{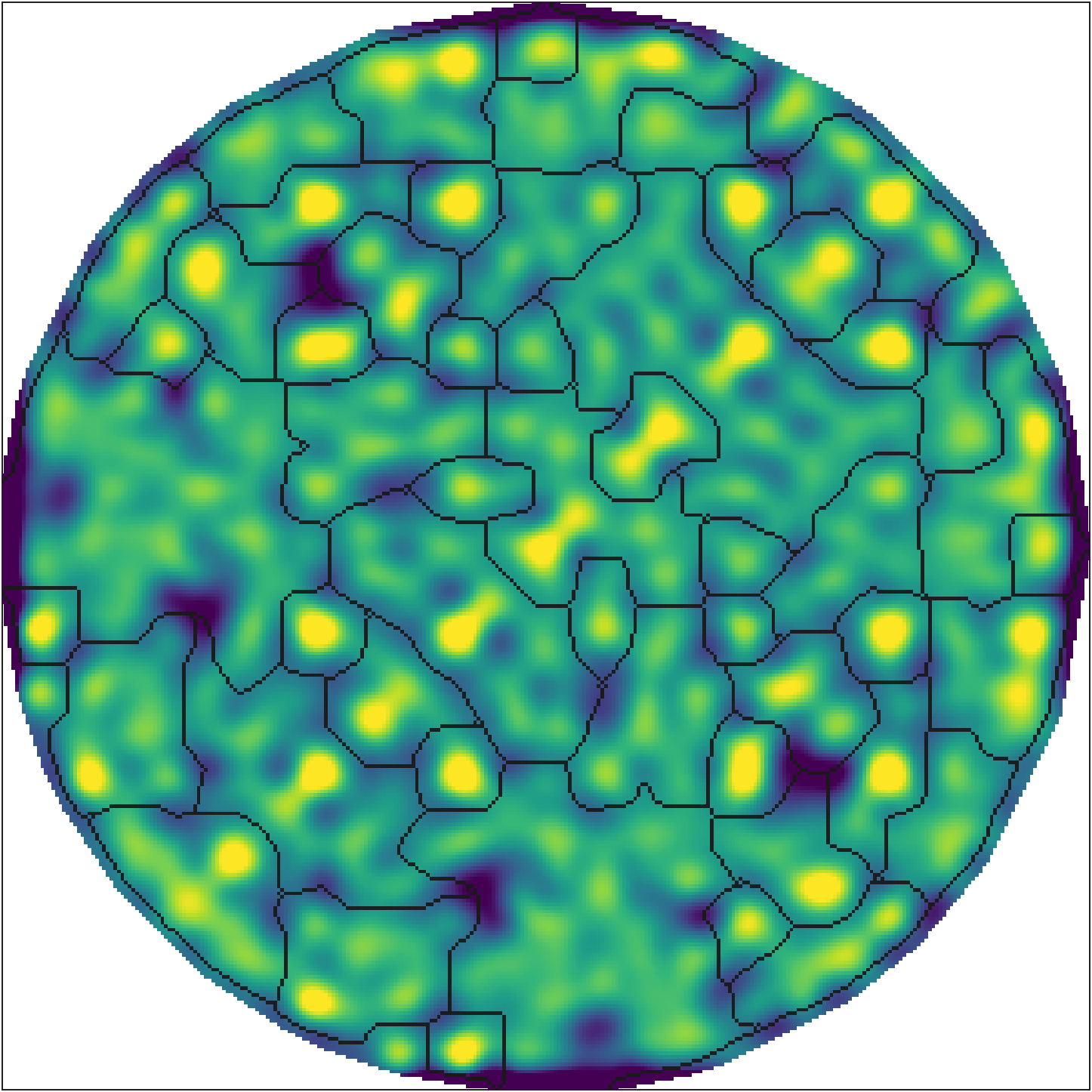}          & \pnl{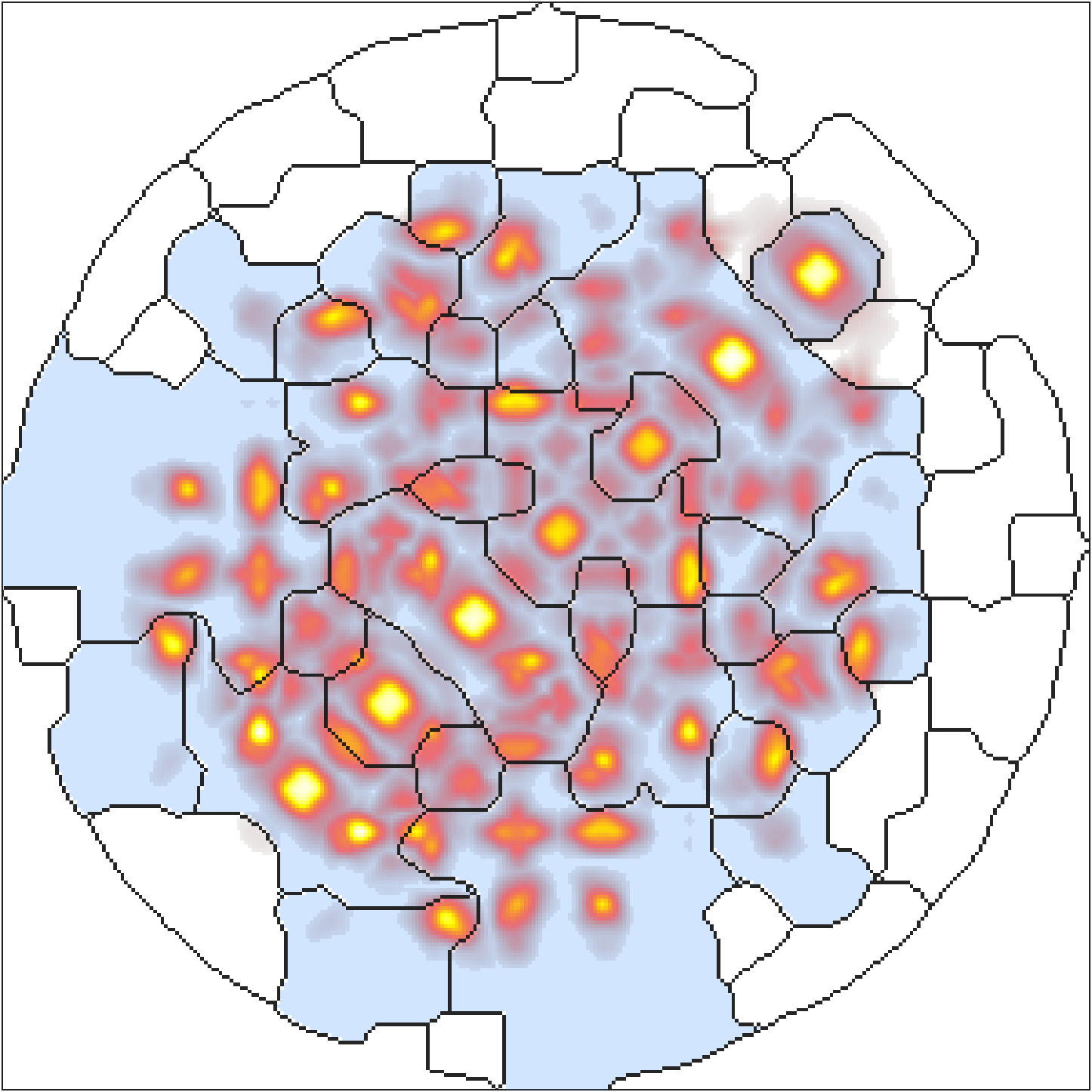}          \\
\rlbl{Eisenstein Prime}   & \pnl{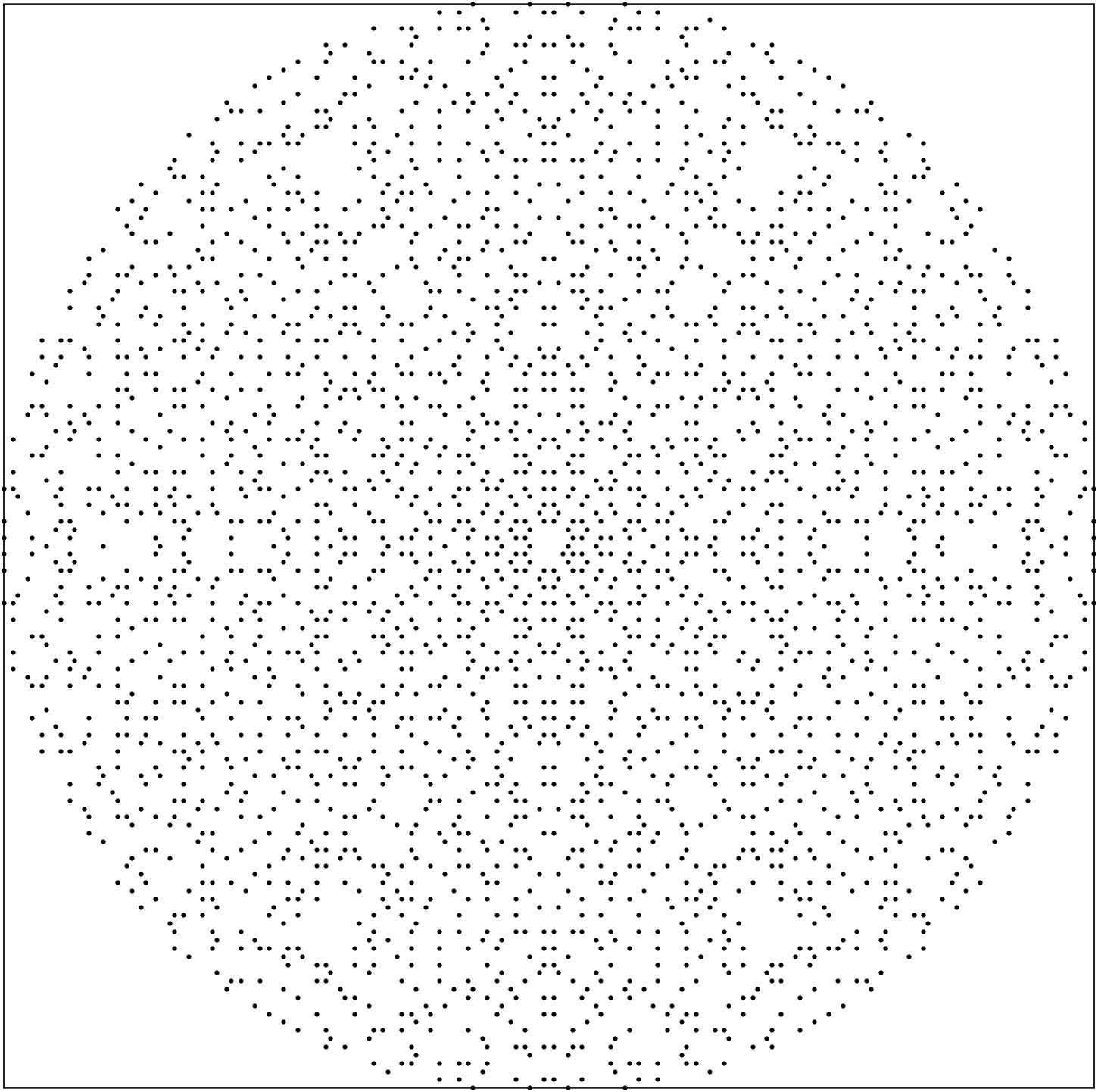} & \pnl{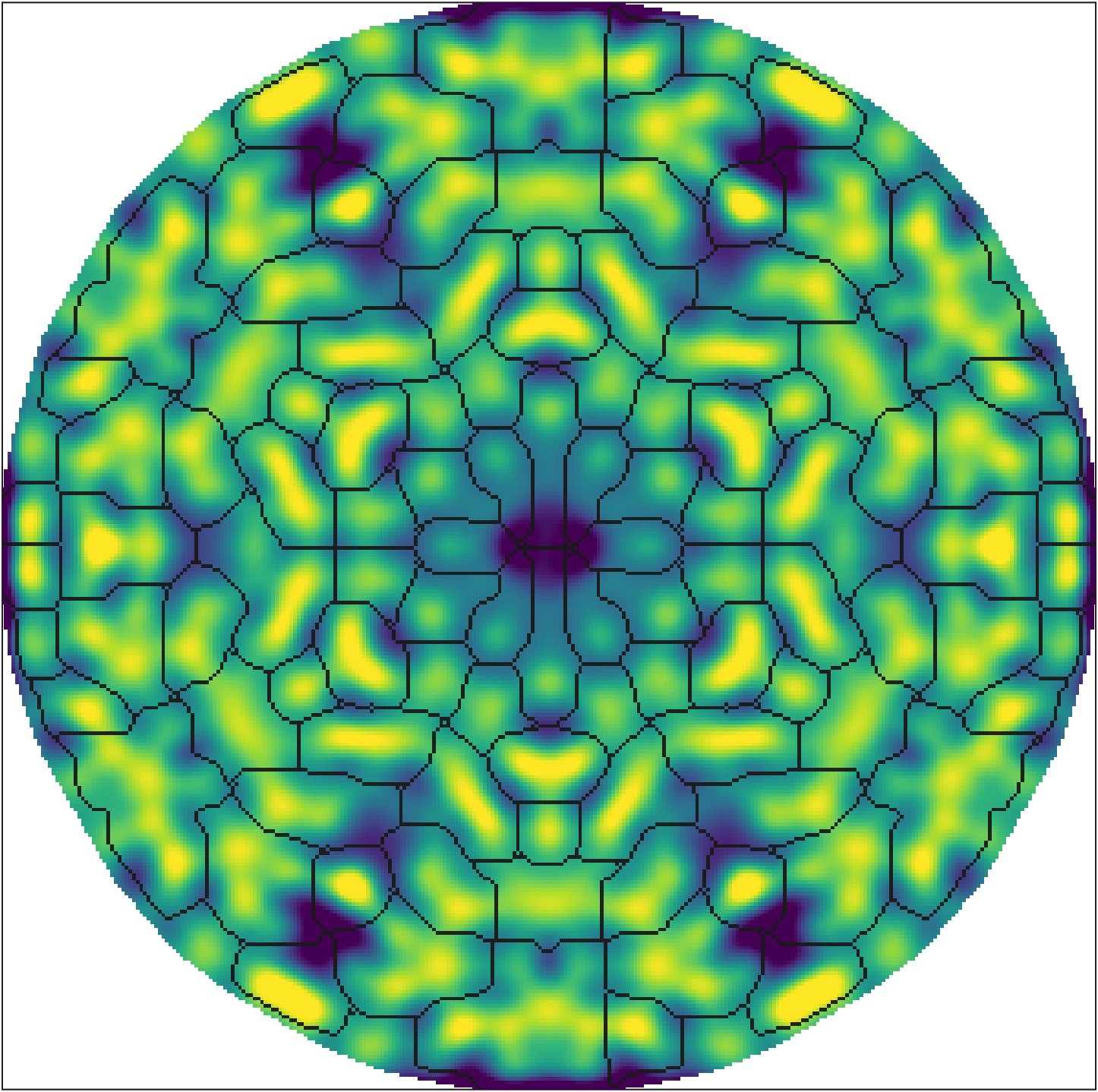} & \pnl{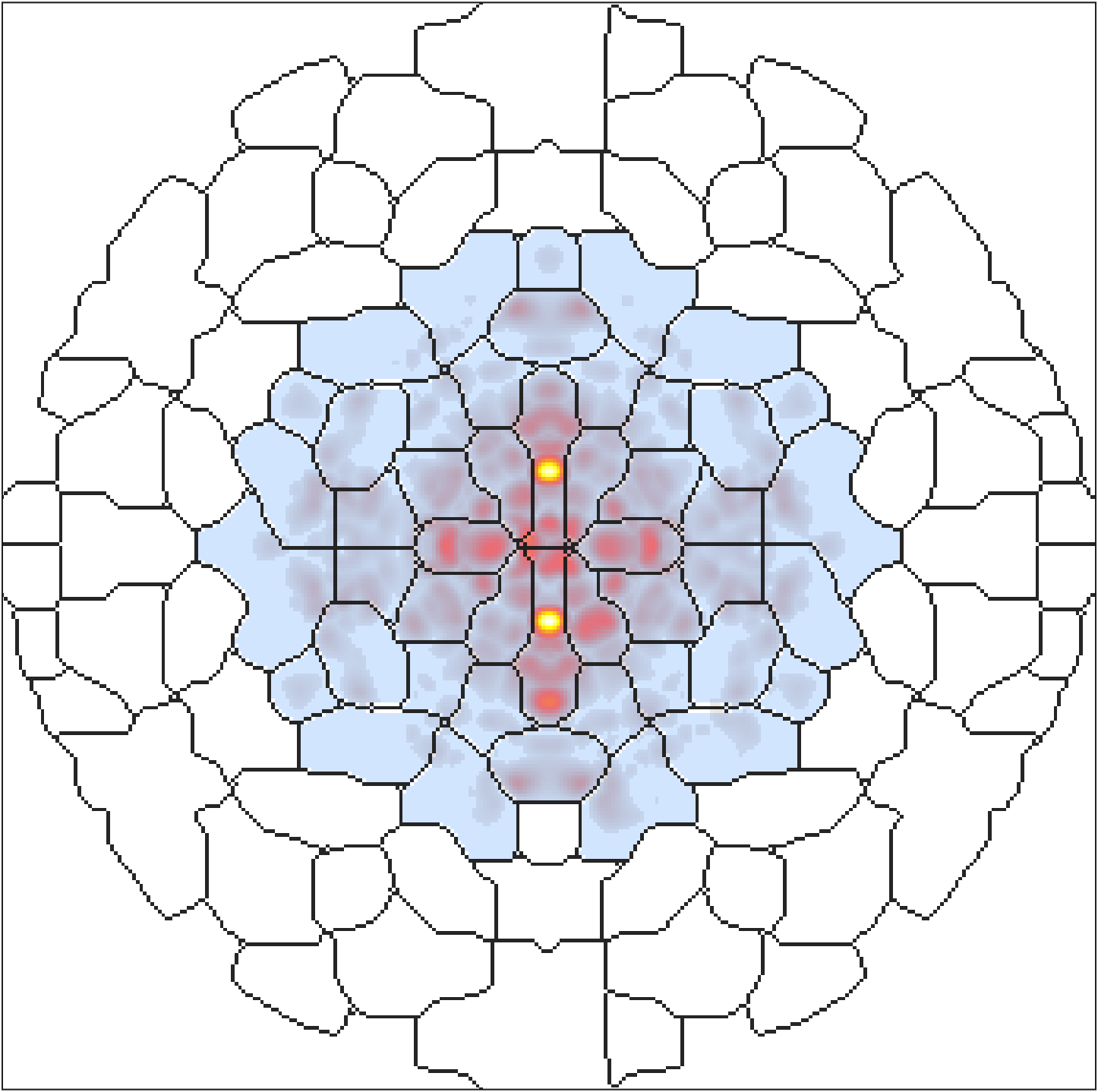} \\
\rlbl{Halton}          & \pnl{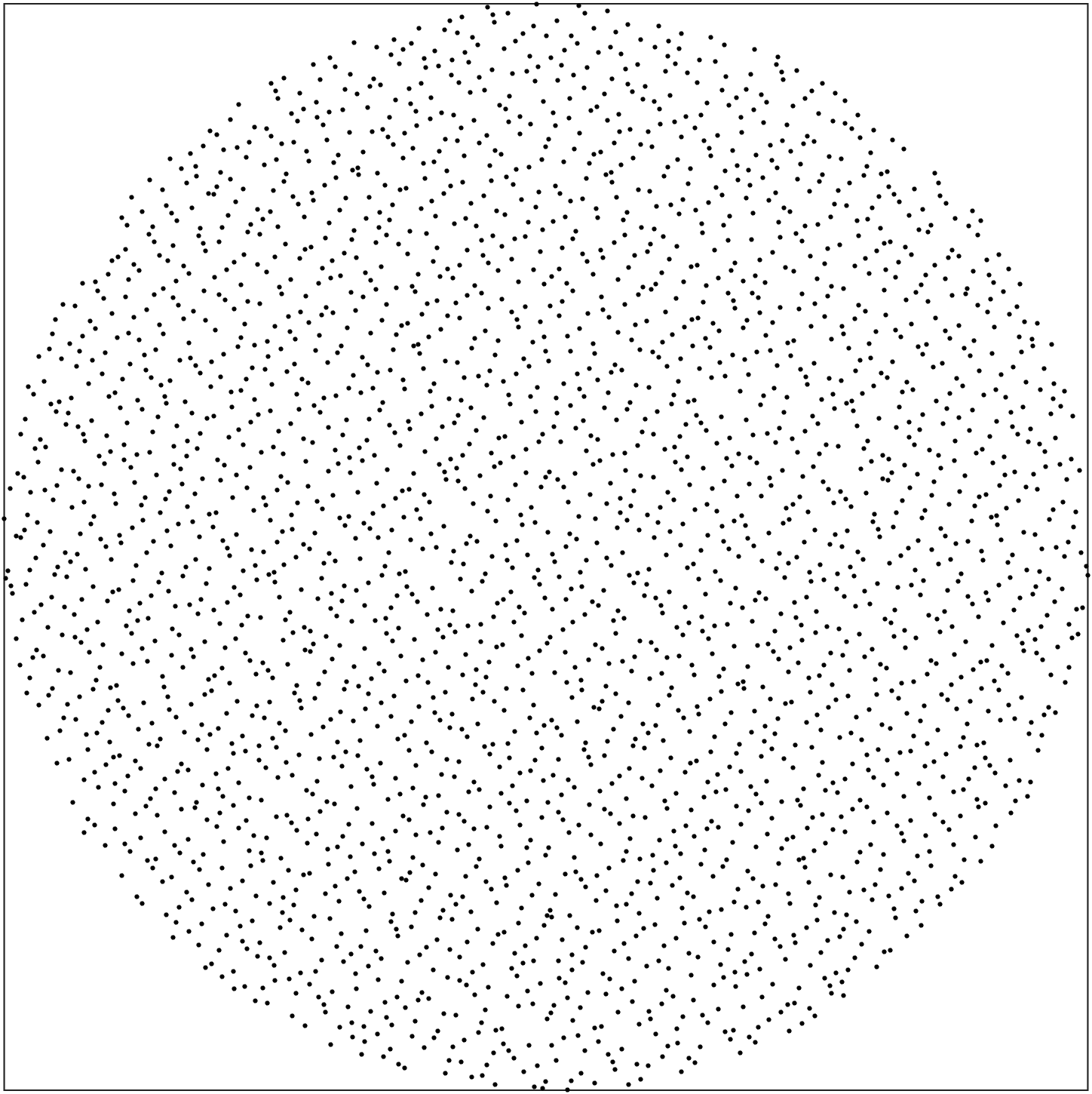}          & \pnl{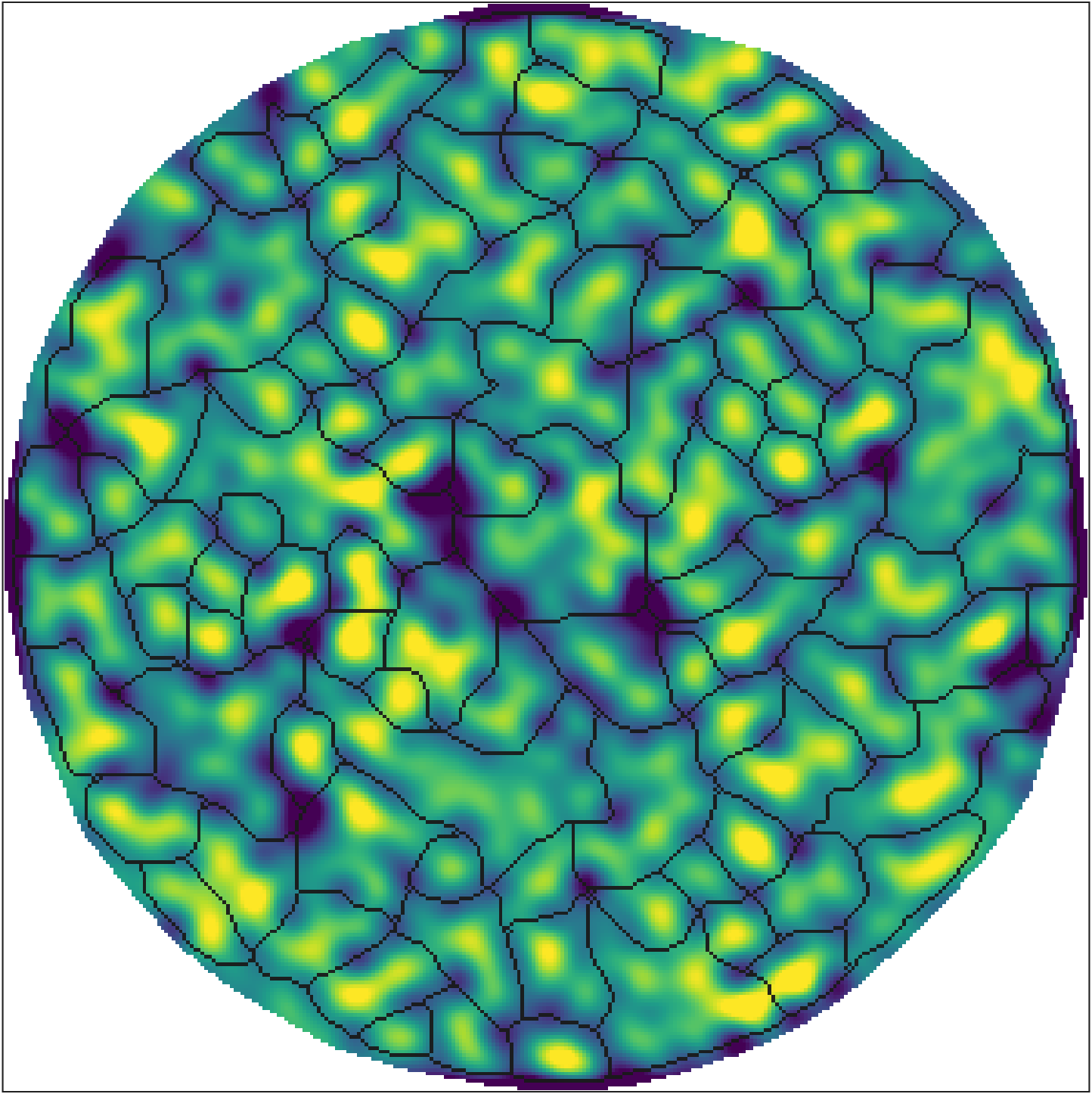}          & \pnl{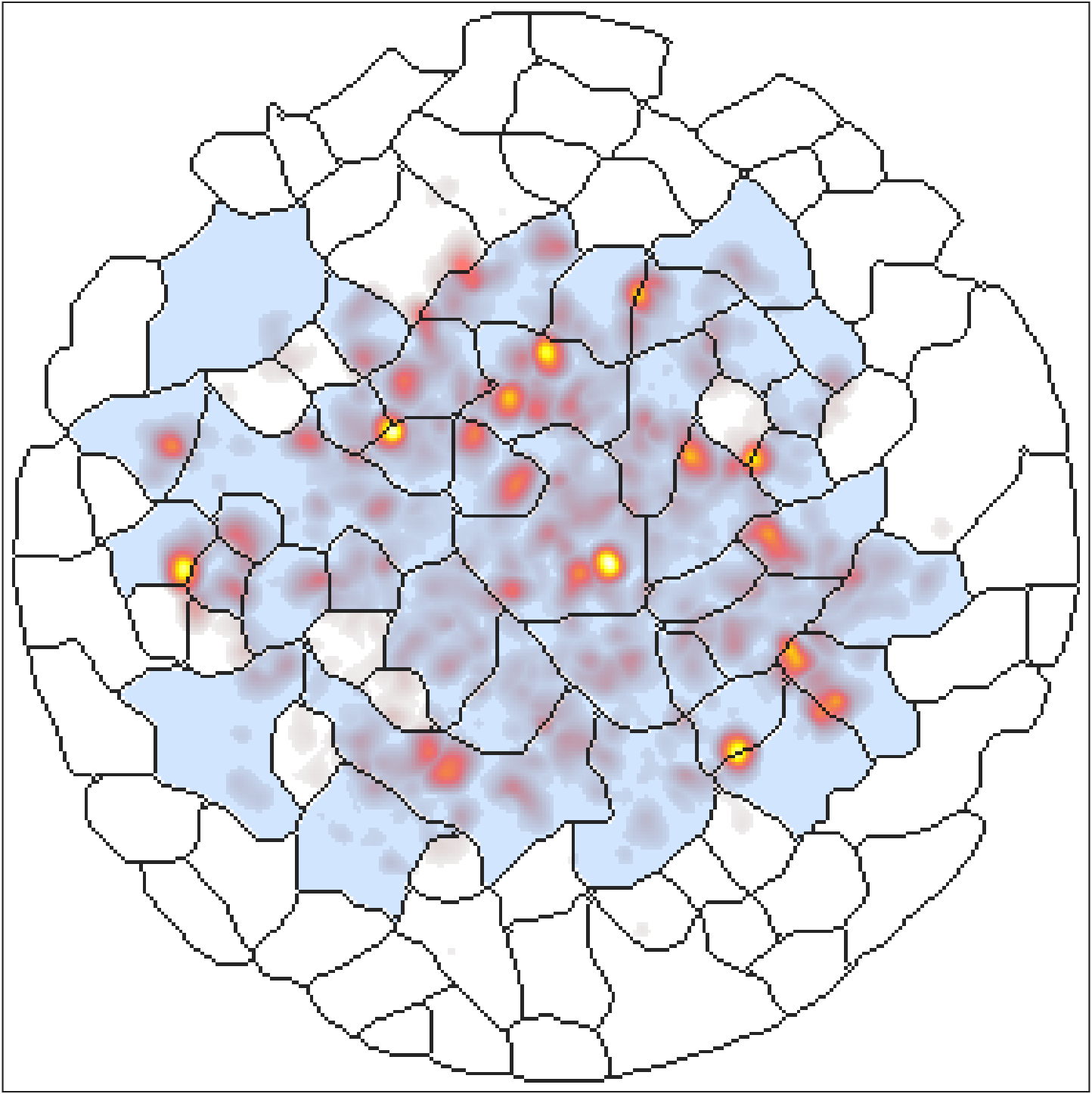}          \\
\rlbl{Rudin-Shapiro}          & \pnl{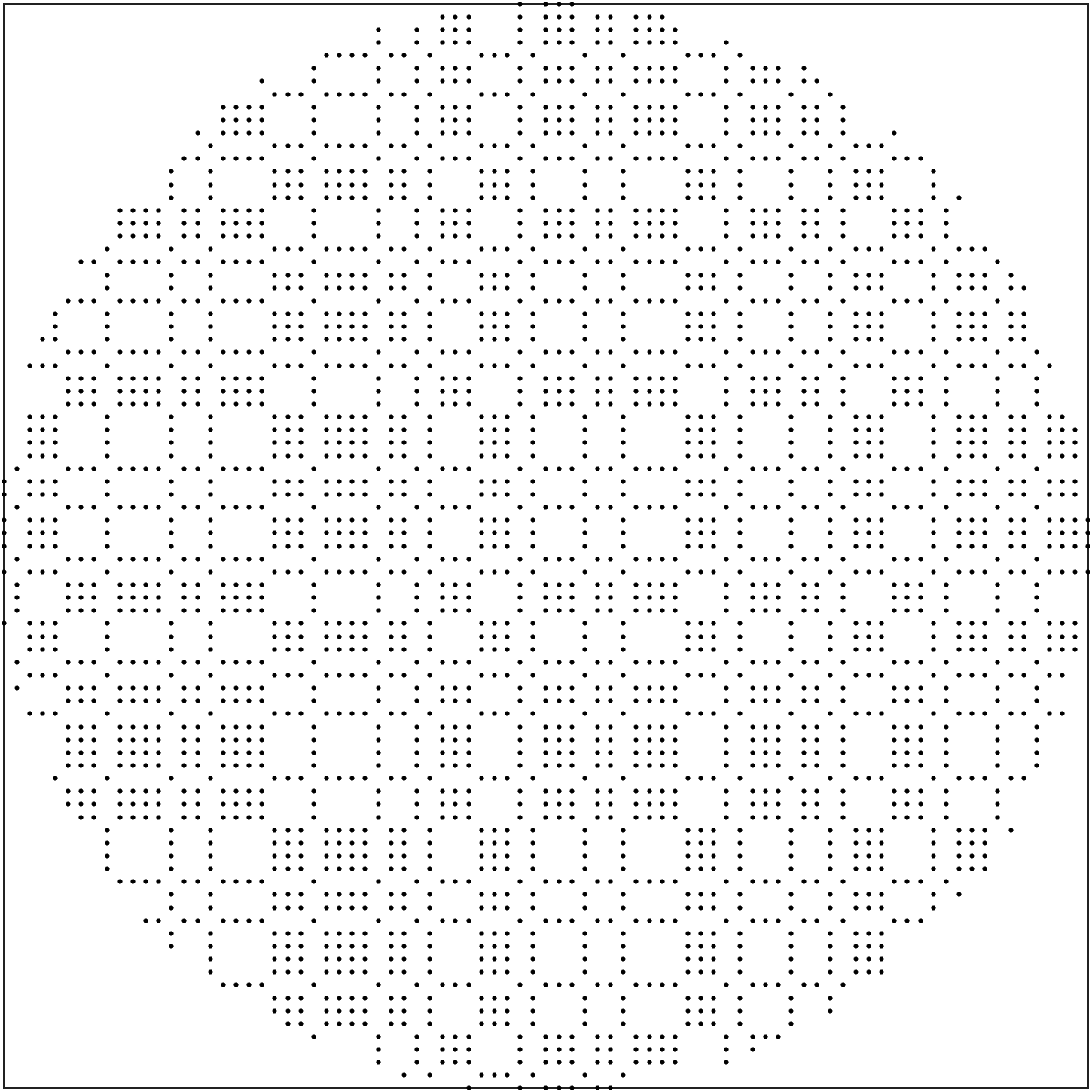}       & \pnl{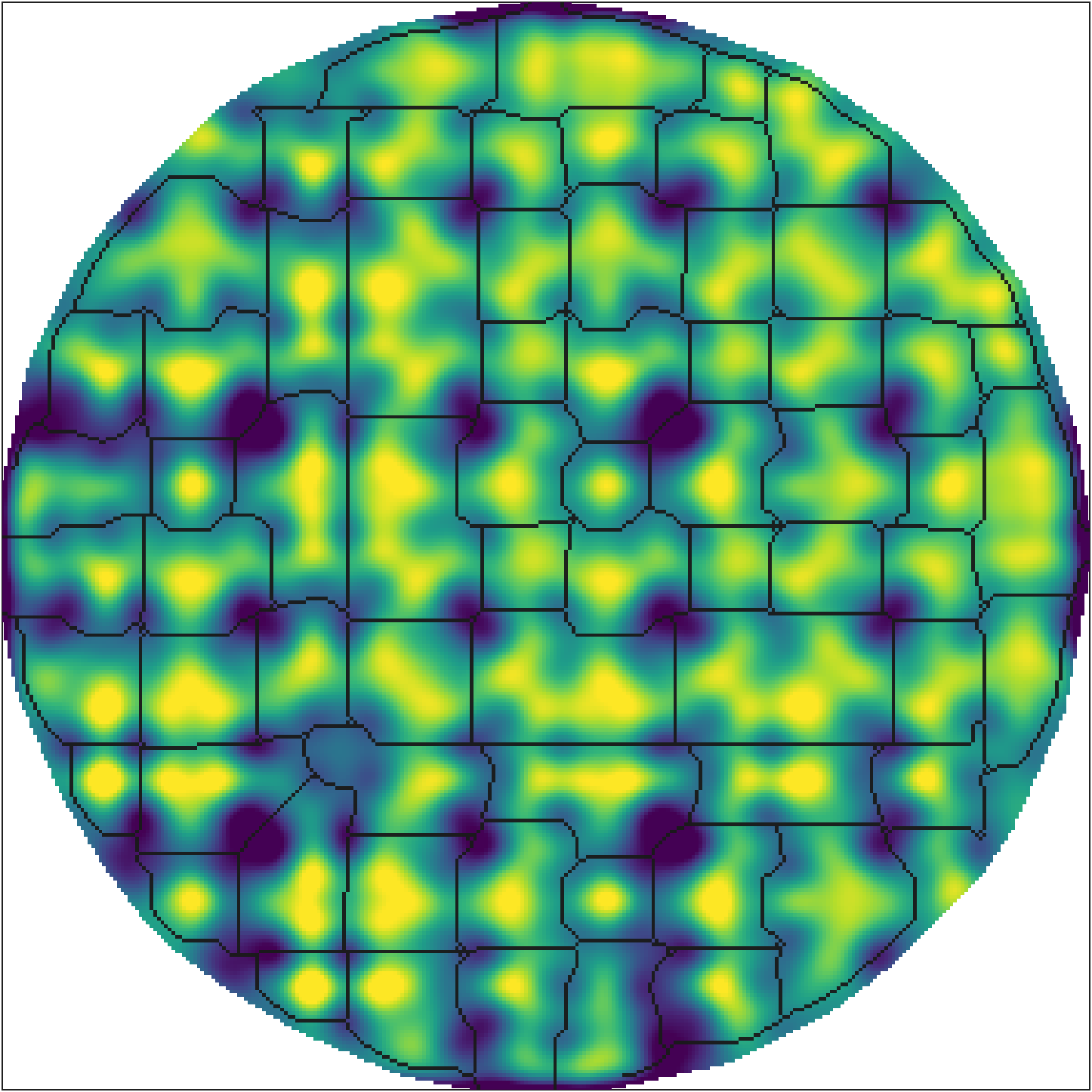}       & \pnl{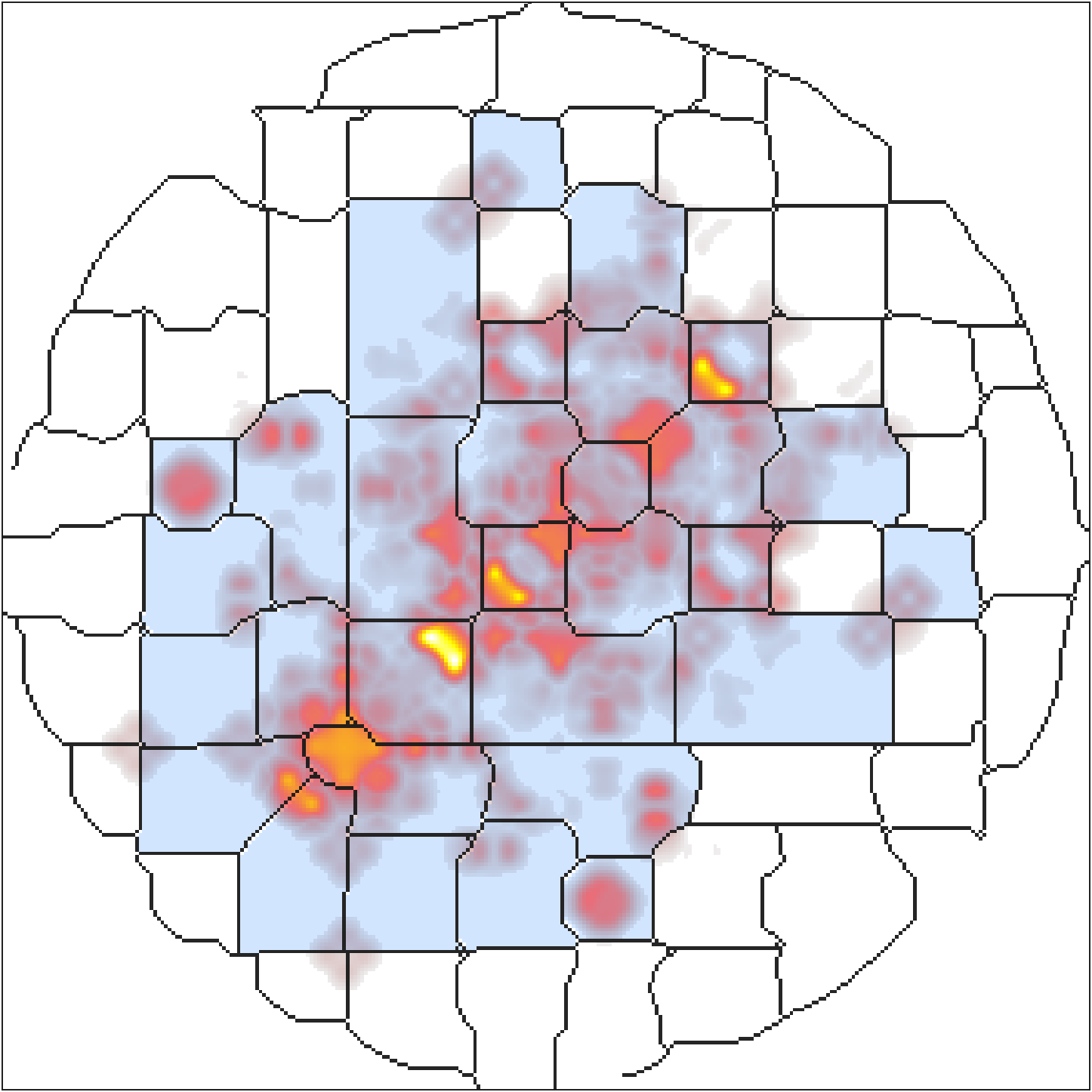}       \\
\end{tabular}
    \caption{Point patterns vs Herm landscape and top-100 mode localization (TE, $N{=}2800$).}
    \label{fig:validation-grid2}
\end{figure}

Figure~\ref{fig:validation-grid2} extends the TM validation gallery across all the structures under the TE polarization, displaying the point patterns, Herm landscape with watershed, and top-100 modes for Bernoulli, Coprime, Eisenstein prime, Halton, and Rudin--Shapiro dipole arrays at $N=2800$. In all panels, the mode envelope remains within the union of basins predicted by the Herm landscape. This qualitative agreement confirms that the polarization-isotropic source adjustment (Appendix~\ref{app:tepol}) successfully reproduces TM results in TE.


\bibliography{references}

\end{document}